\pdfoutput=1
\documentclass[floats,floatfix,showpacs,amssymb,prd,onecolumn,superscriptaddress,nofootinbib]{revtex4-2}

\usepackage{amsfonts,amsmath,amssymb,ascmac,bm,tensor}
\usepackage{fnpct} 
\usepackage{comment}
\usepackage{ifpdf}
\usepackage{slashed}
\usepackage{esint}
\usepackage{color}
\usepackage[mathscr]{eucal}
\usepackage[utf8]{inputenc}
\usepackage{physics}
\usepackage{cancel}
\usepackage{soul}
\usepackage{simpler-wick}

\ifpdf
  \usepackage{graphicx}     %   usepackage without driver option
  \usepackage[bookmarksopen,colorlinks=true,linkcolor=bblue,citecolor=bblue,urlcolor=ppink]{hyperref}
\else     % For (p)LaTeX + dvipdfmx
\fi

\hypersetup{
           breaklinks=true,   % splits links across lines
           colorlinks=true,   % displays links as colored text instead of blocks
           linkcolor=bblue,
           citecolor=bblue,
           urlcolor=ppink
        }

\definecolor{red}{rgb}{1,0,0}
\definecolor{darkred}{rgb}{0.6,0,0}
\definecolor{darkgreen}{rgb}{0.992447,0.623778,0.034597}
\definecolor{ppink}{rgb}{1,0.4,0.4} 
\definecolor{bblue}{rgb}{0.284602,0.317763,0.963947}
\definecolor{purple}{rgb}{0.5 ,0, 0.7}

\newcommand{\zo}{{(0)}}
\newcommand{\fo}{{(1)}}
\renewcommand{\so}{{(2)}}
\newcommand{\tho}{{(3)}}
\newcommand{\foo}{{(4)}}
\newcommand{\no}{{(n)}}

\newcommand{\sw}{\text{sw}}

\newcommand{\tre}{\text{tr} }
\newcommand{\vx}{\text{vx}}

\newcommand{\inte}{\text{int}}

\newcommand{\tmin}{\text{min}}

\newcommand{\ee}{\text{e}}
\newcommand{\tad}{\text{tad}}

\newcommand{\uv}{\text{UV}}
\newcommand{\ir}{\text{IR}}
\renewcommand{\min}{\text{min}}
\newcommand{\rr}{\text{r}}

\newcommand{\ini}{\text{ini}}

\newcommand{\Ci}{\text{Ci}}

\newcommand{\bfk}{\mathbf{k}}

\newcommand{\bfq}{\mathbf{q}}
\newcommand{\bfx}{\mathbf{x}}
\newcommand{\bfy}{\mathbf{y}}

\renewcommand{\Re}{\text{Re}}
\renewcommand{\Im}{\text{Im}}

\makeatletter
\newcommand\footnoteref[1]{\protected@xdef\@thefnmark{\ref{#1}}\@footnotemark}
\makeatother

\allowdisplaybreaks[1]

\begin{document}

%%%%%%%%%%%%%%%%%%%%%%%%%%%
%%%%%%%%%%% Title %%%%%%%%%%%
%%%%%%%%%%%%%%%%%%%%%%%%%%%

\title{
Conservation of superhorizon curvature perturbations at one loop:\\
Backreaction in the in-in formalism and Renormalization
}

\author{Keisuke Inomata}
\affiliation{William H. Miller III Department of Physics and Astronomy, Johns Hopkins University, 3400 N. Charles Street, Baltimore, Maryland, 21218, USA}

\begin{abstract} 
\noindent
We show that the superhorizon-limit curvature perturbations are conserved at one-loop level in single-field inflation models with a transient non-slow-roll period.
We take the spatially-flat gauge, where the backreaction plays a crucial role for the conservation of superhorizon curvature perturbations unless the counter terms are tuned.
We calculate the backreaction with the in-in formalism.
In addition, we explicitly show the renormalization of the UV divergences with the counter terms.
\end{abstract}

\date{\today}
\maketitle

\tableofcontents

%%%%%%%%%%%%%%%%%%%%%%%%%%%%%%%%
\section{Introduction}
%%%%%%%%%%%%%%%%%%%%%%%%%%%%%%%%

Cosmological perturbations serve as crucial imprints of the early Universe.
They arise from quantum fluctuations during inflation and eventually give rise to the cosmic microwave background (CMB) anisotropies and the large-scale structure (LSS), both of which have been observed for decades.
The cosmological perturbations are often characterized by the curvature perturbations because their amplitudes are thought to become constant (conserved) after they exit the horizon in single-field inflation models.
The conservation of the curvature perturbations enables us to access the information at the time when the perturbations exit the horizon during inflation. 
The conservation of the curvature perturbations are shown not only at linear level~\cite{Bardeen:1980kt,Kodama:1984ziu}, but also at one-loop (lowest-order nonlinear) level in single-clock inflation~\cite{Senatore:2009cf,Senatore:2012nq,Pimentel:2012tw}.
Furthermore, the curvature conservation has also been shown non-perturbatively with the separate universe assumption~\cite{Lyth:2004gb}.

Contrary to the previous studies, the conservation of curvature perturbations at one-loop level has been recently questioned in the context of inflation models for primordial black holes (PBHs), which are the candidate of dark matter~\cite{Chapline:1975ojl,Ivanov:1994pa,Yokoyama:1995ex,GarciaBellido:1996qt,Afshordi:2003zb,Frampton:2010sw,Belotsky:2014kca,Carr:2016drx,Inomata:2017okj,Espinosa:2017sgp} and/or the BHs detected by LIGO-Virgo-KAGRA collaborations~\cite{Bird:2016dcv,Clesse:2016vqa,Sasaki:2016jop,Eroshenko:2016hmn,Sasaki:2018dmp,Carr:2020gox,Green:2020jor,Escriva:2022duf}.
In Ref.~\cite{Kristiano:2022maq}, it is claimed that the curvature perturbations are not conserved at one-loop level even on superhorizon limit in a single-field inflation model if it has the transition of slow-roll (SR) period $\to$ ultra slow-roll (USR) period $\to$ SR period.\footnote{See also Ref.~\cite{Cheng:2021lif} for an earlier work that claims the non-conservation of the superhorizon-limit curvature perturbations in single-field models with the use of the Hartree factorization.}
USR period is different from SR period in that the inflaton background evolution is not dominated by the potential gradient, instead is dominated by the Hubble friction~\cite{Kinney:1997ne,Inoue:2001zt,Kinney:2005vj,Martin:2012pe}.
USR period is often considered in inflation models for PBHs because the curvature perturbations are enhanced on small scales during USR period. 
In Ref.~\cite{Kristiano:2022maq}, they claim that the enhanced perturbations on small scales give scale-invariant corrections to the superhorizon curvature power spectrum at one-loop level if the tree-level curvature power spectrum is scale-invariant on superhorizon. 
The claim means that, if we consider the large enhancement of the small-scale perturbations with USR in the context of PBHs, the perturbations on CMB and LSS scales are modified by the enhanced perturbations on small scales.
If this is true, we can constrain the PBH models by imposing the validity of the perturbation theory on the CMB and LSS scales.

This claim has been investigated in Refs.~\cite{Riotto:2023hoz,Choudhury:2023vuj,Kristiano:2023scm,Riotto:2023gpm,Firouzjahi:2023aum,Motohashi:2023syh,Firouzjahi:2023ahg,Franciolini:2023agm,Tasinato:2023ukp,Cheng:2023ikq,Fumagalli:2023hpa,Maity:2023qzw,Tada:2023rgp,Firouzjahi:2023bkt,Davies:2023hhn,Iacconi:2023ggt,Saburov:2024und,Ballesteros:2024zdp,Kristiano:2024vst,Kristiano:2024ngc,Kawaguchi:2024rsv,Fumagalli:2024jzz,Sheikhahmadi:2024peu,Frolovsky:2025qre}.
The scale-invariant one-loop corrections from small-scale perturbations mean the non-conservation of the superhorizon curvature perturbations at one-loop level.
This claim seems inconsistent with the separate universe picture: a sufficiently large-scale region, which is at least larger than the horizon scale, behaves as a (locally) homogeneous and isotropic universe and evolves independently of other regions~\cite{Sasaki:1998ug,Wands:2000dp,Lyth:2003im}.
In fact, the claim of the scale invariance of the one-loop power spectrum has been refuted in Refs.~\cite{Fumagalli:2023hpa,Tada:2023rgp} with cubic interactions considered (though criticized in Ref.~\cite{Firouzjahi:2023bkt}) and in Refs.~\cite{Inomata:2024lud,Kawaguchi:2024rsv,Fumagalli:2024jzz} with both cubic and quartic interactions considered.

To be more specific about the scale invariance of the one-loop power spectrum, let us suppose that the USR period occurs only for $\eta_i < \eta < \eta_e$ with $\eta$ being the conformal time and the other periods are SR.
Then, focusing on the loop correction from small-scale perturbations with $k > k_\ir$ to large-scale perturbations, we express the one-loop correction of the superhorizon-limit curvature power spectrum $\mathcal P_{\zeta,1\text{-}\text{loop}}$ in $\eta<\eta_i$ and $\eta > \eta_e$ as~\cite{Fumagalli:2023hpa}
\begin{align}
  \lim_{q\to 0}\mathcal P_{\zeta,1\text{-}\text{loop}}(q) = \begin{cases}
  0 \qquad& (\eta < \eta_i) \\
  c\, \mathcal P_{\zeta,\tre}(q) \displaystyle\int_{k_\ir} \dd \ln k\, \mathcal P_{\zeta,\tre}(k) \qquad& (\eta > \eta_e)
  \end{cases},
\end{align}
where $\mathcal P_{\zeta,\tre}$ is the tree-level power spectrum and $c$ is independent of $q$.
The claim of Ref.~\cite{Kristiano:2022maq} and many of the following works is that, if we consider a sharp transition between the USR and SR, $c \gtrsim 1$ can be realized even in the de Sitter (or decoupling) limit $\epsilon (\equiv -\dot H/H^2) \to 0$ with $H$ the Hubble parameter and the dot denoting the time derivative.
The sharp transition here means that the transition occurs within less than one e-fold.
On the other hand, the claim in Refs.~\cite{Fumagalli:2023hpa,Tada:2023rgp,Inomata:2024lud,Kawaguchi:2024rsv,Fumagalli:2024jzz} is that $c = 0$ at least in the de Sitter limit.
Namely, Refs.~\cite{Fumagalli:2023hpa,Tada:2023rgp,Inomata:2024lud,Kawaguchi:2024rsv,Fumagalli:2024jzz} show the conservation of superhorizon curvature perturbations at one-loop level in the de Sitter limit.

In this paper, we follow up on our previous work~\cite{Inomata:2024lud}, one of the papers that show the curvature conservation.
The distinct feature of Ref.~\cite{Inomata:2024lud} is that it uses the spatially-flat gauge, while the other papers use the comoving gauge to show the curvature conservation~\cite{Fumagalli:2023hpa,Tada:2023rgp,Kawaguchi:2024rsv,Fumagalli:2024jzz}.
The advantage of the spatially flat gauge is that the calculation becomes quite simple in the de Sitter limit thanks to the decoupling between the inflaton fluctuations and the metric perturbations~\cite{Baumann:2011su,Pajer:2016ieg}.
On the other hand, in the spatially-flat gauge, the curvature power spectrum is given by $\mathcal P_\zeta = H^2 \mathcal P_{\delta \phi}/\langle \dot\phi \rangle^2$ during SR period with $\delta \phi$ the inflaton fluctuation and $\langle\dot{\phi} \rangle^2$ the square of the inflaton background velocity. 
Due to this, we need to calculate the one-loop corrections to $\mathcal P_{\delta \phi}$ and the one-loop backreaction to $\langle \dot{\phi} \rangle$ separately to obtain the one-loop curvature power spectrum.
In particular, the backreaction to $\langle \dot{\phi} \rangle$ generally plays an important role for the curvature conservation~\cite{Inomata:2024lud}.

The main goal of this paper is to show the conservation of the superhorizon curvature power spectrum at one-loop level in detail, based on Ref.~\cite{Inomata:2024lud}.
We not only explicitly show the intermediate steps omitted in Ref.~\cite{Inomata:2024lud} but also perform two new analyses:
1) We calculate the backreaction with the in-in formalism. 
In our previous work~\cite{Inomata:2024lud}, we relied on the equation of motion for the background when we discuss the backreaction, while we used the in-in formalism for the one-loop calculation of $\mathcal P_{\delta \phi}$.
We will see that the same conclusion can be obtained with this new analysis, which reinforces the conclusion of our previous work. 
2) We explicitly include the counter terms in the calculation and perform the renormalization, briefly outlined in our previous work. 
We will explicitly see how the UV divergences are cancelled by the counter terms.

Throughout this work, we consider the case where the inflation has a transient non-slow-roll period: SR $\to$ non-SR $\to$ SR, similar to Ref.~\cite{Inomata:2024lud}.
We here stress that the non-SR period in this paper is still in the inflationary era with $\epsilon \ll 1$. We call this period non-SR in the sense that the SR condition is violated through $|\dot\epsilon/H\epsilon| \gtrsim \mathcal O(1)$ (not through $\epsilon \gtrsim \mathcal O(1)$) during that period.
The transient non-SR period can be anything, such as USR or the parametric resonance period with some oscillatory features in the potential~\cite{Inomata:2022yte}.
We assume that the enhancement of the curvature perturbations is realized by the non-SR period.
This is a typical single-field inflation model for the PBH scenarios.

This paper is organized as follows. 
In Sec.~\ref{sec:in_in}, we summarize the in-in formalism and discuss the evolution of the linear perturbations.
In Sec.~\ref{sec:one_loop_ps}, we show the concrete expressions of the one-loop power spectrum. 
In Sec.~\ref{sec:backreaction}, we calculate the backreaction with the in-in formalism. 
Then, we show the conservation of the curvature perturbations at one-loop level in Sec.~\ref{sec:curv_cons}. 
In Sec.~\ref{sec:renorm}, we explicitly show how the UV divergences from the loop integrals are removed by the counter terms.
We conclude our paper in Sec.~\ref{sec:concl}.

%%%%%%%%%%%%%%%%%%%%%%%%%%%%%%%%
\section{In-in formalism and linear perturbations}
\label{sec:in_in}
%%%%%%%%%%%%%%%%%%%%%%%%%%%%%%%%

Throughout this work, we take the de Sitter limit ($\epsilon \to 0$) and neglect the higher-order terms in $\epsilon$.\footnote{
  In general, we cannot exactly take $\epsilon =0$ when we discuss inflation, which means that $H$ is always time-dependent during inflation, though its time-dependence is suppressed by $\epsilon$.
  Taking the de-Sitter limit here practically means that we consider extremely small (but nonzero) $\epsilon$ and focus on the contributions that are not suppressed by $\epsilon$.
}
This enables us to neglect all the metric perturbations in our analysis~\cite{Baumann:2011su,Pajer:2016ieg}.
Then, the Lagrangian for canonical single-field inflation models can be expressed as\footnote{Unlike in the quantum field theory in Minkowski spacetime, we do not introduce a counter term for the wavefunction renormalization ($Z\partial^\mu \phi \partial_\mu \phi$) in Eq.~(\ref{eq:s_and_l}) because the inflaton fluctuations ($\delta \phi$) asymptote to free fields in the subhorizon limit and their normalization is determined so that they are the Bunch-Davies solution in that limit (Eq.~(\ref{eq:u_k_subh_limit})).}
\begin{align}
  S = \int \dd \eta \, \dd^3 x \, a^4 \mathcal L, \ \  \mathcal L = - \frac{1}{2} \partial^\mu \phi \partial_\mu \phi - V_b(\phi).
  \label{eq:s_and_l}
\end{align}
$V_{b}(\phi)$ is the bare potential, which consists of the finite tree-level potential $V(\phi)$ and the counter term $V_c(\phi)$ as $V_b(\phi) = V(\phi) + V_c(\phi)$.
By definition, $V_c$ is of order of one or higher loops.
We assume that $V_{b}$ and $V$ are smooth functions of $\phi$, otherwise we cannot expand the potential with respect to the perturbation from the background of $\phi$. 
This secures that $V_{c}$ is also smooth. 
We decompose the field into the tree-level background and the perturbation, $\phi(\bfx,\eta) = \bar\phi(\eta) + \delta \phi(\bfx,\eta)$.
$\bar\phi$ follows the following equation of motion:
\begin{align}
  \bar \phi'' + 2 \mathcal H \bar \phi' + a^2 V_\fo(\bar \phi) = 0,
  \label{eq:b_eom}
\end{align}
where $V_{(n)}(\phi) \equiv \dd^n V(\phi)/\dd \phi^n$.
With this definition of $\bar \phi$, we generally get $\expval{\delta \phi(\bfx,\eta)} \neq 0$ at one-loop level, which means that the backreaction appears in $\delta \phi(\bfx,\eta)$.
Note that this definition of $\bar \phi$ is slightly different from that in our previous work~\cite{Inomata:2024lud}. We will come back to this point at the end of Sec.~\ref{sec:one_loop_ps}.

To use the in-in formalism~\cite{Weinberg:2005vy}, we expand the Lagrangian with respect to $\delta \phi$:
\begin{align}
  \mathcal L = \frac{1}{2a^2} \left[(\bar \phi')^2 + 2\bar \phi' \delta \phi' + (\delta \phi')^2 - (\partial_i \delta \phi)^2 \right] - V_b(\bar \phi) - \sum_{n=1}\frac{1}{n!} V_{b,(n)}(\bar \phi) \delta\phi^n.
\end{align}
Considering the $\bar \phi$ as a background variable, we obtain the Lagrangian for $\delta \phi$:
\begin{align}
  \mathcal L_{\delta \phi} &= \frac{1}{2a^2} \left[(\delta \phi')^2 - (\partial_i \delta \phi)^2 \right] + \frac{\bar \phi' \delta \phi'}{a^2} - \sum_{n=1}\frac{1}{n!} V_{b,(n)}(\bar \phi) \delta\phi^n \nonumber \\
  &=  \frac{1}{2a^2} \left[(\delta \phi')^2 - (\partial_i \delta \phi)^2 \right]  - \sum_{n=2}\frac{1}{n!} V_{b,(n)}(\bar \phi) \delta\phi^n - V_{c,\fo}(\bar \phi) \delta\phi + \frac{1}{a^4}\frac{\dd}{\dd \eta} \left[ a^2 \bar \phi' \delta \phi \right],
  \label{eq:l_d_phi}
\end{align}
where we have used Eq.~(\ref{eq:b_eom}) to obtain the second line.
The last term leads to the boundary term of the action, $S_\text{boundary} = \left[\int \dd^3 x \, a^2 \bar \phi' \delta \phi\right]^{\eta_\text{fin}}_{\eta_\text{ini}}$ with $\eta_\text{fin}$ and $\eta_\text{ini}$ being some final and initial time.
Since this boundary term does not include $\delta \phi'$, we can safely neglect it in the following analysis because it commutes with the interaction Hamiltonian $H_\inte$ at equal time, which is defined below (Eq.~(\ref{eq:int_hamil}))~\cite{Arroja:2011yj,Kawaguchi:2024lsw}.\footnote{
We can see this more specifically as follows. 
  We here denote the boundary term by $\mathcal B(\tau) = \int\dd^3 x a^2 \bar \phi' \delta \phi$. Including this boundary term, we can express the time-evolution operator in the interaction picture as 
  \begin{align}
    U(\eta,\eta_\ini) = T\left[\ee^{-i \int^\eta_{\eta_\ini} \dd \eta' (H_\inte(\eta') + \dd \mathcal B(\eta')/\dd \eta')} \right] = \ee^{-i \mathcal B(\eta)}T \left[\ee^{-i \int^\eta_{\eta_\ini} \dd \eta' H_\inte(\eta') }\right]\ee^{i \mathcal B(\eta_\ini)}, \nonumber
  \end{align}
  where $T$ represents the time-ordering product and we have used the fact that $\mathcal B(\eta)$ commutes $H_\inte(\eta)$ at equal time.
  Imposing the Bunch-Davies vacuum condition at $\eta_\ini$, we here neglect the contribution from the initial time surface, $\mathcal B(\eta_\ini)$.
  Then, we can express the expectation value of $Q$ (any product of $\delta \phi$) in the in-in formalism as 
  \begin{align}
    \vev{Q(\eta)} = \bra{0} (T \ee^{-i \int^\eta_{\eta_\ini} \dd \eta' H_\text{int}})^\dagger \ee^{i \mathcal B(\eta)} Q(\eta) \ee^{-i \mathcal B(\eta)} (T \ee^{-i \int^\eta_{\eta_\ini} \dd \eta' H_\text{int}}) \ket{0} = \bra{0} (T \ee^{-i \int^\eta_{\eta_\ini} \dd \eta' H_\text{int}})^\dagger Q(\eta) (T \ee^{-i \int^\eta_{\eta_\ini} \dd \eta' H_\text{int}}) \ket{0}, \nonumber
  \end{align}
  where we have used the fact that $Q(\eta)$ commute $\mathcal B(\eta)$ at equal time.
  This means that the boundary term can be neglected in our analysis.
}
From Eq.~(\ref{eq:l_d_phi}), the Hamiltonian for $\delta \phi$ is given by 
\begin{align}
  \mathcal H &= \frac{1}{2a^2} \left[(\delta \phi')^2 + (\partial_i \delta \phi)^2 \right]  + \sum_{n=2}\frac{1}{n!} V_{b,(n)}(\bar \phi) \delta\phi^n + V_{c,\fo}(\bar \phi) \delta\phi \nonumber \\
  &= \mathcal H_0 + \sum_{n}\mathcal H_{\inte,n}.
\end{align}
We have decomposed the Hamiltonian into the free Hamiltonian
\begin{align}
  \mathcal H_0 \equiv \frac{1}{2a^2} \left[(\delta \phi')^2 + (\partial_i \delta \phi)^2 \right] + \frac{1}{2} V_\so(\bar \phi) \delta \phi^2,
\end{align}
and the interaction Hamiltonian
\begin{align}
  \mathcal H_{\inte,n} \equiv \begin{cases}
  \cfrac{1}{n!} V_{c,(n)}(\bar\phi) \delta\phi^n & \quad (n=1,2) \\[2ex]
  \cfrac{1}{n!} V_{b,(n)}(\bar \phi) \delta\phi^n = \cfrac{1}{n!} \left[V_{(n)}(\bar \phi) + V_{c,(n)}(\bar \phi)\right]\delta\phi^n & \quad (\text{others}) \\  
    \end{cases}.
\end{align}

\begin{figure}
        \centering \includegraphics[width=0.6\columnwidth]{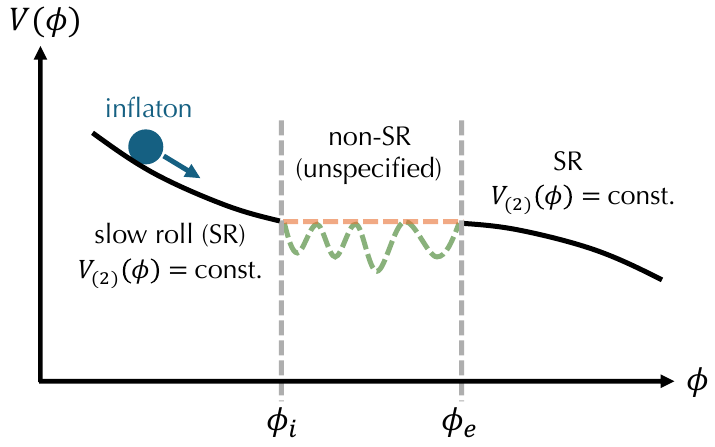}
        \caption{ The schematic picture of the inflaton potential we consider.
        The potential in the non-SR period can have any form, such as the flat region that leads to USR (orange) and the oscillatory region that leads to the parametric resonance of the perturbations (green).
    }
        \label{fig:pot}
\end{figure}

Figure~\ref{fig:pot} shows the inflaton potential we consider in this paper. 
We consider a transient non-SR period within $\phi_i < \phi < \phi_e$. 
We assume that the small-scale curvature perturbations are enhanced during this period, which can be e.g. USR period or a parametric resonance period with oscillatory features. 
Outside the non-SR period, we assume that $V_{(n \geq 3)}$ is zero (or exponentially suppressed), which leads to $V_\so = \text{const.}$ in $\phi< \phi_i$ or $\phi_e < \phi$.
In this setup, $\mathcal H_{\inte}$ is nonzero only in $\phi_i < \phi < \phi_e$.
For later convenience, we define $\eta_i$ and $\eta_e$ through $\phi_i = \bar \phi(\eta_i)$ and $\phi_e = \bar \phi(\eta_e)$.
Then, we can express the expectation value of any product of $\delta \phi$, denoted by $Q$, with the in-in formalism~\cite{Weinberg:2005vy}: 
\begin{align}
  \expval{Q(\eta)} = \bra{0} (T \ee^{-i \int^\eta_{\eta_i} \dd \eta' H_\text{int}})^\dagger Q^I(\eta) (T \ee^{-i \int^\eta_{\eta_i} \dd \eta' H_\text{int}}) \ket{0},
  \label{eq:in_in}
\end{align}
where $\ket{0}$ is the non-interacting vacuum and 
\begin{align}
  H_{\text{int}} = \sum_{n=1} H_{\text{int},n}, \ 
  H_{\text{int},n} \equiv \int \dd^3 x\,\, a^4 \mathcal H_{\inte,n}.
  \label{eq:int_hamil}
\end{align}
Note that we do not use the $i \epsilon$ prescription in the lower bound of the time integrals because, in our setup, the interaction Hamiltonian is negligible in $\eta < \eta_i$ and the vacuum automatically becomes the non-interacting vacuum in $\eta<\eta_i$.
Namely, we do not need to use the Gell-Mann and Low theorem in our setup.\footnote{\label{ft:i_eps}
  We can easily see that introducing the $i \varepsilon$ prescription does not change anything if $H_\inte(\eta)$ is a holomorphic function in $\text{Re}[\eta] < 0$, $|\text{Im}[\eta]| \leq \varepsilon$, and $H_\inte(\eta) = 0$ in $\Re[\eta] < \eta_i$ (this is true in reasonable situations):
  \begin{align}
      \int^\eta_{\eta_i(1 \pm i\varepsilon)}  \dd \eta' H_\text{int}(\eta') = \int^\eta_{\eta_i} \dd \eta' H_\text{int}(\eta') + \int^{\eta_i}_{\eta_i(1 \pm i \varepsilon)} \dd \eta' H_\text{int}(\eta') = \int^\eta_{\eta_i}  \dd \eta' H_\text{int}(\eta') + i\int^{0}_{\pm\varepsilon} \dd \eta'' H_\text{int}(\eta_i + i \eta'') = \int^\eta_{\eta_i}  \dd \eta' H_\text{int}(\eta'), \nonumber
  \end{align}
  where $\eta < 0$ and we have assumed that the integral contour of $\eta'$ is inside the region of $\text{Re}[\eta'] < 0$ and $|\text{Im}[\eta']| \leq \varepsilon$.
  Note that the integral of a holomorphic function only depends on the lower and upper bounds of the integral, independently of the integral contour. 
}

Let us see the evolution of the linear perturbation $\delta \phi$ in detail.
$\delta \phi$ in the interaction picture is given by 
\begin{align}
  &\delta \phi^I(\bfx,\eta) = \int \frac{\dd^3 k}{(2\pi)^3} \ee^{i \bfk \cdot \bfx} \delta \phi^I_{\bfk}(\eta)= \int \frac{\dd^3 k}{(2\pi)^3} \ee^{i \bfk \cdot \bfx} \left[ u_k(\eta) a(\bfk) +  u^{*}_k(\eta) a^{\dagger}(-\bfk) \right],
\end{align}
where the commutation relations of the operators are given by $[a(\bfk), a(\bfk')] = [a^\dagger(\bfk), a^\dagger(\bfk')] = 0$ and $[a(\bfk), a^\dagger(-\bfk')] = (2\pi)^3 \delta(\bfk + \bfk')$.
$u_k$ follows the equation of motion from the free part,
\begin{align}
  \left[\frac{\dd^2}{\dd \eta^2} + 2 \mathcal H \frac{\dd}{\dd \eta} + k^2 + a^2(\eta) V_\so(\eta) \right] u_k(\eta) = 0,
  \label{eq:u_eom}
\end{align}
where $V_\no(\eta) \equiv V_\no(\bar\phi(\eta))$. 
The canonical commutation relation $a^2[\delta \phi^I(\bfx),\delta \phi^{I\prime}(\bfy)] = i \delta(\bfx-\bfy)$ leads to the Wronskian condition for $u_k$~\cite{Inomata:2022yte},
\begin{align}
  \Im[u_k(\eta) u^{*\prime}_k(\eta)] = \frac{1}{2a^2(\eta)}.
  \label{eq:wronskian}
\end{align}
Since $V_\so = \text{const.}$ in $\eta < \eta_i$, we can solve the equation of motion Eq.~(\ref{eq:u_eom}) in $\eta < \eta_i$ as 
\begin{align}
  \label{eq:u_k_ini}
  u_k(\eta) = -\frac{H k\eta}{\sqrt{2k^3}}\ee^{i\frac{\left(2\nu + 1\right)\pi}{4}} \sqrt{\frac{-\pi k\eta}{2}} H_\nu^\fo(-k\eta) \ \  \text{for } \eta < \eta_i,
\end{align}
where $\nu = \sqrt{9/4 - m_i^2/H^2}$ with $m_i^2 \equiv V_\so(\eta_i)$ and we have used $\mathcal H = aH = -1/\eta$.
Throughout this work, we assume $|m_i^2/H^2| \ll 1$ during the SR inflation periods, but do not neglect it for completeness.
Note that we can take the de Sitter limit ($\epsilon \to 0$) with $|m_i^2/H^2|$ fixed.
The normalization in Eq.~(\ref{eq:u_k_ini}) is determined so that $u_k$ becomes the Bunch-Davies vacuum solution on deep subhorizon scales:
\begin{align}
  u_k(\eta) &\simeq -\frac{H k \eta}{\sqrt{2k^3}}\ee^{-ik\eta} \left( 1 + i \frac{4\nu^2 -1}{8(-k\eta)} - \frac{16\nu^4 -40\nu^2 + 9}{128(-k\eta)^2} \right) \ \text{ for } |k\eta| \gg 1 \text{ and } \eta < \eta_i,
  \label{eq:u_k_subh_limit}
\end{align}
where we have explicitly shown the terms up to $\mathcal O((-k\eta)^{-2})$ because we use them in Sec.~\ref{sec:renorm} when we discuss the next-leading UV divergence. 
We note that, even for $\eta > \eta_i$, we can express $u_k(\eta) \simeq -Hk\eta/\sqrt{2 k^3} \ee^{-ik\eta} (1 + i A(\eta)/(-k\eta))$ with $\Im[A(\eta)]=0$ in $|k\eta| \gg 1$ and $A(\eta)$ independent of $k$.\footnote{We can see $\Im[A(\eta)] = 0$ by reexpressing Eq.~(\ref{eq:u_eom}) as $\left[\frac{\dd^2}{\dd x^2} + 1 + \frac{V_\so(\eta)/H^2-2}{x^2} \right] \left(\frac{u_k(\eta)}{x}\right) = 0$
where $x = k\eta$. 
We expand this with $x \gg 1$ limit and obtain $u_k(\eta) \propto x\,  \ee^{-ix} (1 + i A(\eta)/x + \mathcal O(x^{-2}) )$.
By substituting this solution into this reexpressed equation of motion, we can easily see that $\Im[A(\eta)] = 0$ must be satisfied because the numerator $(V_\so(\eta)/H^2 -2)$ is real.
}
This means that, in the subhorizon limit ($|k\eta| \gg 1$), $u_k \simeq -Hk\eta/\sqrt{2 k^3} \ee^{-ik\eta}$ is always satisfied (even during the non-SR period).

On the other hand, the superhorizon-limit expression is given by 
\begin{align}
  u_q(\eta) &\simeq -i \Gamma(\nu)\frac{\sqrt{2}H}{\sqrt{\pi q^3}}\ee^{i\frac{\left(2\nu + 1\right)\pi}{4}} \left(\frac{-q\eta}{2}\right)^{-\nu+\frac{3}{2}}\left(1 + \frac{(-q\eta)^2}{4(\nu - 1)} + i \frac{\pi}{\Gamma(\nu) \Gamma(\nu+1)} \left(\frac{-q\eta}{2}\right)^{2\nu} \right) \ \text{ for } |q\eta| \ll 1 \text{ and } \eta < \eta_i,
  \label{eq:u_k_sh_limit}
\end{align}
where we have kept the terms up to $\mathcal O( (-q\eta)^{2\nu})$ with $2\nu \simeq 3$ in $|m_i^2/H^2| \ll 1$.
Note that, throughout this work, we use $\bfq$ (and $q$) to denote superhorizon modes.
Using Eq.~(\ref{eq:u_k_sh_limit}), we can express the following quantity, which appears later, in the superhorizon limit:
\begin{align}
  \Im\left[ u_q(\eta) u^*_q(\eta') \right] = \frac{H^2}{4 \nu} (\eta \eta')^{-\nu + \frac{3}{2}} \left[ (-\eta)^{2\nu} - (-\eta')^{2\nu} \right] \ \ \text{ for } \eta' < \eta, \ |q\eta'| \ll 1, \text{ and } \eta < \eta_i.
  \label{eq:im_uu}
\end{align}
We can see that this quantity does not depend on the scale $q$. 
Actually, the independence of $q$ remains even for $\eta \geq \eta_i$ and/or $\eta' \geq \eta_i$.
To see this, we express Eq.~(\ref{eq:im_uu}) as (see Eq.~(62) in Ref.~\cite{Inomata:2022yte})
\begin{align}
  \Im\left[ u_q(\eta) u^*_q(\eta') \right] &= -|u_q(\eta)||u_q(\eta')| \sin \left[ \int^\eta_{\eta'} \dd \eta'' \frac{1}{2 a^2(\eta'') |u_q(\eta'')|^2} \right] \nonumber \\
  &\simeq -|u_q(\eta)||u_q(\eta')| \int^\eta_{\eta'} \dd \eta'' \frac{1}{2 a^2(\eta'') |u_q(\eta'')|^2}  \ \ \text{ for } |q\eta| \ll 1 \text{ and } |q\eta'| \ll 1,
  \label{eq:im_uu_2}
\end{align}
where we have used the Wronskian condition Eq.~(\ref{eq:wronskian}).
To proceed, we note that the time dependence of $|u_q|$ follows (from Eq.~(\ref{eq:u_eom}))
\begin{align}
   \left[\frac{\dd^2}{\dd \eta^2} + 2 \mathcal H \frac{\dd}{\dd \eta} + a^2(\eta) V_\so(\eta) \right] |u_q(\eta)| = 0\  \text{ for } |q\eta| \ll 1.
\end{align}
We here recall the leading term $|u_q(\eta)| = C_q (-\eta)^{-\nu + \frac{3}{2}}$ for $\eta < \eta_i$ with $C_q$ some time-independent function with $C_q \propto q^{-\nu}$. 
From this, we can express $|u_q(\eta)| = C_q B(\eta)$ for $\eta \geq \eta_i$ with $B(\eta)$ some function of $\eta$ that satisfies the boundary condition at $\eta_i$, consistently with $\propto (-\eta)^{-\nu + \frac{3}{2}}$.
The $q$-dependence appears only in $C_q$.
Substituting these expressions, we can see that Eq.~(\ref{eq:im_uu_2}) does not depend on $q$ even for $\eta > \eta_i$ and/or $\eta'>\eta_i$ as long as $|q\eta| \ll 1$ and $|q\eta'| \ll 1$ are satisfied. 
For later convenience, we define the $q$-independent expression as 
\begin{align}
  \Im\left[ u_0(\eta) u^*_0(\eta') \right] \equiv \lim_{q\to 0}\Im\left[ u_q(\eta) u^*_q(\eta') \right].
\end{align}

\begin{figure}
        \centering \includegraphics[width=0.8\columnwidth]{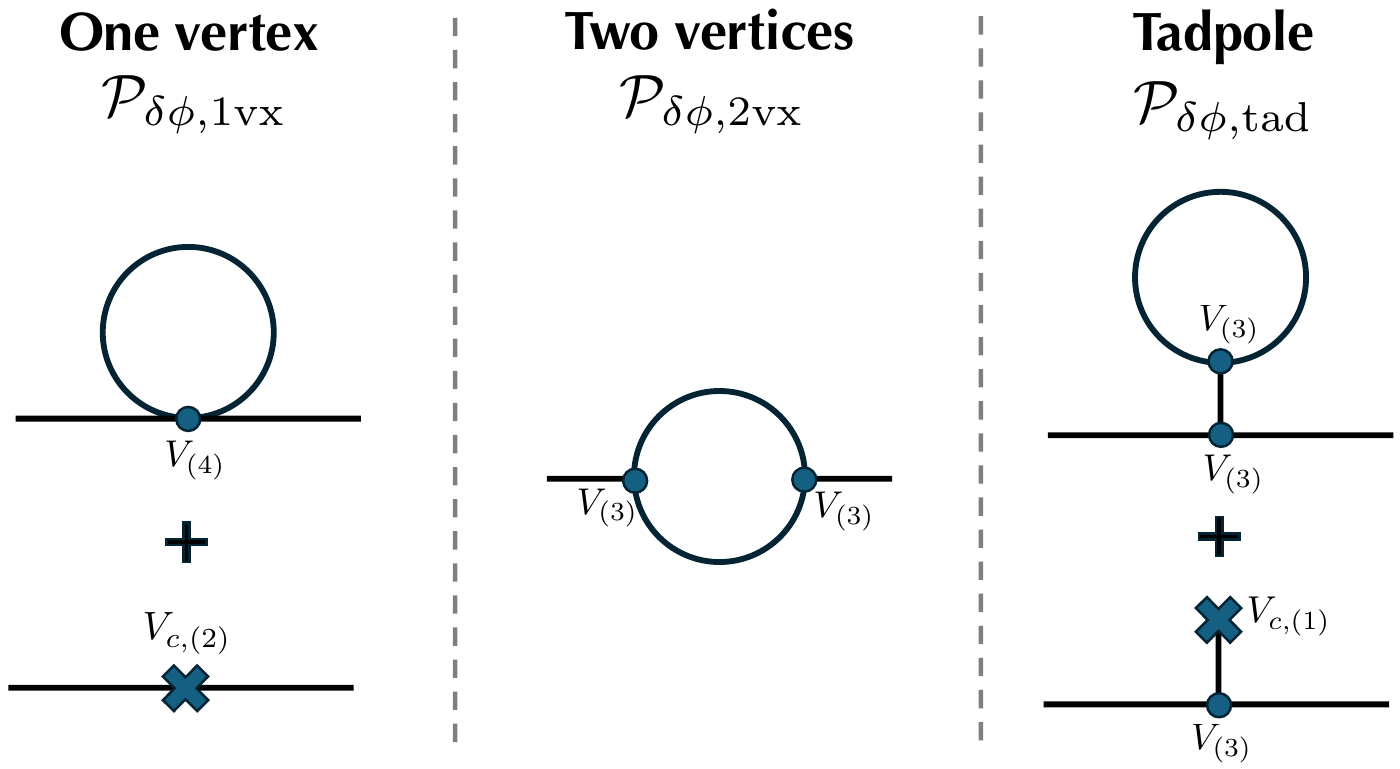}
        \caption{ The Feynman diagrams of the one-loop contributions in Eq.~(\ref{eq:two_vx_one_loop}).
        The circles and crosses denote the vertices that come from the derivatives of the tree-level potential and the counter terms, respectively. 
    }
        \label{fig:loops}
\end{figure}

Substituting $Q(\eta) = \delta \phi_{\bfq}(\eta) \delta \phi_{\bfq'}(\eta)$ into Eq.~(\ref{eq:in_in}), we obtain the two-point correlation function up to one-loop level:\footnote{
  Note that
  \begin{align}
    \int^\eta_{\eta_i}\dd \eta' \int^{\eta'}_{\eta_i}\dd \eta'' \vev{ \left( H_{\text{int},1}(\eta')\delta \phi^I_{\bfq}(\eta) \delta \phi^I_{\bfq'}(\eta) - \delta \phi^I_{\bfq}(\eta) \delta \phi^I_{\bfq'}(\eta) H_{\inte,1}(\eta') \right) H_{\text{int},3}(\eta'')} = 0 \ \  \text{ for } \bfq,\bfq' \neq 0. \nonumber
  \end{align}
}
\begin{align}
  &\expval{\delta \phi_{\bfq}(\eta) \delta \phi_{\bfq'}(\eta)} = \vev{\delta \phi^I_{\bfq}(\eta) \delta \phi^I_{\bfq'}(\eta) } + 2\, \Im\left[\int^\eta_{\eta_i} \dd \eta'\vev{ \delta \phi^I_{\bfq}(\eta) \delta \phi^I_{\bfq'}(\eta) (H_{\text{int},4}(\eta') + H_{\inte,2}(\eta'))}\right] \nonumber \\
  &+ 2\, \Re\left[ \int^\eta_{\eta_i}\dd \eta' \int^{\eta'}_{\eta_i}\dd \eta'' \vev{ \left(H_{\text{int},3}(\eta') \delta \phi^I_{\bfq}(\eta) \delta \phi^I_{\bfq'}(\eta) - \delta \phi^I_{\bfq}(\eta) \delta \phi^I_{\bfq'}(\eta) H_{\inte,3}(\eta') \right) (H_{\text{int},3}(\eta'') + H_{\text{int},1}(\eta''))}\right] \nonumber \\
  &= (2\pi)^3 \delta(\bfq + \bfq') \frac{2\pi^2}{q^3} \left[ \mathcal P_{\delta \phi,\tre}(q,\eta) + \mathcal P_{\delta \phi, 1\vx}(q,\eta) + \mathcal P_{\delta \phi, 2\vx}(q,\eta) + \mathcal P_{\delta \phi,\tad}(q,\eta) \right], 
  \label{eq:two_vx_one_loop}
\end{align}
where note that $H_{\inte,1}$ and $H_{\inte,2}$ come from the counter terms at one-loop level and therefore are of the same order as $H_{\inte,3}\sim \delta \phi^3$ and $H_{\inte,4}\sim \delta \phi^4$, respectively.
$\mathcal P_{\delta \phi,\tre}$ is the tree level power spectrum without any $H_\inte$ (the first term in the first equality), $\mathcal P_{\delta \phi, 1\vx}$ is the one-loop contribution with one vertex of $H_{\inte,2}$ or $H_{\inte,4}$ (the second term), and $\mathcal P_{\delta \phi, 2\vx}$ is the one-loop contribution with two vertices of two $H_{\inte,3}$ except for the tadpole contribution $\mathcal P_{\delta \phi, \tad}$ (the second line).
See Figure~\ref{fig:loops} for the Feynman diagrams of the one-loop contributions.

The tree level power spectrum is given by 
\begin{align}
   \mathcal P_{\delta \phi,\tre}(q,\eta) = \frac{q^3}{2\pi^2}|u_q(\eta)|^2.
   \label{eq:1vx_g}
\end{align}
In the next section, we derive the concrete expressions of the one-loop power spectrum.

%%%%%%%%%%%%%%%%%%%%%%%%%%%%%%%%
\section{One-loop power spectrum}
\label{sec:one_loop_ps}
%%%%%%%%%%%%%%%%%%%%%%%%%%%%%%%%

%%%%%%%%%%%%%
\subsection{One-vertex and two-vertex contributions}

The one-vertex contribution in Eq.~(\ref{eq:two_vx_one_loop}) is given by
\begin{align}
  \mathcal P_{\delta \phi, 1\vx}(q,\eta) &= \frac{q^3}{2\pi^2} \Im\left[ u^2_q(\eta) \int^\eta_{\eta_i} \dd \eta' a^4(\eta') (u^*_q(\eta'))^2 \left(V_\foo(\eta')\int\frac{\dd^3 k}{(2\pi)^3} |u_k(\eta')|^2 + 2V_{c,\so}(\eta')\right) \right] \nonumber \\
  &=  \frac{q^3}{\pi^2} \int^\eta_{\eta_i} \dd \eta' a^4(\eta') \Im\left[ u_q(\eta) u^*_q(\eta') \right] \Re\left[ u_q(\eta) u^*_q(\eta') \right] \left(V_\foo(\eta')\int\frac{\dd^3 k}{(2\pi)^3} |u_k(\eta')|^2 + 2V_{c,\so}(\eta')\right).
  \label{eq:1vx}
\end{align}

The two-vertex contribution in Eq.~(\ref{eq:two_vx_one_loop}) is given by 
\begin{align}
  \mathcal P_{\delta \phi, 2\vx}(q,\eta) 
  &= -\frac{q^3}{\pi^2} \int^\eta_{\eta_i} \dd \eta' \int^{\eta'}_{\eta_i} \dd \eta'' a^4(\eta')  a^4(\eta'') V_\tho(\eta') V_\tho(\eta'') \Re\left[ [u_q(\eta) u_q^*(\eta')- u^*_q(\eta) u_q(\eta')] 
   u_q(\eta) u_q^*(\eta'') \phantom{\int \frac{\dd^3 k}{(2\pi)^3}} \right. \nonumber \\
  &\qquad \qquad \qquad \left. \times \int \frac{\dd^3 k}{(2\pi)^3} u_k(\eta')u^*_k(\eta'')u_{|\bfq - \bfk|}(\eta')u^*_{|\bfq - \bfk|}(\eta'') \right] \nonumber \\
  &= \mathcal P^a_{\delta \phi, 2\vx}(q,\eta) + \mathcal P^b_{\delta \phi, 2\vx}(q,\eta).
  \label{eq:2vx}
\end{align}
$\mathcal P^a_{\delta \phi, 2\vx}$ and $\mathcal P^b_{\delta \phi, 2\vx}$ are given by\footnote{
  $\mathcal P^a_{\delta \phi, 2\vx}$ and $\mathcal P^b_{\delta \phi, 2\vx}$ correspond to the power spectrum of $\expval{\delta \phi^\so \delta \phi^\so}$ and $\expval{\delta \phi^\fo \delta \phi^\tho} + \expval{\delta \phi^\tho \delta \phi^\fo}$ in Ref.~\cite{Inomata:2024lud}, respectively. See also Appendix of Ref.~\cite{Inomata:2022yte}.
}
\begin{align}
  \label{eq:p_a}
  \mathcal P^a_{\delta \phi, 2\vx}(q,\eta) &= \frac{2q^3}{\pi^2}  \int^\eta_{\eta_i} \dd \eta' \int^{\eta'}_{\eta_i} \dd \eta'' a^4(\eta')  a^4(\eta'') V_\tho(\eta') V_\tho(\eta'') \Im[u_q(\eta) u_q^*(\eta')] \Im[u_q(\eta) u_q^*(\eta'')] \nonumber \\
  &\qquad \qquad \qquad \times \int \frac{\dd^3 k}{(2\pi)^3} \Re[u_k(\eta')u^*_k(\eta'')u_{|\bfq - \bfk|}(\eta')u^*_{|\bfq - \bfk|}(\eta'')] \\
  %%%%
  \label{eq:p_b}  
  \mathcal P^b_{\delta \phi, 2\vx}(q,\eta) &= \frac{2q^3}{\pi^2}  \int^\eta_{\eta_i} \dd \eta' \int^{\eta'}_{\eta_i} \dd \eta'' a^4(\eta')  a^4(\eta'') V_\tho(\eta') V_\tho(\eta'') \Im[u_q(\eta) u_q^*(\eta')] \Re[u_q(\eta) u_q^*(\eta'')] \nonumber \\
  &\qquad \qquad \qquad \times \int \frac{\dd^3 k}{(2\pi)^3} \Im[u_k(\eta')u^*_k(\eta'')u_{|\bfq - \bfk|}(\eta')u^*_{|\bfq - \bfk|}(\eta'')] \nonumber \\
  &= \frac{2q^3}{\pi^2} \int^\eta_{\eta_i} \dd \eta' \int^{\eta'}_{\eta_i} \dd \eta'' a^4(\eta') 
  a^4(\eta'') V_\tho(\eta') V_\tho(\eta'') \Im[u_q(\eta) u_q^*(\eta')] \Re[u_q(\eta) u_q^*(\eta'')] \nonumber \\
  &\qquad \times  \int \frac{\dd^3 k}{(2\pi)^3} (\Im[u_k(\eta')u^*_k(\eta'')] \Re[u_{|\bfq - \bfk|}(\eta')u^*_{|\bfq - \bfk|}(\eta'')] + (k \leftrightarrow |\bfq - \bfk|) ).
\end{align}
Note that $\mathcal P^a_{\delta \phi,2\vx}(q,\eta) \propto q^3$ in the superhorizon limit ($|q\eta| \ll 1$) if we focus on the loop contributions from $k \gg q$.
Approximating $u_{|\bfq - \bfk|} \to u_{k}$ in the integrand in Eq.~(\ref{eq:p_a}), we can easily see $\mathcal P^a_{\delta \phi,2\vx}(q,\eta) \propto q^3$ in the superhorizon limit. Note that $\Im[u_q(\eta) u_q^*(\eta')]$ and $\Im[u_q(\eta) u_q^*(\eta'')]$ do not depend on $q$ in that limit (see the discussion below Eq.~(\ref{eq:im_uu})).

\subsection{Tadpole contribution}

The tadpole contribution in Eq.~(\ref{eq:two_vx_one_loop}) is given by 
\begin{align}
  \mathcal P_{\delta \phi, \tad}(q,\eta) 
  &= \frac{2q^3}{\pi^2} \int^\eta_{\eta_i} \dd \eta' \int^{\eta'}_{\eta_i} \dd \eta''a^4(\eta') a^4(\eta'') V_\tho(\eta') \Im\left[ u_q(\eta) u^*_q(\eta') \right] \Re\left[ u_q(\eta) u^*_q(\eta') \right] \nonumber \\
  &\qquad \times \Im[ u_{0}(\eta') u_{0}^*(\eta'')]  \left( V_\tho(\eta'') \int \frac{\dd^3 k}{(2\pi)^3} |u_{k}(\eta'')|^2 + 2 V_{c,\fo}(\eta'')\right).
  \label{eq:tad}
\end{align}
This can be rewritten as 
\begin{align}
  \mathcal P_{\delta \phi, \tad}(q,\eta) 
  &= \frac{q^3}{\pi^2} \int^\eta_{\eta_i} \dd \eta' a^4(\eta') V_\tho(\eta')\Im\left[ u^2_q(\eta) (u^*_q(\eta'))^2 \right] \expval{\delta \phi(\bfx, \eta')}.
  \label{eq:tad0}
\end{align}
$\expval{\delta \phi(\bf x,\eta)}$ is the backreaction at one-loop level, which can be calculated with the in-in formalism as 
\begin{align}
  \expval{\delta \phi(\bf x,\eta)}
  &= 2 \int^\eta_{\eta_i} \dd \eta' \, \Im\left[ \vev{ \delta \phi^I(\bfx, \eta)  (H_{\text{int},3}(\eta')+ H_{\text{int},1}(\eta') )} \right]  \nonumber \\
  &= 2 \int \frac{\dd^3 q}{(2\pi)^3} \ee^{i \bf q \cdot \bf x} \int^\eta_{\eta_i} \dd \eta' \, \Im\left[ \vev{ \delta \phi^I_{\mathbf q}(\eta)  (H_{\text{int},3}(\eta')+ H_{\text{int},1}(\eta') )} \right]  \nonumber \\
  &=  \int \dd^3 q\, \delta(\mathbf q) \ee^{i \mathbf q \cdot \mathbf x} \int^\eta_{\eta_i} \dd \eta' a^4(\eta') \text{Im} [ u_q(\eta) u_q^*(\eta')] \left(V_\tho(\eta') \int \frac{\dd^3 k}{(2\pi)^3} |u_{k}(\eta')|^2 + 2 V_{c,\fo}(\eta')\right) \nonumber \\
  &=  \int^\eta_{\eta_i} \dd \eta' a^4(\eta') \Im[ u_0(\eta) u_0^*(\eta')]\left(V_\tho(\eta') \int \frac{\dd^3 k}{(2\pi)^3} |u_{k}(\eta')|^2 + 2 V_{c,\fo}(\eta')\right). 
  \label{eq:one_pt}
\end{align}
Note that the right-hand side (RHS) of Eq.~(\ref{eq:tad0}) finally becomes independent of $\bfx$.

We stress that the value of $\expval{\delta \phi(\bf x,\eta)}$ depends on the choice of the counter term, $V_{c,\fo}$.
This means that, if we want to concretely calculate the one-loop corrections in some models, we need to introduce a guiding principle to determine the counter term.
Note that, once we fix $V_{c,\fo}$, $V_{c,\so} (= \dd V_{c,\fo}/\dd \phi)$ is automatically fixed. 
Fixing the counter term determines the physical meaning of the tree-level potential. 
Actually, the specification of $V_{c,\fo}$ is unnecessary for the proof of the curvature conservation and we will not specify $V_{c,\fo}$ throughout this work for generality.
That said, for reference, let us here mention some possible choices of $V_{c,\fo}$.
One possible choice is to tune $V_{c,\fo}$ to cancel the loop contributions from the subhorizon perturbations in the Bunch-Davies vacuum states, which correspond to the subtraction of the vacuum loop contributions.
This choice enables us to regard the tree-level potential as the renormalized potential after the subtraction of the vacuum loops.
In this choice, we generally find $\expval{\delta \phi} \neq 0$ and the one-loop corrections come only from the superhorizon perturbations and the excited states (amplified perturbations) on subhorizon scales (if a parametric resonance occurs~\cite{Inomata:2022yte}).
Another possible choice is to tune $V_{c,\fo}$ to cancel all the loop contributions by imposing $\expval{\delta \phi} =0$.
This can be realized if we take
\begin{align}
  V_{c,\fo}(\eta) = - \frac{1}{2}V_{\tho}(\eta) \int\frac{\dd^3 k}{(2\pi)^3} |u_k(\eta)|^2 \quad (\text{for } \expval{\delta \phi} = 0. \text{ Note: We do not impose this for any other arguments}).
  \label{eq:vcfo_zero_tad}
\end{align}
In this choice, the tree-level potential can be regarded as the one-loop effective potential, where the background evolution with the one-loop backreaction is fully described with Eq.~(\ref{eq:b_eom}).
In other words, the one-loop backreaction is implicitly included in the definition of $V(\phi)$ and the (explicit) backreaction disappears in this choice. 
In Sec.~\ref{sec:curv_cons}, we will see that the scale-independent part (or the part that remains in $q \to 0$ limit) of the one-loop power spectrum is proportional to $\langle \delta \dot \phi \rangle$, which means that the scale-independent part goes to zero in this choice of $V_{c,\fo}$.
Meanwhile, the scale-dependent part of the one-loop power spectrum remains nonzero even in this choice.
Note again that Eq.~(\ref{eq:vcfo_zero_tad}) is just one of the examples of the choice of the counter term and, for generality, we do not impose it in the following.

This tuning of the counter term must be distinguished from the redefinition of the perturbation.
In principle, by redefining the perturbation (and the tree-level background) as $\delta \phi_\rr \equiv \delta \phi - \expval{\delta \phi}$ with $\bar \phi_\rr \equiv \bar \phi + \expval{\delta \phi}$, we can always realize $\expval{\delta \phi_\rr} = 0$ without changing the counter term $V_{c,\fo}$.
However, this does not change the evolution of the whole background ($\bar \phi_\rr = \bar \phi + \expval{\delta \phi}$) and the two point correlation function (that is, $\expval{\delta \phi_{\rr,\bfq} \delta \phi_{\rr,\bfq'}} = \expval{\delta \phi_{\bfq} \delta \phi_{\bfq'}}$ because $\expval{\delta \phi_{\bfq} \expval{\delta \phi}} = 0$).
In our previous work~\cite{Inomata:2024lud}, we considered $\delta \phi_\rr$ (without tuning $V_{c,\fo}$) and discussed the backreaction for the whole background $\bar \phi_\rr$.
Specifically, we obtained the following equation of motion for $\bar \phi_\rr$~\cite{Inomata:2024lud},
\begin{align}
  \bar \phi_\rr''(\eta) + 2 \mathcal H \bar \phi_\rr'(\eta) + a^2 V_\fo(\bar \phi_\rr(\eta)) = -\frac{a^2}{2} \left[V_\tho(\bar \phi_\rr(\eta)) \int\frac{\dd^3 k}{(2\pi)^3} |u_{\rr,k}(\eta)|^2 + 2V_{c,\fo}(\bar \phi_\rr(\eta))\right],
  \label{eq:b_eom_r}
\end{align}
where $u_{\rr,k}$ is the $u_k$ for $\delta \phi_\rr$.
Unless Eq.~(\ref{eq:vcfo_zero_tad}) is satisfied, the RHS is nonzero and $\bar \phi_\rr$ gets backreaction from $\delta \phi_\rr$.
This is not surprising because the perturbation redefinition does not change the form of the tree-level potential.
Note that the backreaction is defined as the deviation from the tree-level evolution determined by Eq.~(\ref{eq:b_eom}).
See Appendix~\ref{app:tad} for a detailed comparison with Ref.~\cite{Inomata:2024lud}.

\section{Inflaton background velocity with backreaction}
\label{sec:backreaction}

In this section, we discuss the evolution of inflaton background velocity, $\langle \dot{\phi}\rangle \equiv \frac{\dd}{\dd t} \left( \bar \phi + \expval{\delta \phi} \right)$. 
In particular, $\langle \delta \dot \phi \rangle$ plays an important role for the curvature conservation, which we will see in Sec.~\ref{sec:curv_cons}.

We first consider the tree-level inflaton background.
Taking the physical time derivative of Eq.~(\ref{eq:b_eom}), we obtain 
\begin{align}
  \left[ \frac{\dd^2}{\dd \eta^2} + 2 \mathcal H \frac{\dd}{\dd \eta} + a^2(\eta) V_\so(\eta) \right] \dot{\bar \phi}(\eta) = 0,
  \label{eq:back_eom0}
\end{align}
where we have neglected $\dot H$ because we are considering the de Sitter limit $\epsilon \to 0$.
This equation of motion is the same as Eq.~(\ref{eq:u_eom}) with the superhorizon limit:
\begin{align}
  \left[\frac{\dd^2}{\dd \eta^2} + 2 \mathcal H \frac{\dd}{\dd \eta} + a^2(\eta) V_\so(\eta) \right] u_q(\eta)|_{q \to 0} = 0. 
  \label{eq:u_eom_0}
\end{align}
In the superhorizon limit, we can neglect the decaying mode in $u_q$ and $\dot {\bar \phi}$ and hence can see $u_q/{\dot {\bar \phi}} = \text{const.}$
This physically means the conservation of linear curvature perturbation, which is given by $\zeta = -H \delta \phi/{\dot {\bar \phi}}$ in the spatially flat gauge. 
For later convenience, we here define $\zeta_q \equiv -H u_q/{\dot {\bar \phi}}$.

Next, we calculate the backreaction to the inflaton velocity with the in-in formalism. 
For convenience, we first reexpress Eq.~(\ref{eq:one_pt}) with the physical time $t = \int^\eta \dd \eta' a(\eta')$:
\begin{align}
  \expval{\delta \phi(\bfx,t)}
  &=  \int^t_{t_i} \dd t' a^3(t') \Im[ u_0(t) u_0^*(t')] \left[ V_\tho(t') \int \frac{\dd^3 k}{(2\pi)^3} |u_{k}(t')|^2 + 2 V_{c,\fo}(t') \right],
  \label{eq:dt_one_pt}
\end{align}
where $f(t) \equiv f(\eta(t))$ with $f$ being any function and $t_i$ is the physical time that corresponds to $\eta_i$.
With the physical time, we can reexpress the equation of motion of $u_k$ Eq.~(\ref{eq:u_eom}) as 
\begin{align}
  \left[ \frac{\dd^2}{\dd t^2} + 3H \frac{\dd}{\dd t} + \frac{k^2}{a^2(t)} + V_\so(t) \right] u_k(t) = 0.
\end{align}
To proceed, we here obtain the retarded Green function for this equation:
\begin{align}
  \label{eq:g_tt}
  g_k(t;t') &= \Theta(t-t')\frac{u_k(t) u_k^*(t') - u^*_k(t) u_k(t')}{\dot u_k(t') u^{*}_k(t') - u_k(t') \dot u^*_k(t')} \nonumber \\
  &= -2\Theta(t-t')a^3(t') \Im[u_k(t) u_k^*(t')].
\end{align}
This $g_k(t;t')$ satisfies
\begin{align}
  \left[ \frac{\dd^2}{\dd t^2} + 3H \frac{\dd}{\dd t} + \frac{k^2}{a^2(t)} + V_\so(t) \right] g_k(t;t') = \delta(t - t').
  \label{eq:green_eom}
\end{align}
Using this, we can reexpress Eq.~(\ref{eq:dt_one_pt}) as 
\begin{align}
  \expval{\delta \phi(\bfx,t)}
  &=  -\frac{1}{2} \int^t_{t_i} \dd t' g_0(t;t')\left[ V_\tho(t') \int \frac{\dd^3 k}{(2\pi)^3} |u_{k}(t')|^2 + 2 V_{c,\fo}(t') \right],
  \label{eq:one_pt_g}
\end{align}
where $g_0(t;t') \equiv \lim_{q \to 0} g_q(t;t')$.

We are interested in the backreaction to the inflaton velocity: 
\begin{align}
  \langle \delta \dot \phi(\bfx,t) \rangle
  &=  -\frac{1}{2} \int^t_{t_i} \dd t' \frac{\dd g_0(t;t')}{\dd t}\left[ V_\tho(t') \int \frac{\dd^3 k}{(2\pi)^3} |u_{k}(t')|^2 + 2 V_{c,\fo}(t') \right].
  \label{eq:dt_one_pt_g}
\end{align}
To proceed, we take a time derivative of Eq.~(\ref{eq:green_eom}),
\begin{align}
  \left[ \frac{\dd^2}{\dd t^2} + 3H \frac{\dd}{\dd t} + \frac{k^2}{a^2(t)} + V_\so(t) \right] \frac{\dd g_k(t;t')}{\dd t} = \frac{\dd}{\dd t}\delta(t - t') + \left[ 2 H\frac{k^2}{a^2(t)} - V_\tho(t) \dot {\bar\phi}(t) \right] g_k(t;t').
\end{align}
Solving this with the Green function method, we obtain 
\begin{align}
  \frac{\dd g_k(t;t')}{\dd t} &= \int^t_{-\infty} \dd t''\, g_k(t;t'') \left[ \frac{\dd}{\dd t''}\delta(t'' - t') + \left[ 2 H\frac{k^2}{a^2(t'')} - V_\tho(t'') \dot {\bar\phi}(t'') \right] g_k(t'';t') \right] \nonumber \\
  &= -\frac{\dd g_k(t;t')}{\dd t'}  + \int^t_{t'} \dd t''\, g_k(t;t'') \left[ 2 H\frac{k^2}{a^2(t'')} - V_\tho(t'') \dot {\bar\phi}(t'') \right] g_k(t'';t'),
\end{align}
where we have used the condition of the retarded Green function, $g_k(t;t') = \dd g_k(t;t')/\dd t = 0$ for $t<t'$.
Substituting this into Eq.~(\ref{eq:dt_one_pt_g}), we obtain 
\begin{align}
  \langle \delta \dot \phi(\bfx,t) \rangle
  &=  -\frac{1}{2} \int^t_{t_i} \dd t' \left[ -\frac{\dd g_0(t;t')}{\dd t'} + \int^t_{t'} \dd t''\, g_0(t;t'') \left[ - V_\tho(t'') \dot {\bar\phi}(t'') \right] g_0(t'';t') \right] \left[ V_\tho(t') \int \frac{\dd^3 k}{(2\pi)^3} |u_{k}(t')|^2 + 2 V_{c,\fo}(t') \right] \nonumber \\
   &= \langle \delta \dot \phi(\bfx,t) \rangle_\tad + \langle \delta \dot \phi(\bfx,t) \rangle_{1\vx} + \langle \delta \dot \phi(\bfx,t) \rangle_{2\vx}.
  \label{eq:dt_one_pt_g2}
\end{align}
$\langle \delta \dot \phi(\bfx,t) \rangle_\tad$ is defined as 
\begin{align}
 &\langle \delta \dot \phi(\bfx,t) \rangle_\tad \equiv \frac{1}{2} \int^t_{t_i} \dd t'\int^t_{t'} \dd t''\, g_0(t;t'') V_\tho(t'') \dot {\bar\phi}(t'') g_0(t'';t') \left[ V_\tho(t') \int \frac{\dd^3 k}{(2\pi)^3} |u_{k}(t')|^2 + 2 V_{c,\fo}(t') \right] \nonumber \\
& = 2 \int^\eta_{\eta_i} \dd \eta'\int^{\eta'}_{\eta_i} \dd \eta''\, a^4(\eta') a^4(\eta'')\Im[u_0(\eta) u^*_0(\eta')] V_\tho(\eta') \dot {\bar\phi}(\eta') \Im[u_0(\eta')u^*_0(\eta'')]
 \left[ V_\tho(\eta'') \int \frac{\dd^3 k}{(2\pi)^3} |u_{k}(\eta'')|^2 + 2 V_{c,\fo}(\eta'') \right],
 \label{eq:tad_back}
\end{align}
where we have changed the variables as $\eta' = \int^{t''} \dd \bar t/a(\bar t)$ and $\eta'' = \int^{t'} \dd \bar t/a(\bar t)$ in the second line. 
$\langle \delta \dot \phi(\bfx,t) \rangle_{1\vx}$ is defined as
\begin{align}
  \langle \delta \dot \phi(\bfx,t) \rangle_{1\vx} &\equiv -\frac{1}{2} \int^t_{t_i} \dd t' g_0(t;t') \dot{\bar \phi}(t') \left[V_\foo(t') \int \frac{\dd^3 k}{(2\pi)^3} |u_{k}(t')|^2 + 2V_{c,\so}(t') \right] \nonumber \\
  &= \int^\eta_{\eta_i} \dd \eta' a^4(\eta') \Im[u_0(\eta)u_0^*(\eta')]\dot{\bar \phi}(\eta') \left[ V_\foo(\eta') \int \frac{\dd^3 k}{(2\pi)^3} |u_{k}(\eta')|^2 + 2V_{c,\so}(\eta') \right].
  \label{eq:br_1vx}
\end{align}
$\langle \delta \dot \phi(\bfx,t) \rangle_{2\vx}$ is defined as 
\begin{align}
  \label{eq:2vx_tad}
  \langle \delta \dot \phi(\bfx,t) \rangle_{2\vx} &\equiv -\frac{1}{2} \int^t_{t_i} \dd t' g_0(t;t') V_\tho(t') \frac{\dd}{\dd t'} \int \frac{\dd^3 k}{(2\pi)^3} |u_{k}(t')|^2.
\end{align}
This expression of $\langle \delta \dot \phi(\bfx,t) \rangle_{2\vx}$ needs further calculation before changing the physical time to the conformal time.
Let us focus on the time derivative term:
\begin{align}
  \frac{\dd}{\dd t'} \int \frac{\dd^3 k}{(2\pi)^3} |u_k(t')|^2 &= \frac{\dd}{\dd t'} \int^{k_\uv \frac{a(t_\min)}{a_i}}_{k_\ir} \frac{k^2 \dd k}{2\pi^2} |u_k(t')|^2 \nonumber \\
  &= H \mathcal P_{\delta \phi,\tre}\left( k_\uv \frac{a(t')}{a_i},t' \right) + \int^{k_\uv \frac{a(t_\min)}{a_i}}_{k_\ir} \frac{k^2 \dd k}{2\pi^2} \left(\frac{\dd u_k(t')}{\dd t'} u_k^*(t') + u_k(t')\frac{\dd u^*_k(t')}{\dd t'} \right),
  \label{eq:u_k2_dt}
\end{align}
where we have introduced the UV cutoff with the physical scale, $k_\uv (a(t)/a_i)$ with $a_i \equiv a(t_i)$, and the IR cutoff with the comoving scale $k_\ir$, similar to Refs.~\cite{Senatore:2009cf,Giddings:2010nc,Byrnes:2010yc,Senatore:2012nq}.
We set the IR cutoff scale ($1/k_\ir$) to be much smaller than the scales where the perturbations are enhanced during the non-SR period, while $k_\ir$ can be on superhorizon as long as $q \ll k_\ir$ is satisfied when we discuss the one-loop power spectrum.
We set $t_\min = t'$ except for the term dependent on $u_k(t'')$ with $t'' < t'$, which appears in the expression of $\dd u_k(t')/\dd t'$ (in Eq.~(\ref{eq:du_dt})).
Hence, the time derivative acts on the upper bound on the $k$ integral, which leads to the first term in the second line of Eq.~(\ref{eq:u_k2_dt}).
For the term dependent on $u_k(t'')$ with $t'' < t'$, we set $t_\min = t''$.
This is because, if the loop integral includes perturbations with different times, the physical UV cutoff must be fixed with the smallest $k_\uv a(t)/a_i$, which is determined by the perturbation at the earliest time~\cite{Senatore:2009cf}.
Note that, even if we use $a(t')$ instead of $a(t_\min)$ in Eq.~(\ref{eq:u_k2_dt}), we can obtain the same expression (Eq.~(\ref{eq:u_k2_dt_2})) by assuming that $u_k$ is switched on at the physical cutoff off scale with the Bunch-Davies vacuum. See Appendix~\ref{app:uv_bound} for details.

To obtain the expression of $\dd u_k(t)/\dd t$, we take the time derivative of Eq.~(\ref{eq:dt_one_pt}):
\begin{align}
  \left[\frac{\dd^2}{\dd t^2} + 3 H \frac{\dd}{\dd t} + \frac{k^2}{a^2} + V_\so(t) \right] \frac{\dd u_k(t)}{\dd t} = \left(2 H \frac{k^2}{a^2} - V_\tho(t) \dot{\bar \phi} \right)u_k(t).
  \label{eq:u_eom_dt}
\end{align}
To proceed, we also take the $\ln k$ derivative on Eq.~(\ref{eq:dt_one_pt}):
\begin{align}
  \left[\frac{\dd^2}{\dd t^2} + 3 H \frac{\dd}{\dd t} + \frac{k^2}{a^2} + V_\so(t) \right] \frac{\dd u_k(t)}{\dd \ln k} = -\frac{2 k^2}{a^2}u_k(t).
  \label{eq:u_eom_dk}
\end{align}
Combining this and Eq.~(\ref{eq:u_eom_dt}), we obtain
\begin{align}
  \left[\frac{\dd^2}{\dd t^2} + 3 H \frac{\dd}{\dd t} + \frac{k^2}{a^2} + V_\so(t) \right] \left(\frac{\dd u_k(t)}{\dd t} + H \frac{\dd u_k(t)}{\dd \ln k}\right) = - V_\tho(t) \dot{\bar \phi}\, u_k(t).
  \label{eq:u_eom_dt2}
\end{align}
Using the Green function method, we get 
\begin{align}
  \frac{\dd u_k(t)}{\dd t} &= -\int^t_{t_i} \dd t' g_k(t;t') V_\tho(t') \dot {\bar \phi}(t') u_k(t') - H \frac{\dd u_k(t)}{\dd \ln k} + C H u_k(t) + D H u_k^*(t) \nonumber \\
  &= 2\int^\eta_{\eta_i} \dd \eta' a^4(\eta') V_\tho(\eta') \dot {\bar \phi}(\eta') \Im[u_k(\eta) u^*_k(\eta')] u_k(\eta') - H \frac{\dd u_k(\eta)}{\dd \ln k} + C H u_k(\eta) + D H u_k^*(\eta),
  \label{eq:du_dt}
\end{align}
where $C$ and $D$ are some constants, determined by the initial condition, and $H$ is put in front of them to make $C$ and $D$ dimensionless.
In the second line, we have changed the physical time to the comoving time and used Eq.~(\ref{eq:g_tt}).
Let us here determine $C$ and $D$ by using the expression of $u_k$ in $\eta<\eta_i$, Eq.~(\ref{eq:u_k_ini}).
In $\eta < \eta_i$, Eq.~(\ref{eq:du_dt}) can be reexpressed as 
\begin{align}
  \label{eq:u_k_eta}
  \frac{\dd u_k(\eta)}{\dd \ln \eta}
  &= \frac{\dd u_k(\eta)}{\dd \ln k} - C u_k(\eta) - D u_k^*(\eta) \quad \text{ for } \eta< \eta_i,
\end{align}
where we have used $\dd/\dd t = (1/a)\dd/\dd \eta$ and $a H = -1/\eta$.
Before substituting Eq.~(\ref{eq:u_k_ini}) into this, we reexpress Eq.~(\ref{eq:u_k_ini}) as 
\begin{align}
  u_k(\eta) = k^{-3/2} h(k\eta) \quad \text{ for } \eta < \eta_i,
\end{align}
where 
\begin{align}
  h(k\eta) \equiv -\frac{H k\eta}{2}\ee^{i\frac{\left(2\nu + 1\right)\pi}{4}} \sqrt{-\pi k\eta} H_\nu^\fo(-k\eta).
\end{align}
From this expression, we can easily see 
\begin{align}
  \frac{\dd u_k(\eta)}{\dd \ln \eta} -  \frac{\dd u_k(\eta)}{\dd \ln k} = \frac{3}{2}u_k(\eta) \quad \text{ for } \eta < \eta_i.
\end{align}
Substituting this into Eq.~(\ref{eq:u_k_eta}), we can determine $C = -3/2$ and $D=0$.
Then, we finally obtain 
\begin{align}
  \frac{\dd u_k(t)}{\dd t} &= 2\int^\eta_{\eta_i} \dd \eta' a^4(\eta') V_\tho(\eta') \dot {\bar \phi}(\eta') \Im[u_k(\eta) u^*_k(\eta')] u_k(\eta') - H k^{-3/2}\frac{\dd (k^{3/2} u_k(\eta))}{\dd \ln k}.
  \label{eq:du_dt_f}
\end{align}
Substituting this into Eq.~(\ref{eq:u_k2_dt}), we obtain
\begin{align}
  \frac{\dd}{\dd t'} \int \frac{\dd^3 k}{(2\pi)^3} |u_k(t')|^2 
  &= H \mathcal P_{\delta \phi,\tre}\left( k_\uv \frac{a(\eta')}{a_i},\eta' \right) - H \int^{k_\uv \frac{a(\eta')}{a_i}}_{k_\ir} \dd \ln k \frac{1}{2\pi^2} \frac{\dd (k^3|u_k(\eta')|^2)}{\dd \ln k} \nonumber \\
  &\quad + 4 \int^{\eta'}_{\eta_i} \dd \eta'' a^4(\eta'') V_\tho(\eta'') \dot {\bar \phi}(\eta'') \int^{k_\uv \frac{a(\eta'')}{a_i}}_{k_\ir} \frac{k^2 \dd k}{2\pi^2}  \Im[u_k(\eta') u^*_k(\eta'')] \Re[u_k(\eta') u^*_k(\eta'')] \nonumber \\
  &=  H \mathcal P_{\delta \phi,\tre}\left( k_\ir,\eta' \right) + 4 \int^{\eta'}_{\eta_i} \dd \eta'' a^4(\eta'') V_\tho(\eta'') \dot {\bar \phi}(\eta'') \int^{k_\uv \frac{a(\eta'')}{a_i}}_{k_\ir} \frac{k^2 \dd k}{2\pi^2}\Im[u_k(\eta') u^*_k(\eta'')] \Re[u_k(\eta') u^*_k(\eta'')],
  \label{eq:u_k2_dt_2}
\end{align}
where note again that the physical UV cutoff scale is determined by the earliest perturbations involved in the loop integral.
Using this equation, we can reexpress Eq.~(\ref{eq:2vx_tad}) with the comoving time as 
\begin{align}
  \label{eq:ddot_2vx}
  &\langle \delta \dot \phi(\bfx,t) \rangle_{2\vx} = \int^\eta_{\eta_i} \dd \eta' a^4(\eta')\Im[u_0(\eta) u_0^*(\eta')] H  V_\tho(\eta') \mathcal P_{\delta \phi,\tre}\left( k_\ir,\eta' \right) \nonumber \\
  & +4 \int^\eta_{\eta_i} \dd \eta' \int^{\eta'}_{\eta_i} \dd \eta'' a^4(\eta') a^4(\eta'') V_\tho(\eta') V_\tho(\eta'') \dot{\bar \phi}(\eta'')\Im[u_0(\eta) u_0^*(\eta')]\int^{k_\uv \frac{a(\eta'')}{a_i}}_{k_\ir} \frac{k^2 \dd k}{2\pi^2} \Im[u_k(\eta') u^*_k(\eta'')] \Re[u_k(\eta') u^*_k(\eta'')] \nonumber \\
  &= \langle \delta \dot \phi(\bfx,t) \rangle_{\ir} + \langle \delta \dot \phi(\bfx,t) \rangle_{\overline{2\vx}},
\end{align}
where $\langle \delta \dot \phi(\bfx,t) \rangle_{\ir}$ is the first term in the first equality, which comes from the IR cutoff, and $\langle \delta \dot \phi(\bfx,t) \rangle_{\overline{2\vx}}$ is the second term.

%%%%%%%%%%%%%%%%%%%
\section{Conservation of curvature perturbations}
\label{sec:curv_cons}
%%%%%%%%%%%%%%%%%%%

Using the results in the previous sections, let us see the conservation of curvature perturbations at one-loop level.
We first investigate the time-dependence of the following quantity, which will be connected to the curvature power spectrum later (Eq.~(\ref{eq:zeta_cons_fin})):
\begin{align}
 \frac{\expval{\delta \phi_{\bfq}(\eta) \delta \phi_{\bfq'}(\eta)}}{\langle \dot{\phi}(\eta) \rangle^2} = \frac{(2\pi)^3 \delta(\bfq + \bfq') \frac{2\pi^2}{q^3}\mathcal P_{\delta \phi}(q,\eta)}{\left(\dot{\bar \phi}(\eta) + \langle \delta \dot \phi(\eta) \rangle\right)^2} = \frac{(2\pi)^3 \delta(\bfq + \bfq') \frac{2\pi^2}{q^3}\mathcal P_{\delta \phi,\tre}(q,\eta) \left(1 + \frac{\mathcal P_{\delta \phi}(q,\eta) - \mathcal P_{\delta \phi,\tre}(q,\eta)}{\mathcal P_{\delta \phi,\tre}(q,\eta)} \right)}{ \left(\dot{\bar \phi}(\eta)\right)^2 \left(1 + 2\frac{\langle \delta \dot \phi(\eta) \rangle}{\dot {\bar \phi}(\eta)}\right)}, 
 \label{eq:zeta_cons}
\end{align}
where $\mathcal P_{\delta \phi}$ is the sum of the tree and the loop contributions (the last line in Eq.~(\ref{eq:two_vx_one_loop}) up to one-loop level) and we have neglected the higher-order loop contributions in the second equality.
We focus on the conservation of the curvature perturbations whose scales satisfy $q \ll -1/\eta_i$ (superhorizon limit). 
On such scales, we can regard $\zeta_q (\equiv -H u_q/\dot{\bar \phi})$ as a constant.
From this, we can see $\mathcal P_{\delta \phi,\tre}(q,\eta)/(\dot {\bar \phi}(\eta))^2 = \mathcal P_{\zeta,\tre}(q)/H^2 = \text{const.}$ with $\mathcal P_{\zeta,\tre}(q) = q^3 |\zeta_q|^2/(2\pi^2)$.
Then, the time-dependence in Eq.~(\ref{eq:zeta_cons}) is determined only by the following two quantities:
\begin{align}
  \frac{\mathcal P_{\delta \phi}(q,\eta) - \mathcal P_{\delta \phi, \tre}(q,\eta)}{\mathcal P_{\delta \phi,\tre}(q,\eta)} &= \frac{\mathcal P_{\delta \phi, 1\vx}(q,\eta) + \mathcal P_{\delta \phi, 2\vx}(q,\eta) + \mathcal P_{\delta \phi,\tad}(q,\eta)}{\mathcal P_{\delta \phi,\tre}(q,\eta)}, \\
  \frac{\langle \delta \dot \phi(\eta) \rangle}{\dot {\bar \phi}(\eta)} &= \frac{\langle \delta \dot \phi(\eta) \rangle_\tad + \langle \delta \dot \phi(\eta) \rangle_{1\vx} + \langle \delta \dot \phi(\eta) \rangle_{2\vx}}{\dot {\bar \phi}(\eta)},
  \label{eq:cons_cond}
\end{align}
where we have neglected the higher-order loop contributions.
In the following, we see how these two quantities are related to each other.

From Eqs.~(\ref{eq:1vx}) and (\ref{eq:br_1vx}), we can obtain
\begin{align}
  &\lim_{q \to 0}\frac{\mathcal P_{\delta \phi, 1\vx}(q,\eta)}{\mathcal P_{\delta \phi,\tre}(q,\eta)} = 2 \frac{\langle \delta \dot \phi(\eta) \rangle_{1\vx}}{\dot {\bar \phi}(\eta)} \nonumber \\
  &= 2 \int^\eta_{\eta_i} \dd \eta' a^4(\eta') \Im[u_0(\eta)u_0^*(\eta')]\frac{\dot{\bar \phi}(\eta')}{\dot{\bar \phi}(\eta)}\left[ V_\foo(\eta') \int \frac{\dd^3 k}{(2\pi)^3} |u_{k}(\eta')|^2 + 2V_{c,\so}(\eta') \right].
  \label{eq:1vx_cond}
\end{align}
where we have used 
\begin{align}
  \Re[u_q(\eta)u^*_q(\eta')] = \frac{\dot{\bar \phi}(\eta) \dot{\bar \phi}(\eta')}{H^2} |\zeta_q|^2, \ \ 
  \mathcal P_{\delta \phi,\tre}(q,\eta) = \frac{q^3}{2\pi^2} \left(\frac{\dot{\bar \phi}(\eta)}{H}\right)^2 |\zeta_q|^2,
\end{align}
where note again that $\zeta_q$ is constant in the superhorizon limit.
Similarly, from Eqs.~(\ref{eq:tad}) and (\ref{eq:tad_back}), we can obtain
\begin{align}
  &\lim_{q \to 0}\frac{\mathcal P_{\delta \phi, \tad}(q,\eta)}{\mathcal P_{\delta \phi,\tre}(q,\eta)} = 2 \frac{\langle \delta \dot \phi(\eta) \rangle_{\tad}}{\dot {\bar \phi}(\eta)}\nonumber \\
  & =  4 \int^\eta_{\eta_i} \dd \eta'\int^{\eta'}_{\eta_i} \dd \eta''\, a^4(\eta') a^4(\eta'')\Im[u_0(\eta) u^*_0(\eta')] V_\tho(\eta') \frac{\dot {\bar\phi}(\eta')}{\dot{\bar \phi}(\eta)} \Im[u_0(\eta')u^*_0(\eta'')]
 \left[ V_\tho(\eta'') \int \frac{\dd^3 k}{(2\pi)^3} |u_{k}(\eta'')|^2 + 2 V_{c,\fo}(\eta'') \right].
  \label{eq:tad_cond}
\end{align}
For the two-vertex contributions, we can neglect $\mathcal P^a_{\delta \phi,2\vx}$ because it is suppressed by $q^3$.
Then, from Eqs.~(\ref{eq:p_b}) and (\ref{eq:ddot_2vx}), we can obtain
\begin{align}
  &\lim_{q \to 0}\frac{\mathcal P_{\delta \phi, 2\vx}(q,\eta)}{\mathcal P_{\delta \phi,\tre}(q,\eta)} = 2 \frac{\langle \delta \dot \phi(\eta) \rangle_{\overline{2\vx}}}{\dot {\bar \phi}(\eta)} \nonumber \\
  &= 8 \int^\eta_{\eta_i} \dd \eta' \int^{\eta'}_{\eta_i} \dd \eta'' a^4(\eta') a^4(\eta'') V_\tho(\eta') V_\tho(\eta'') \frac{\dot{\bar \phi}(\eta'')}{\dot{\bar \phi}(\eta)}
  \Im[u_0(\eta) u_0^*(\eta')] \int^{k_\uv \frac{a(\eta'')}{a_i}}_{k_\ir} \frac{k^2 \dd k}{2\pi^2} \Im[u_k(\eta') u^*_k(\eta'')] \Re[u_k(\eta') u^*_k(\eta'')].
  \label{eq:2vx_cond}
\end{align}

From Eqs.~(\ref{eq:1vx_cond}), (\ref{eq:tad_cond}), and (\ref{eq:2vx_cond}), we can see 
\begin{align}
  \lim_{q \to 0}\frac{\mathcal P_{\delta \phi}(q,\eta) - \mathcal P_{\delta \phi, \tre}(q,\eta)}{\mathcal P_{\delta \phi,\tre}(q,\eta)} &= 2 \frac{\langle \delta \dot \phi(\eta) \rangle - \langle \delta \dot \phi(\eta) \rangle_\ir}{\dot {\bar \phi}(\eta)}.
  \label{eq:zeta_cons2}
\end{align}
Note that this equation means that, if we tune $V_{c,\fo}$ to satisfy $\expval{\delta \phi} = 0$ (and therefore $\langle \delta \dot \phi \rangle = 0$), the sum of the one-loop power spectra automatically satisfies $\lim_{q \to 0} (\mathcal P_{\delta \phi,1\vx}(q) + \mathcal P_{\delta \phi,2\vx}(q)) = -2 \mathcal P_{\delta \phi, \tre}(q)(\langle \delta \dot \phi(\eta) \rangle_\ir/\dot {\bar \phi}(\eta))$.\footnote{$\mathcal P_{\delta \phi,\tad} = 0$ when $\expval{\delta \phi} = 0$. Meanwhile, $\langle \delta \dot \phi \rangle_\ir$ can be nonzero even if we impose $\langle \delta \dot \phi \rangle = \langle \delta \dot \phi \rangle_{1\vx} + \langle \delta \dot \phi \rangle_{2\vx} + \langle \delta \dot \phi \rangle_\tad + \langle \delta \dot \phi \rangle_\ir = 0$.}
Substituting Eq.~(\ref{eq:zeta_cons2}) into Eq.~(\ref{eq:zeta_cons}), we obtain
\begin{align}
   \lim_{q \to 0}\frac{\expval{\delta \phi_{\bfq}(\eta) \delta \phi_{\bfq'}(\eta)}}{\langle \dot{\phi}(\eta) \rangle^2} = (2\pi)^3 \delta(\bfq + \bfq') \frac{2\pi^2}{q^3} \frac{\mathcal P_{\zeta,\tre}(q)}{H^2} \left( 1 - 2\frac{\langle \delta \dot \phi(\eta) \rangle_\ir}{\dot {\bar \phi}(\eta)} \right).
   \label{eq:del_phi_cons}
\end{align}
If we neglect the IR cutoff contribution $\langle \delta \dot \phi(\eta) \rangle_\ir$, this quantity is conserved. 
Setting the IR cutoff scale ($1/k_\ir$) to be much larger than the scales where the perturbations are enhanced, we neglect its contribution in the following. 
See Appendix~\ref{app:ir} for the physical interpretation of the IR cutoff contribution.

We here connect the left-hand side of Eq.~(\ref{eq:del_phi_cons}) to the curvature power spectrum during the SR periods ($\eta<\eta_i$ or $\eta > \eta_e$).
We here use the separate universe assumption with the smoothing scale of $1/k_\ir (\ll 1/q)$ during the SR periods~\cite{Lyth:2004gb}.
Then, we can use the $\delta N$ formalism for the perturbations on the larger scales ($> 1/k_\ir$):
\begin{align}
  \zeta = \delta N = H\frac{\delta \phi}{\langle\dot \phi \rangle} + \mathcal O\left( \frac{m_{\phi}^2}{H^2} \left(H\frac{\delta \phi}{\langle\dot \phi \rangle}\right)^2 \right) \quad \text{for } \eta < \eta_i \text{ or } \eta > \eta_e,
  \label{eq:zeta_def}
\end{align}
where $m^2_{\phi} = V_\so(\eta_i)$ in $\eta<\eta_i$ and $= V_\so(\eta_e)$ in $\eta>\eta_e$.
This $\zeta$ corresponds to the gauge-invariant curvature perturbations in the uniform density gauge, which becomes identical to the gauge-invariant curvature perturbations in the comoving gauge in the superhorizon limit~\cite{Lyth:2004gb}.
Note that $V_\so=\text{const.}$ in $\eta<\eta_i$ or $\eta > \eta_e$ and $V_\so(\eta_i) \neq V_\so(\eta_e)$ in general.
The background velocity, $\langle\dot \phi \rangle$, includes the backreaction from the perturbations whose scales are smaller than $1/k_\ir$, which we have calculated in Sec.~\ref{sec:backreaction}.
We assume $|m^2_{\phi}/H^2| \ll 1$ and neglect the higher-order contributions in Eq.~(\ref{eq:zeta_def}).
Substituting this into Eq.~(\ref{eq:del_phi_cons}), we obtain 
\begin{align}
   \lim_{q \to 0}\expval{\zeta_{\bfq}(\eta) \zeta_{\bfq'}(\eta)} = \lim_{q \to 0}\frac{H^2\expval{\delta \phi_{\bfq}(\eta) \delta \phi_{\bfq'}(\eta)}}{\langle \dot{\phi}(\eta) \rangle^2} = (2\pi)^3 \delta(\bfq + \bfq') \frac{2\pi^2}{q^3} \mathcal P_{\zeta,\tre}(q) \quad  \text{for } \eta < \eta_i \text{ or } \eta > \eta_e,
   \label{eq:zeta_cons_fin}
\end{align}
where note again we have neglected the IR cutoff contribution.
This shows that the superhorizon-limit curvature perturbations in $\eta < \eta_i$ and $\eta>\eta_e$ are the same, which means the conservation of curvature perturbations at one-loop level.

Before closing this section, let us briefly mention the curvature perturbations during the non-SR period ($\eta_i < \eta < \eta_e$).
If we use the separate universe assumption even during the non-SR period, we can connect the inflaton fluctuations to curvature perturbations on $q < k_\ir$ as Eq.~(\ref{eq:zeta_def}). 
Unlike during SR periods, the higher-order contributions, which correspond to the $m_\phi^2/H^2$ term in Eq.~(\ref{eq:zeta_def}), are non-negligible during the non-SR period.
However, the leading order contributions from higher-order contributions in Eq.~(\ref{eq:zeta_def}) to the two-point correlation function $\expval{\zeta_{\bfq} \zeta_{\bfq'}}$ is of $\mathcal O(X \mathcal P_{\zeta,\tre}(k_1) \mathcal P_{\zeta,\tre}(k_2))$ with $k_1, k_2 < k_\ir$ and $X$ determined by the detail of the potential. 
This is because the modes with $k> k_\ir$ are already smoothed out in Eq.~(\ref{eq:zeta_def}).
Unless $X$ is extremely large, the contributions of $\mathcal O(X \mathcal P_{\zeta,\tre}(k_1) \mathcal P_{\zeta,\tre}(k_2))$ are much smaller than the small-scale enhanced contributions by the definition of $k_\ir$.
If we neglect the contributions of $\mathcal O(X \mathcal P_{\zeta,\tre}(k_1) \mathcal P_{\zeta,\tre}(k_2))$ (consistently with the introduction of $k_\ir$ for the loop integrals), we can see the conservation of the curvature perturbations even during $\eta_i < \eta < \eta_e$.

\section{Renormalization}
\label{sec:renorm}

In the previous sections, we have shown the conservation of curvature perturbations without discussing the UV renormalization.
The point is that, even if the UV divergences were not cancelled by the counter terms, the UV divergences for $\expval{\delta \phi_{\bfq}\delta \phi_{\bfq'}}$ and $\langle \dot \phi^2 \rangle$ would appear in the same way and be cancelled when we discuss the curvature power spectrum with $\expval{\zeta_{\bfq}\zeta_{\bfq'}} = H^2\expval{\delta \phi_{\bfq}\delta \phi_{\bfq'}}/\langle \dot \phi^2 \rangle$. This is why we did not need to explicitly discuss the UV renormalization at least for the proof of the curvature conservation.
However, if the UV divergences could not be cancelled by the counter terms, each of $\expval{\delta \phi_\bfq \delta \phi_{\bfq'}}$ and $\langle \dot {\phi}^2 \rangle$ would become unphysical at one-loop level.
In this section, we see how the counter terms cancel the UV divergences of $\expval{\delta \phi_\bfq \delta \phi_{\bfq'}}$ and $\langle \dot {\phi}^2 \rangle$.

Before seeing the concrete loop contributions, let us first focus on the UV behavior of the variance $\expval{\delta \phi^2} = \int \dd^3 k/(2\pi)^3 |u_k(\eta)|^2$, the origin of the UV divergences in the loop power spectrum.
Substituting Eq.~(\ref{eq:u_k_subh_limit}) into the loop integral, we can separate the UV cutoff dependent contributions as 
\begin{align}
  \label{eq:u2_dk}
  \int \frac{\dd k^3}{(2\pi)^3} |u_k(\eta)|^2 = \int^{k_\uv \frac{a(\eta)}{a_i}}_{k_\ir}\frac{k^2 \dd k}{2\pi^2} |u_k(\eta)|^2 = \frac{H^2}{8\pi^2} \left( \frac{k_\uv}{a_iH}\right)^2 + E(\eta) \ln \left(\frac{k_\uv}{a_i H}\right) + F(\eta),
\end{align}
where $E$ and $F$ are functions of time.
Note that there are no terms of $\mathcal O(k_\uv)$ because $u_k \propto (1 + i A(\eta)/(-k\eta))$ with $\Im[A(\eta)]=0$ (see below Eq.~(\ref{eq:u_k_subh_limit})).
Strictly speaking, we can move the contribution $\propto \ln(1/(a_iH))$ to the $k_\uv$ independent function $F$, but we separate it from $F$ to make the term inside the logarithm dimensionless.

Let us obtain the concrete expression of $E(\eta)$.
In $\eta < \eta_i$, we can obtain $E(\eta)$ by using Eq.~(\ref{eq:u_k_subh_limit}),
\begin{align}
  \label{eq:e_ini}
  E(\eta) = \frac{H^2}{32\pi^2} (4\nu^2-1) \ \ \text{ for } \eta < \eta_i.
\end{align}
To determine the form of $E$ in $\eta > \eta_i$, we use the result of Eq.~(\ref{eq:u_k2_dt_2}):
\begin{align}
  \frac{\dd}{\dd \eta} \int \frac{\dd k^3}{(2\pi)^3} |u_k(\eta)|^2 
  &= -\frac{\mathcal P_{\delta \phi,\tre}\left( k_\ir,\eta \right)}{\eta} + 4a(\eta) \int^{\eta}_{\eta_i} \dd \eta' a^4(\eta') V_\tho(\eta') \dot {\bar \phi}(\eta') \int^{k_\uv \frac{a(\eta')}{a_i}}_{k_\ir} \frac{k^2 \dd k}{2\pi^2} \Im[u_k(\eta) u^*_k(\eta')] \Re[u_k(\eta) u^*_k(\eta')] \nonumber \\
  &= 2a(\eta) \int^{\eta}_{\eta_i} \dd \eta' a^4(\eta') V_\tho(\eta') \dot {\bar \phi}(\eta') \int^{k_\uv \frac{a(\eta')}{a_i}}_{k_* \frac{a(\eta')}{a_i}} \frac{k^2 \dd k}{2\pi^2} \Im[ (u_k(\eta) u^*_k(\eta'))^2] + \tilde F(\eta,k_*),
  \label{eq:u_k2_dt_ren}
\end{align}
where $\tilde F$ is some function of $\eta$ and we have introduced the lower cutoff $k_*$, which can be any scale as long as it satisfies $\max[a_i H, a_i/\Delta t_V] \ll k_* < k_\uv$. 
$\Delta t_V$ is the shortest timescale during which the potential varies by $|\Delta V_\no/V_\no|_{n \leq 4} \simeq \mathcal O(1)$.
For example, if we consider a sudden transition between the SR and non-SR periods, $\Delta t_V$ becomes small.
Note that we can fix $k_*$ once the potential is fixed.
We are interested in the $k_\uv$ dependence in the limit of $k_\uv \to \infty$ with $k_*$ fixed.
We introduce $k_*$ to use the subhorizon limit expression, given in Eq.~(\ref{eq:u_k_subh_limit}).\footnote{
  In $k< a_i/\Delta t_V$, $u_k$ can deviate from the Bunch-Davies solution (Eq.~(\ref{eq:u_k_subh_limit})) even on the subhorizon scales due to the particle production associated with a non-adiabatic change of the background evolution~\cite{Inomata:2021uqj,Inomata:2021tpx}.
}
Substituting Eq.~(\ref{eq:u_k_subh_limit}) into this expression and taking $k_\uv \to \infty$ limit, we obtain 
\begin{align}
  \label{eq:u2_dk_deta}
  \frac{\dd}{\dd \eta} \int \frac{\dd k^3}{(2\pi)^3} |u_k(\eta)|^2 
  &= -2a(\eta) \int^{\eta}_{\eta_i} \dd \eta' a^4(\eta') V_\tho(\eta') \dot {\bar \phi}(\eta') \int^{k_\uv \frac{a(\eta')}{a_i}}_{k_* \frac{a(\eta')}{a_i}} \frac{\dd k}{8\pi^2}  \frac{\sin\left[ 2k(\eta - \eta')\right]}{a^2(\eta) a^2(\eta')} + \tilde F(\eta ,k_*) \nonumber \\
  &= 2a(\eta) \int^{\eta-\eta_i}_{0} \dd \eta_- \frac{a^2(\eta')}{a^2(\eta)} V_\tho(\eta') \dot {\bar \phi}(\eta') \frac{\cos[2 k_\uv \frac{a(\eta')}{a_i} \eta_- ] -\cos[2 k_* \frac{a(\eta')}{a_i}\eta_-]}{16 \pi^2 \eta_-}+ \tilde F(\eta,k_*) \nonumber \\
  &= -\frac{1}{8\pi^2} a(\eta) V_\tho(\eta) \dot {\bar \phi}(\eta) \ln \left(\frac{k_\uv}{k_*} \right)+ \tilde F(\eta ,k_*),
\end{align}
where $\eta_- = \eta-\eta'$ and we have used $\int^\infty_{0} \dd z (\cos(a z) - \cos(b z))/z = -\ln(a/b)$.
Since we are interested in the $k_\uv$ dependent term, we express $\ln(k_\uv/k_*) = \ln(k_\uv/(a_iH)) + \ln(a_iH/k_*)$ and add the latter term to $\tilde F(\eta,k_*)$.
Note that $F'(\eta) = \tilde F(\eta,k_*) + \ln(a_iH/k_*)$ and the $k_*$ dependence is cancelled for $F'(\eta)$. This can be seen from the fact the left hand side of Eq.~(\ref{eq:u2_dk_deta}) does not depend on $k_*$, which can be set arbitrarily under the condition $\max[a_i H, a_i/\Delta t_V] \ll k_* < k_\uv$.
Then, comparing Eqs.~(\ref{eq:u2_dk}) and (\ref{eq:u2_dk_deta}), we obtain 
\begin{align}
  E'(\eta) = -\frac{1}{8\pi^2} V_\tho(\eta) {\bar \phi}'(\eta).
  \label{eq:e_dash}
\end{align}
With the initial condition given by Eq.~(\ref{eq:e_ini}), we obtain the complete expression of $E(\eta)$:
\begin{align}
  E(\eta) &=  -\frac{1}{8\pi^2} \int^{\eta}_{\eta_i} \dd \eta' V_\tho(\eta') {\bar \phi}'(\eta') + \frac{H^2}{32\pi^2} (4\nu^2-1) \nonumber \\
  &= \frac{H^2}{8\pi^2} \left( -\frac{V_\so(\eta) - V_\so(\eta_i) }{H^2} + \frac{4\nu^2-1}{4} \right) \nonumber \\
  &= \frac{H^2}{8\pi^2} \left( 2-\frac{V_\so(\eta)}{H^2} \right),
\end{align}
where we have used $4\nu^2 - 1 = 8 - 4V_\so(\eta_i)/H^2$ in the final line.

In the following, we calculate the UV divergences that appear in the one-loop power spectra.

\subsection{Tadpole and one-vertex contributions}

We begin with the tadpole and one-vertex contributions.
The $k_\uv$ dependent term in the tadpole contribution, Eq.~(\ref{eq:tad}), is given by\footnote{
Note that, although we define $\mathcal P^\uv$ as the $k_\uv$ dependent terms in each loop contribution throughout this section, $\mathcal P^\uv$ also includes $k_\uv$ independent terms (e.g. the terms proportional to $\ln(a_i H)$ or $\ln k_*$, as we will see in Eq.~(\ref{eq:p_b_ren})), which are arbitrary.
This arbitrariness does not matter in our UV renormalization procedure because we are only interested in whether the $k_\uv$ dependence can be removed by the counter terms.
Note that the $k_*$ dependence disappears when we see the whole power spectrum, $\mathcal P^\uv + (\text{finite power spectrum})$. This is ensured by the fact that the whole loop power spectrum does not depend on $k_*$, which we introduce just for convenience.
}
\begin{align}
   \mathcal P^\uv_{\delta \phi, \tad}(q,\eta) &= \frac{2q^3}{\pi^2} \int^\eta_{\eta_i} \dd \eta' \int^{\eta'}_{\eta_i} \dd \eta''a^4(\eta') a^4(\eta'') V_\tho(\eta') \Im\left[ u_q(\eta) u^*_q(\eta') \right] \Re\left[ u_q(\eta) u^*_q(\eta') \right] \nonumber \\
  &\qquad \times \Im[ u_{0}(\eta') u_{0}^*(\eta'')]  \left( V_\tho(\eta'') \left[\frac{H^2}{8\pi^2} \left( \frac{k_\uv}{a_iH}\right)^2 + E(\eta'') \ln \left(\frac{k_\uv}{a_i H}\right)\right]+ 2 V^\uv_{c,\fo}(\eta'')\right),
  \label{eq:tad_ren}  
\end{align}
where $V^{\uv}_{c,\no}$ is the $k_\uv$ dependent term in $V_{c,\no}$.
To make $\mathcal P_{\delta \phi,\tad}$ independent of $k_\uv$, we set
\begin{align}
  V^\uv_{c,\fo}(\eta) = -\frac{V_\tho(\eta)}{2} \left[\frac{H^2}{8\pi^2} \left( \frac{k_\uv}{a_iH}\right)^2 + E(\eta) \ln \left(\frac{k_\uv}{a_i H}\right)\right].
  \label{eq:v_c1_uv}
\end{align}
Note that, once $V_{c,\fo}$ is fixed, $V_{c,\so}$ is automatically fixed as
\begin{align}
  V^\uv_{c,\so}(\eta) = -\frac{V_\foo(\eta)}{2} \left[\frac{H^2}{8\pi^2} \left( \frac{k_\uv}{a_iH}\right)^2 + E(\eta) \ln \left(\frac{k_\uv}{a_i H}\right)\right] - \frac{V_\tho(\eta) E'(\eta) \ln \left(\frac{k_\uv}{a_i H}\right) }{2\bar \phi'(\eta)}.
\end{align}
Using this, we can express the $k_\uv$ dependent terms in the one-vertex contribution, Eq.~(\ref{eq:1vx}), as 
\begin{align}
  \mathcal P^\uv_{\delta \phi, 1\vx}(q,\eta)
  &=  \frac{q^3}{\pi^2} \int^\eta_{\eta_i} \dd \eta' a^4(\eta') \Im\left[ u_q(\eta) u^*_q(\eta') \right] \Re\left[ u_q(\eta) u^*_q(\eta') \right] \left(V_\foo(\eta') \left[ \frac{H^2}{8\pi^2} \left( \frac{k_\uv}{a_iH}\right)^2 + E(\eta') \ln \left(\frac{k_\uv}{a_i H}\right)\right] + 2V^\uv_{c,\so}(\eta')\right) \nonumber \\
  &=  \frac{q^3}{\pi^2} \int^\eta_{\eta_i} \dd \eta' \frac{1}{8\pi^2} a^4(\eta')  V^2_\tho(\eta') \Im\left[ u_q(\eta) u^*_q(\eta') \right] \Re\left[ u_q(\eta) u^*_q(\eta') \right] \ln \left(\frac{k_\uv}{a_i H}\right),
  \label{eq:1vx_ren}
\end{align}
where we have used Eq.~(\ref{eq:e_dash}).
We will see in the next subsection that this $k_\uv$ dependent term in $\mathcal P_{\delta \phi,1\vx}$ is cancelled by the $k_\uv$ dependent term in $\mathcal P_{\delta \phi,2\vx}$.

\subsection{Two-vertex contributions}

Let us first focus on the $\mathcal P^a_{2\vx}$ contribution, Eq.~(\ref{eq:p_a}):
\begin{align}
  \label{eq:p_a_ren}
  \mathcal P^a_{\delta \phi, 2\vx}(q,\eta)  &= \frac{2q^3}{\pi^2}  \int^\eta_{\eta_i} \dd \eta' \int^{\eta'}_{\eta_i} \dd \eta'' a^4(\eta')  a^4(\eta'') V_\tho(\eta') V_\tho(\eta'') \Im[u_q(\eta) u_q^*(\eta')] \Im[u_q(\eta) u_q^*(\eta'')] \nonumber \\
  &\qquad \qquad \qquad \times \int \frac{\dd \Omega_k}{4\pi} \int^{k_\uv \frac{a(\eta'')}{a_i}}_{k_\ir} \frac{k^2 \dd k}{2\pi^2} \Re[u_k(\eta')u^*_k(\eta'')u_{|\bfq - \bfk|}(\eta')u^*_{|\bfq - \bfk|}(\eta'')],
\end{align}
where $\int \dd \Omega_k$ is the integral over the solid angle of $\bfk$.
Substituting the UV limit expression of $u_k$, Eq.~(\ref{eq:u_k_subh_limit}), into this, we find
\begin{align}
  \label{eq:p_a_ren1}
  \mathcal P^{a,\uv}_{\delta \phi, 2\vx}(q,\eta)  &= \frac{2q^3}{\pi^2}  \int^\eta_{\eta_i} \dd \eta' \int^{\eta'}_{\eta_i} \dd \eta'' a^4(\eta')  a^4(\eta'') V_\tho(\eta') V_\tho(\eta'') \Im[u_q(\eta) u_q^*(\eta')] \Im[u_q(\eta) u_q^*(\eta'')] \nonumber \\
  &\qquad \qquad \qquad \times \int \frac{\dd \Omega_k}{4\pi} \int^{k_\uv \frac{a(\eta'')}{a_i}}_{k_*\frac{a(\eta'')}{a_i}} \frac{k^2 \dd k}{2\pi^2} \frac{H^4 {\eta'}^2 {\eta''}^2}{4k |\bfk - \bfq|} \cos\left[ (k+|\bfk - \bfq|)(\eta' - \eta'') \right].
\end{align}
In the limit of $q \ll k_*$, we obtain 
\begin{align}
  \label{eq:p_a_ren2}
  \mathcal P^{a,\uv}_{\delta \phi, 2\vx}(q,\eta)  &= \frac{2q^3}{\pi^2}  \int^\eta_{\eta_i} \dd \eta' \int^{\eta'}_{\eta_i} \dd \eta'' a^2(\eta')  a^2(\eta'') V_\tho(\eta') V_\tho(\eta'') \Im[u_q(\eta) u_q^*(\eta')] \Im[u_q(\eta) u_q^*(\eta'')] \nonumber \\
  &\qquad \qquad \qquad \times \frac{1}{8\pi^2} \int^{k_\uv \frac{a(\eta'')}{a_i}}_{k_*\frac{a(\eta'')}{a_i}} \dd k \cos\left[ 2k(\eta' - \eta'') \right] \nonumber \\
  &= \frac{2q^3}{\pi^2}  \int^\eta_{\eta_i} \dd \eta' \int^{\eta'}_{\eta_i} \dd \eta'' a^2(\eta')  a^2(\eta'') V_\tho(\eta') V_\tho(\eta'') \Im[u_q(\eta) u_q^*(\eta')] \Im[u_q(\eta) u_q^*(\eta'')] \nonumber \\
  &\qquad \qquad \qquad \times \frac{1}{8\pi^2} \frac{\sin[2 k_\uv \frac{a(\eta'')}{a_i} (\eta' - \eta'')] -\sin[2 k_* \frac{a(\eta'')}{a_i}(\eta' - \eta'')]}{2(\eta' - \eta'')} \nonumber \\
  &= \frac{2q^3}{\pi^2} \int^{\eta}_{\eta_i} \dd \eta_+ \int^{\min[2\eta_+ -2\eta_i, 2\eta- 2\eta_+]}_{0} \dd \eta_-\, a^2(\eta')  a^2(\eta'') V_\tho(\eta') V_\tho(\eta'') \Im[u_q(\eta) u_q^*(\eta')] \Im[u_q(\eta) u_q^*(\eta'')] \nonumber \\
  &\qquad \qquad \qquad \times \frac{\sin[2 k_\uv \frac{a(\eta'')}{a_i} \eta_- ] -\sin[2 k_* \frac{a(\eta'')}{a_i}\eta_-]}{16 \pi^2\eta_-},
\end{align}
where $\eta_+ = (\eta' + \eta'')/2$ and $\eta_- = \eta' - \eta''$.
In the $k_\uv \to \infty$ limit, the $k_\uv$ dependence disappears because\footnote{\label{foot:time_uv_cut}
 In some papers, the UV cutoff scale is introduced even for the time integrals as $\int^{\eta}_{\eta_i} \dd \eta' \int^{\eta'}_{\eta_i} \dd \eta'' \to \int^{\eta}_{\eta_i} \dd \eta' \int^{\eta'-\frac{a_i}{a(\eta')k_\uv}}_{\eta_i} \dd \eta''$~\cite{Senatore:2009cf,Ballesteros:2024zdp}.
 This UV time cutoff only changes the lower bound of the $\eta_-$ integral in Eqs.~(\ref{eq:p_a_ren2}) and (\ref{eq:p_b_ren}) as $\int_0 \dd \eta_- \to \int_{\frac{a_i}{a(\eta_+)k_\uv}} \dd \eta_-$, while $\mathcal P^{a,\uv}_{\delta \phi,2\vx}$ remains convergent and $\mathcal P^{b,\uv}_{\delta \phi,2\vx}$ remains logarithmically divergent in $k_\uv \to \infty$.
} 
%%%%
\begin{align}
  &\lim_{k_\uv \to \infty} \int^{\eta}_{\eta_i} \dd \eta_+ \int^{\min[\eta_+ -2\eta_i, 2\eta- \eta_+]}_{0} \dd \eta_-\, a^4(\eta')  a^4(\eta'') V_\tho(\eta') V_\tho(\eta'') \Im[u_q(\eta) u_q^*(\eta')] \Im[u_q(\eta) u_q^*(\eta'')]  
   \frac{\sin[2 k_\uv \frac{a(\eta'')}{a_i} \eta_- ]}{16 \pi^2\eta_-} \nonumber \\
  &= \int^{\eta}_{\eta_i} \dd \eta_+ \frac{a^4(\eta_+) V^2_\tho(\eta_+)}{32\pi} \left(\Im[u_q(\eta) u_q^*(\eta_+)] \right)^2,
\end{align}
%%%%
where we have used $\int^\infty_0 (\sin z/z) \dd z = \pi/2$ for the equality.
We can also see that the $k_*$ dependent term in Eq.~(\ref{eq:p_a_ren2}) is finite.
Moreover, if we expand the integrand of Eq.~(\ref{eq:p_a_ren1}) with respect to $q/k (\ll 1)$ and perform the integrals, we can easily see that no $k_\uv$ divergences appear from the higher order contributions of $q/k$.
For these reasons, there is no need to renormalize the $\mathcal P^a_{\delta \phi, 2\vx}$ contribution.

Next, we focus on the $\mathcal P^b_{2\vx}$ contribution, Eq.~(\ref{eq:p_b}):
\begin{align}
    \label{eq:p_b_2}
  \mathcal P^b_{\delta \phi, 2\vx}(q,\eta)  &= \frac{2q^3}{\pi^2}  \int^\eta_{\eta_i} \dd \eta' \int^{\eta'}_{\eta_i} \dd \eta'' a^2(\eta')  a^2(\eta'') V_\tho(\eta') V_\tho(\eta'') \Im[u_q(\eta) u_q^*(\eta')] \Re[u_q(\eta) u_q^*(\eta'')] \nonumber \\
  &\qquad \qquad \qquad \times \int \frac{\dd \Omega_k}{4\pi} \int^{k_\uv \frac{a(\eta'')}{a_i}}_{k_\ir} \frac{k^2 \dd k}{2\pi^2} \Im[u_k(\eta')u^*_k(\eta'')u_{|\bfq - \bfk|}(\eta')u^*_{|\bfq - \bfk|}(\eta'')].
\end{align}
Similar to Eq.~(\ref{eq:p_a_ren1}), we substitute the UV limit expression of $u_k$ and then obtain 
\begin{align}
  \label{eq:p_b_uv}
  \mathcal P^{b,\uv}_{\delta \phi, 2\vx}(q,\eta)  &= -\frac{2q^3}{\pi^2}  \int^\eta_{\eta_i} \dd \eta' \int^{\eta'}_{\eta_i} \dd \eta'' a^4(\eta')  a^4(\eta'') V_\tho(\eta') V_\tho(\eta'') \Im[u_q(\eta) u_q^*(\eta')] \Re[u_q(\eta) u_q^*(\eta'')] \nonumber \\
  &\qquad \qquad \qquad \times \int \frac{\dd \Omega_k}{4\pi} \int^{k_\uv \frac{a(\eta'')}{a_i}}_{k_*\frac{a(\eta'')}{a_i}} \frac{k^2 \dd k}{2\pi^2} \frac{H^4 {\eta'}^2 {\eta''}^2}{4k |\bfk - \bfq|} \sin\left[ (k+|\bfk - \bfq|)(\eta' - \eta'') \right].
\end{align}
In the limit of $q \ll k_*$, we obtain
\begin{align}
  \label{eq:p_b_ren}
  \mathcal P^{b,\uv}_{\delta \phi, 2\vx}(q,\eta)  &= -\frac{2q^3}{\pi^2}  \int^\eta_{\eta_i} \dd \eta' \int^{\eta'}_{\eta_i} \dd \eta'' a^2(\eta')  a^2(\eta'') V_\tho(\eta') V_\tho(\eta'') \Im[u_q(\eta) u_q^*(\eta')] \Re[u_q(\eta) u_q^*(\eta'')] \nonumber \\
  &\qquad \qquad \qquad \times \frac{1}{8\pi^2} \int^{k_\uv \frac{a(\eta'')}{a_i}}_{k_*\frac{a(\eta'')}{a_i}} \dd k \sin\left[ 2k(\eta' - \eta'') \right] \nonumber \\
  &= \frac{2q^3}{\pi^2} \int^{\eta}_{\eta_i} \dd \eta_+ \int^{\min[2\eta_+ -2\eta_i, 2\eta- 2\eta_+]}_{0} \dd \eta_-\, a^2(\eta')  a^2(\eta'') V_\tho(\eta') V_\tho(\eta'') \Im[u_q(\eta) u_q^*(\eta')] \Re[u_q(\eta) u_q^*(\eta'')] \nonumber \\
  &\qquad \qquad \qquad \times \frac{\cos[2 k_\uv \frac{a(\eta'')}{a_i} \eta_- ] -\cos[2 k_* \frac{a(\eta'')}{a_i}\eta_-]}{16 \pi^2 \eta_-} \nonumber \\
  &= -\frac{q^3}{\pi^2} \int^{\eta}_{\eta_i} \dd \eta_+ \frac{1}{8 \pi^2} a^4(\eta_+) V^2_\tho(\eta_+) \Im[u_q(\eta) u_q^*(\eta_+)] \Re[u_q(\eta) u_q^*(\eta_+)]\ln\left( \frac{k_\uv}{k_*}\right),
\end{align}
where we have used $\int^\infty_0 \dd t (\cos(a t) - \cos(b t))/t = -\ln(a/b)$.
$\mathcal P^b_{\delta \phi, 2\vx}$ has a logarithmic divergence, unlike $\mathcal P^a_{\delta \phi, 2\vx}$.
We can expand the integrand in Eq.~(\ref{eq:p_b_uv}) with respect to $q/k$ and find $\mathcal O(q^2)$ corrections as the next-leading terms after performing the $\Omega_k$ integral.
These next-leading terms do not give any $k_\uv$ dependent terms in $k_\uv \to \infty$ limit.\footnote{
The next-leading $\mathcal O(q^2)$ terms are proportional to $\int \dd k (q^2 \eta_-/k) \cos[2 k\eta_-]$, $\int \dd k (q^2 \eta_-^2) \sin[2 k\eta_-]$, or $\int \dd k (q^2/k^2) \sin[2 k\eta_-]$. 
For the first two terms, the additional $\eta_-$('s) make them converge in $k_\uv \to \infty$ after the $\eta_-$ integral.
For the last term, we find
  \begin{align}
   \int^{k_\uv \frac{a(\eta'')}{a_i}}_{k_*\frac{a(\eta'')}{a_i}} \dd k \frac{q^2}{k^2} \sin [2k\eta_-] = 2q^2 \eta_- \left( \Ci \left[2 k_\uv \frac{a(\eta'')}{a_i} \eta_- \right] -\Ci\left[2 k_* \frac{a(\eta'')}{a_i}\eta_- \right] - \frac{\sin \left[2 k_\uv \frac{a(\eta'')}{a_i} \eta_- \right]}{2k_\uv \eta_-} + \frac{\sin \left[2 k_* \frac{a(\eta'')}{a_i} \eta_- \right]}{2k_* \eta_-} \right). \nonumber 
  \end{align}
  With $\lim_{x\to 0} x\Ci[x] = 0$ and $x \Ci[x]|_{x \gg 1} \simeq \sin x$, we can see that the integral of this over $\eta_-$ gives only $\mathcal O(q^2/k_\uv^2)$ terms.
}
We can see that the $\ln k_\uv$ dependences in Eqs.~(\ref{eq:1vx_ren}) and (\ref{eq:p_b_ren}) are cancelled.
This is not surprising because $\mathcal P^{b}_{\delta \phi,2\vx}$ includes the contribution that can be expressed with $\frac{\dd}{\dd t} \int \frac{\dd^3 k}{(2\pi)^3} |u_k(t)|^2$ (see Eqs.~(\ref{eq:ddot_2vx}) and (\ref{eq:2vx_cond})).
From these observations, we can see that the counter term given by Eq.~(\ref{eq:v_c1_uv}) successfully cancels all the $k_\uv$ dependence in the one-loop power spectrum.
Note that the counter term given by Eq.~(\ref{eq:v_c1_uv}) also cancels the UV divergences in $\langle \delta \dot \phi \rangle$, which can be seen from Eq.~(\ref{eq:zeta_cons2}).

\section{Conclusion}
\label{sec:concl}

In this paper, we have shown that the superhorizon-limit curvature perturbations are conserved at one-loop level in single-field inflation models with a transient non-slow-roll period, following up on our previous work~\cite{Inomata:2024lud}.
We have explicitly showed the intermediate steps omitted in the previous work and also performed the two new analyses: 
1) We have calculated the backreaction with the in-in formalism, while we previously relied on the equation of motion of the background when we calculated the backreaction. 
The results in this paper are consistent with those in the previous work. 
Our new analysis confirms the validity of the previous analysis.
2) We have explicitly renormalized the UV divergences with the counter terms.
We have explicitly taken into account the counter terms throughout the loop calculation and seen how the UV divergences are cancelled by them.

The conclusion of this paper is consistent with our previous work~\cite{Inomata:2024lud}: the superhorizon-limit curvature perturbations are not affected by the small-scale perturbations at one-loop level. 
This means that we can consider the enhancement of small-scale perturbations, often considered in the context of PBH scenarios, without being concerned about the one-loop corrections to the curvature power spectrum on the CMB/LSS scales.
On the other hand, we still need to be careful about the one- or higher-loop corrections or the non-perturbative effects around the small scales where the perturbations are enhanced~\cite{Inomata:2022yte,Iacconi:2023slv,Fumagalli:2023loc,Caravano:2024tlp,Caravano:2024moy}.

It is worth noting that there are two major differences between this paper and Ref.~\cite{Ballesteros:2024zdp}, which appeared after our previous work~\cite{Inomata:2024lud} and discusses the one-loop power spectrum with the same gauge (spatially-flat gauge):
\begin{itemize}
\item The backreaction plays an important role in our analysis, while it is not taken into account in Ref.~\cite{Ballesteros:2024zdp}.
As discussed at the end of Sec.~\ref{sec:one_loop_ps}, we can neglect the backreaction only if we tune $V_{c,\fo}$ so that $\expval{\delta \phi} = 0$ is always satisfied.
If we determine $V_{c,\fo}$ as this, $V_{c,\so}$ is automatically fixed.
If we change $V_{c,\so}$, $V_{c,\fo}$ also changes and the backreaction becomes nonzero in general.
This is inconsistent with the reference, where $V_{c,\so}$ is changed independently while the backreaction is ignored.
We believe that this is why it is found in Ref.~\cite{Ballesteros:2024zdp} that whether the superhorizon curvature perturbations are conserved or not depends on the choice of $V_{c,\so}$.
As we have seen in this paper, if the backreaction is taken into account, the superhorizon curvature perturbations are conserved independently of the choice of $V_{c,\so}$ (and $V_{c,\fo}$).
Note that the redefinition of the perturbations does not remove the backreaction in general (see Sec.~\ref{sec:one_loop_ps} and Appendix~\ref{app:tad}).

\item We have found the UV divergence in the two-vertex contribution in Eq.~(\ref{eq:p_b_ren}), while it is not found in Ref.~\cite{Ballesteros:2024zdp}.
In addition, we have shown that the UV renormalization can be done without the $i\varepsilon$ prescription (see also footnote~\ref{ft:i_eps}), while the importance of it for the UV convergence is stressed in Ref.~\cite{Ballesteros:2024zdp}.
These are due to the difference in the relation between the UV cutoff scale $k_\uv$ and the transition timescale between the SR and the non-SR periods, which we here denote by $\Delta \eta_t$.
We take the limit of $k_\uv \gg 1/\Delta \eta_t$ when we discuss the UV contributions, while the sudden transition limit $k_\uv \ll 1/\Delta \eta_t$ is taken in Ref.~\cite{Ballesteros:2024zdp}.
We believe that we should take the $k_\uv \to \infty$ limit after we fix the potential, which leads to our limit $k_\uv \gg 1/\Delta \eta_t$. 
See Appendix~\ref{app:diff_previous} for a detailed discussion.
\end{itemize}

Finally, let us mention possible future directions. 
In this paper, we have taken the spatially-flat gauge. It is worthwhile to study the connection between our result and the literature that uses the comoving gauge.
Apart from the gauge, we have taken the de Sitter limit and neglected the terms suppressed by $\epsilon$ for simplicity in this paper. It would be meaningful to discuss the one-loop corrections without assuming the de Sitter limit. 
Besides, we have assumed that the hierarchy of the scales between the large-scale perturbations and the small-scale perturbations that affect the large-scale perturbations. Studying the one-loop corrections sourced by the large-scale perturbations would be an interesting future direction, which would be related to the discussion of the IR divergence in the one-loop power spectrum~\cite{Tanaka:2013caa}.
Also, the generalization of our analysis to higher-order loop contributions is left for future work.

\vspace{5pt}
%%%%%%%%%%%%%%%%%%%%%%%%%%%%%%%%
\noindent
\acknowledgments
%%%%%%%%%%%%%%%%%%%%%%%%%%%%%%%%
The author thanks Cheng-Jun Fang and Jacopo Fumagalli for useful discussions and Jason Kristiano for his comment on Eq.~(\ref{eq:u_k2_dt}), which motivated the author to add Appendix~\ref{app:uv_bound}, and for pointing out that the condition of $\expval{\delta \phi} = 0$ is different from $\expval{\zeta} = 0$ due to the nonlinear relation between them.
The author is grateful to the organizers and the attendees of ``Looping in the Primordial Universe'' workshop at CERN for fruitful discussions.
The author was supported by JSPS Postdoctoral Fellowships for Research Abroad.

\appendix
\section{Formalism without tadpole contribution}
\label{app:tad}

In this appendix, we show the formalism with the redefined perturbation $\delta \phi_\rr \equiv \delta \phi - \expval{\delta \phi}$ and the redefined background $\bar \phi_\rr \equiv \bar \phi + \expval{\delta \phi}$.
In particular, we will clarify the consistency between this paper and our previous work~\cite{Inomata:2024lud}, where $\delta \phi_\rr$ and $\bar \phi_\rr$ are used.

First, let us explicitly show $\expval{\delta \phi_{\rr,\bfq} \delta \phi_{\rr,\bfq'}} = \expval{\delta \phi_{\bfq} \delta \phi_{\bfq'}}$.
With the redefined quantities, we modify the equation of motion for the free part, Eq.~(\ref{eq:u_eom}), as 
\begin{align}
    \left[\frac{\dd^2}{\dd \eta^2} + 2 \mathcal H \frac{\dd}{\dd \eta} + k^2 + a^2 V_\so(\bar \phi_\rr(\eta)) \right] u_{\rr,k}(\eta) = 0,
    \label{eq:u_eom_app}
\end{align}
where $u_{\rr}$ is defined as 
\begin{align}
  &\delta \phi^I_\rr(\bfx,\eta) = \int \frac{\dd^3 k}{(2\pi)^3} \ee^{i \bfk \cdot \bfx} \delta \phi^I_{\rr,\bfk}(\eta)= \int \frac{\dd^3 k}{(2\pi)^3} \ee^{i \bfk \cdot \bfx} \left[ u_{\rr,k}(\eta) a(\bfk) +  u^{*}_{\rr,k}(\eta) a^{\dagger}(-\bfk) \right].
\end{align}
Since there is no contribution from the tadpole of $\delta \phi_\rr$, the power spectrum up to one-loop level is given by 
\begin{align}
  \expval{\delta \phi_{\rr,\bfq}(\eta) \delta \phi_{\rr,\bfq'}(\eta)} = (2\pi)^3 \delta(\bfq + \bfq') \frac{2\pi^2}{q^3} \left( \mathcal P_{\delta \phi_\rr,\tre}(q,\eta) + \mathcal P_{\delta \phi_\rr,1\vx}(q,\eta) + \mathcal P_{\delta \phi_\rr,2\vx}(q,\eta) \right).
  \label{eq:d_phi_r}
\end{align}
Given that the difference between $u_k$ and $u_{\rr,k}$ does not appear at tree level (because $\expval{\delta \phi}$ is at one-loop level), we can see $\mathcal P_{\delta \phi_\rr,1\vx} = \mathcal P_{\delta \phi,1\vx}$ and $\mathcal P_{\delta \phi_\rr,2\vx} = \mathcal P_{\delta \phi,2\vx}$.
In the following, we show
\begin{align}
  \mathcal P_{\delta \phi_\rr,\tre}(q,\eta) = \mathcal P_{\delta \phi,\tre}(q,\eta) + \mathcal P_{\delta \phi,\tad}(q,\eta).
  \label{eq:p_tad_r}
\end{align}
To this end, we first expand Eq.~(\ref{eq:u_eom_app}) as 
\begin{align}
    \left[\frac{\dd^2}{\dd \eta^2} + 2 \mathcal H \frac{\dd}{\dd \eta} + k^2 + a^2 V_\so(\bar \phi(\eta)) \right] u_{\rr,k}(\eta) = -a^2 V_\tho(\bar \phi(\eta)) \expval{\delta \phi(\bfx,\eta)} u_{\rr,k}(\eta).
    \label{eq:u_eom_app2}
\end{align}
Solving this equation of motion with the initial condition $u_{\rr,k}(\eta) = u_k(\eta)$ in $\eta < \eta_i$, we obtain
\begin{align}
  u_{\rr,k}(\eta) = u_k(\eta) - \int^\eta_{\eta_i} \dd \eta' g_k(\eta;\eta') a^2(\eta') V_\tho(\bar \phi(\eta')) \expval{\delta \phi(\bfx, \eta')} u_k(\eta'),
\end{align}
where we have changed $u_{\rr,k} \to u_k$ in the time integral by neglecting the higher-order contributions.
Then, we can obtain Eq.~(\ref{eq:p_tad_r}) as 
\begin{align}
  \mathcal P_{\delta \phi_\rr,\tre}(q,\eta) &= \frac{q^3}{2\pi^2} |u_{\rr,q}(\eta)|^2 \nonumber \\
  &= \mathcal P_{\delta \phi,\tre}(q,\eta) - \frac{q^3}{\pi^2} \int^\eta_{\eta_i} \dd \eta' a^2(\eta') g_q(\eta;\eta') \Re[u_q(\eta) u^*_q(\eta')] V_\tho(\bar \phi(\eta')) \expval{\delta \phi(\bfx,\eta')} \nonumber \\
  &= \mathcal P_{\delta \phi,\tre}(q,\eta) + \frac{2q^3}{\pi^2} \int^\eta_{\eta_i} \dd \eta' a^4(\eta') \Im[u_q(\eta) u^*_q(\eta')] \Re[u_q(\eta) u^*_q(\eta')] V_\tho(\bar \phi(\eta')) \expval{\delta \phi(\bfx,\eta')} \nonumber \\
  &= \mathcal P_{\delta \phi,\tre}(q,\eta)  + \mathcal P_{\delta \phi,\tad}(q,\eta),
\end{align}
where see Eq.~(\ref{eq:tad0}) for the tadpole expression.
From this and Eq.~(\ref{eq:d_phi_r}), we can see 
\begin{align}
  \expval{\delta \phi_{\rr,\bfq}(\eta) \delta \phi_{\rr,\bfq'}(\eta)} = \expval{\delta \phi_{\bfq}(\eta) \delta \phi_{\bfq'}(\eta)}.
\end{align}

For the background, in Ref.~\cite{Inomata:2024lud}, we have expressed $\dot{\bar \phi}_\rr = \bar \Pi^\zo + \bar \Pi^\so$, where $\bar \Pi^\zo$ is the modified tree-level background velocity and $\bar \Pi^\so$ is the correction due to $\mathcal O(\delta \phi_\rr^2)$ backreaction.
Specifically, they follow~\cite{Inomata:2024lud}
\begin{align}
  \label{eq:pi_zo}
  \left[\frac{\dd^2}{\dd \eta^2} + 2 \mathcal H \frac{\dd}{\dd \eta} + a^2 V_\so(\bar \phi_\rr(\eta)) \right] \bar \Pi^\zo(\eta) &= 0, \\
  \left[\frac{\dd^2}{\dd \eta^2} + 2 \mathcal H \frac{\dd}{\dd \eta} + a^2 V_\so(\bar \phi_\rr(\eta)) \right] \bar \Pi^\so(\eta) &= - \frac{a^2}{2} \left( 2V_{c,\so} \bar \Pi^\zo + V_\tho \expval{\delta \phi_\rr^2}^{\bm{\cdot}} + V_\foo \expval{\delta \phi_\rr^2} \bar \Pi^\zo \right),
  \label{eq:pi_so}
\end{align}
where $\expval{\delta \phi_\rr^2} = \int\frac{\dd^3 k}{(2\pi)^3} |u_{\rr,k}(\eta)|^2$, and we have omitted the arguments of $\bar\Pi(\eta)$ and $V_\no(\bar \phi_\rr(\eta))$ on the RHS.
We can neglect the backreaction only when the RHS of Eq.~(\ref{eq:pi_so}) is always zero (if so, we do not need to separate $\bar\Pi^\zo$ and $\bar\Pi^\so$ anymore).
Note again that the backreaction is defined as the deviation from the tree-level evolution determined by Eq.~(\ref{eq:pi_zo}).
The zero-backreaction can be realized by tuning $V_{c,\so}$ as 
\begin{align}
  2V_{c,\so} \bar \Pi^\zo + V_\tho \expval{\delta \phi_\rr^2}^{\bm{\cdot}} + V_\foo \expval{\delta \phi_\rr^2} \bar \Pi^\zo = 0,
  \label{eq:vc_so_r}
\end{align}
where we can change $\delta \phi_\rr \to \delta \phi$ in this equation by neglecting higher order contributions.
Actually, this is automatically satisfied if we impose $\expval{\delta \phi} = 0$ for the original perturbation by tuning $V_{c,\fo}$ as Eq.~(\ref{eq:vcfo_zero_tad}).

In Ref.~\cite{Inomata:2024lud}, we have finally shown
\begin{align}
  \frac{\expval{\delta \phi_{\rr,\bfq}(\eta) \delta \phi_{\rr,\bfq'}(\eta)}}{\dot {\bar \phi}^2_\rr(\eta)} = \frac{\expval{\delta \phi_{\rr,\bfq}(\eta) \delta \phi_{\rr,\bfq'}(\eta)}}{(\bar\Pi^\zo(\eta))^2\left(1 + 2 \frac{\bar \Pi^\so(\eta)}{\bar \Pi^\zo(\eta))} \right)} = \text{const}.
\end{align}
This is exactly the same as what we have shown in the main text of this paper:
\begin{align}
  \frac{\expval{\delta \phi_{\bfq}(\eta) \delta \phi_{\bfq'}(\eta)}}{ \left(\dot {\bar \phi}(\eta) + \langle \delta \dot \phi(\eta) \rangle \right)^2} = \text{const}.
\end{align}

\section{Time dependence of the UV boundary in the time integral}
\label{app:uv_bound}

In this Appendix, we derive Eq.~(\ref{eq:u_k2_dt_2}) by taking into account the switch-on time of each mode instead of using $a(t_\tmin)$.
First, let us recall the expression of $\dd u_k/\dd t$: 
\begin{align}
  \label{eq:du_dt_f_infty}
  \frac{\dd u_k(t)}{\dd t} &= 2\int^\eta_{-\infty} \dd \eta' a^4(\eta') V_\tho(\eta') \dot {\bar \phi}(\eta') \Im[u_k(\eta) u^*_k(\eta')] u_k(\eta') - H k^{-3/2}\frac{\dd (k^{3/2} u_k(\eta))}{\dd \ln k}, 
\end{align}
where this solution implicitly assumes that the adiabatic solution at $\eta \to -\infty$, which we will modify in Eq.~(\ref{eq:du_eta_sw}).
To understand the physical meaning of the first term, let us do some experiment. 
We consider the time when $m^2 (\equiv V_\so) = \text{const.}$ and substitute the following adiabatic (Bunch-Davies vacuum) solution (Eq.~(\ref{eq:u_k_ini})) into Eq.~(\ref{eq:du_dt_f_infty}):
\begin{align}
  u_k(\eta) = -\frac{H k\eta}{\sqrt{2k^3}}\ee^{i\frac{\left(2\nu + 1\right)\pi}{4}} \sqrt{\frac{-\pi k\eta}{2}} H_\nu^\fo(-k\eta), \tag{\ref{eq:u_k_ini}}
\end{align}
where $\nu = \sqrt{9/4-m^2/H^2}$.
Then, we find 
\begin{align}
  \label{eq:adiabatic_cond}
  \frac{\dd u_k(t)}{\dd t}  = - H k^{-3/2}\frac{\dd (k^{3/2} u_k(\eta))}{\dd \ln k} \ \ \text{(adiabatic mode)}.
\end{align}
Then, let us consider the transition from $m^2_i$ to $m^2_f$ which occurs only within $\eta_i < \eta < \eta_f$. 
That is, $V_\so = m_i^2 (=\text{const.})$ in $\eta < \eta_i$ and $V_\so = m_f^2 (=\text{const.})$ in $\eta > \eta_f$.
If $u_k$ remains in the adiabatic mode (without excitation), Eq.~(\ref{eq:adiabatic_cond}) remains true even in $\eta > \eta_f$.
However, if $m^2 (\equiv V_\so)$ changes, $u_k$ gets excited and deviates from the adiabatic solution because the adiabatic condition is not exact during the transition.
This excitement of $u_k$ can be understood as the particle production and is caught by the first term in Eq.~(\ref{eq:du_dt_f_infty}). 
Note that, although we have considered the time when $m^2 = \text{const.}$ to define the adiabatic mode in the above case for simplicity, this remains a good approximation if the evolution of $m^2$ is much slower than the oscillation of the perturbations, which is the case for the perturbations around the UV boundary scales.
Under this approximation, we can always define the adiabatic mode Eq.~(\ref{eq:u_k_ini}) at $\eta$ with $m^2 = V_\so(\eta)$.

Next, let us focus on the field variance:
\begin{align}
  \int^{k_\uv \frac{a(t)}{a_i}}_{k_\ir} \frac{k^2 \dd k}{2\pi^2} |u_k(t)|^2,
  \label{eq:vari}
\end{align}
where we here put the UV bound as $k_\uv a(t)/a_i$, instead of $k_\uv a(t_\min)/a_i$.
The key question is: what is $u_k$?
If we assume the adiabatic solution (Eq.~(\ref{eq:u_k_ini})) at $t$, the time derivative of this field variance becomes
\begin{align}
  \frac{\dd}{\dd t} \int^{k_\uv \frac{a(t)}{a_i}}_{k_\ir} \frac{k^2 \dd k}{2\pi^2} |u_k(t)|^2 = H \mathcal P_{\delta \phi,\tre}(k_\ir, t) \ \ (\text{adiabatic mode}).
\end{align}
Let us here recall the meaning of the physical cutoff: $u_k(t)$ is switched on at $k = k_\uv a(t)/a_i$ with the adiabatic solution (without excitation).
This means that $u_k(t)$ in Eq.~(\ref{eq:vari}) can be different from the adiabatic solution because it is switched on before $t$.
We here define the switch-on time by $\eta_\sw(k)$ ($k = k_\uv a(\eta_\sw(k))/a_i$).
Once we take into account the particle production from $\eta_\sw$, the time derivative of $u_k$ becomes
\begin{align}
  \label{eq:du_eta_sw}
  \frac{\dd u_k(t)}{\dd t} &= 2\int^\eta_{\eta_\sw(k)} \dd \eta' a^4(\eta') V_\tho(\eta') \dot {\bar \phi}(\eta') \Im[u_k(\eta) u^*_k(\eta')] u_k(\eta') - H k^{-3/2}\frac{\dd (k^{3/2} u_k(\eta))}{\dd \ln k}. 
\end{align}
Note that we have modified the lower bound of the time integral of Eq.~(\ref{eq:du_dt_f_infty}).
Using this, we obtain
\begin{align}
  &\frac{\dd}{\dd t} \int^{k_\uv \frac{a(t)}{a_i}}_{k_\ir} \frac{k^2 \dd k}{2\pi^2} |u_k(t)|^2 -H \mathcal P_{\delta \phi,\tre}(k_\ir, t) \nonumber \\
  &= 4 \int^{k_\uv \frac{a(\eta)}{a_i}}_{k_\ir} \frac{k^2 \dd k}{2\pi^2} \int^{\eta}_{\eta_\sw(k)} \dd \eta' a^4(\eta') V_\tho(\eta') \dot {\bar \phi}(\eta')  \Im[u_k(\eta) u^*_k(\eta')] \Re[u_k(\eta) u^*_k(\eta')] \nonumber \\ 
  &= 4 \int^{\eta}_{\eta_\ir} \dd \eta' a^4(\eta') V_\tho(\eta') \dot {\bar \phi}(\eta') \int^{k_\uv \frac{a(\eta')}{a_i}}_{k_\ir} \frac{k^2 \dd k}{2\pi^2}\Im[u_k(\eta) u^*_k(\eta')] \Re[u_k(\eta) u^*_k(\eta')],
  \label{eq:d_u_squ}
\end{align}
where $\eta_\ir$ is the time when $k_\ir = k_\uv a(\eta)/a_i$. 
Since we are focusing on the case where $V_\tho(\eta) = 0$ in $k_\uv a(\eta)/a_i < k_\ir$ throughout this work, we can change the lower bound of the time integral of the last line to $-\infty$.
Then, we can see that Eq.~(\ref{eq:d_u_squ}) is consistent with Eq.~(\ref{eq:u_k2_dt_2}).

\section{IR cutoff contribution}
\label{app:ir}

In this appendix, we physically interpret the IR cutoff contribution $\langle \delta \dot \phi(\bfx,t) \rangle_{\ir}$, defined in Eq.~(\ref{eq:ddot_2vx}). 
From Eqs.~(\ref{eq:del_phi_cons}) and (\ref{eq:zeta_cons_fin}), we can see that, if we do not neglect the IR cutoff contribution, the curvature perturbations are not conserved at one-loop level.
Although we neglect it in the main text given that it is negligibly small in our setup, we discuss the origin of the IR cutoff contribution in this appendix.
More specifically, we will see that the spatially homogeneous IR cutoff scale, taken in this paper, leads to the violation of the separate universe picture.

From Eq.~(\ref{eq:one_pt_g}), we can express the backreaction, as
\begin{align}
  \expval{\delta \phi(\bfx,t)}
  &=  -\frac{1}{2} \int^t_{t_i} \dd t' g_0(t;t')\left[ V_\tho(t') \int^{k_\uv \frac{a(t')}{a_i}}_{k_\ir} \frac{k^2 \dd k }{2\pi^2} |u_{k}(t')|^2 + 2 V_{c,\fo}(t') \right].
  \label{eq:one_pt_g_app}
\end{align}
It is natural that we consider the spatially homogeneous physical UV cutoff in the spatially flat gauge because the physical UV cutoff scale must be determined by the energy scale itself, independent of the background field value.
On the other hand, we need to be careful about our choice of the homogeneous comoving $k_\ir$.
The homogeneous $k_\ir$ leads to the homogeneous one-loop backreaction. 
However, if the separate universe picture holds, the clock of each universe must be determined by the local inflaton field value, which can spatially fluctuate if we compare different universes. 
Namely, the separate universe picture prefers the inhomogeneous (field-value dependent) cutoff, which leads to the inhomogeneous backreaction that conserves the separate universe picture. 
This is why the homogeneous $k_\ir$ violates the separate universe picture.\footnote{
  The violation of the separate universe picture can be regarded as the breaking of the spatial rescaling symmetry, $\bfx \to \ee^{\zeta}\bfx$ in the comoving gauge~\cite{Maldacena:2002vr,Hinterbichler:2012nm}.
  A similar symmetry breaking due to the choice of cutoff scales can be seen in the one-loop corrections of the photon propagator.
  The regularization with a naive UV cutoff breaks the gauge symmetry, while the Pauli-Villars regularization or the dimensional regularization conserves it~\cite{Peskin:1995ev}.
}

To clarify this physical interpretation, let us see how the inhomogeneous $k_\ir$ can restore the separate universe picture, though we cannot justify this choice of $k_\ir$.
We here redefine the IR comoving cutoff whose scale is independent of the space (homogeneous) in the time slice of $\delta \phi = 0$.
With this definition, $k_\ir$ depends on the space in the time slice of the spatially-flat gauge with $\delta \phi \neq 0$. 
As a result, $k_\ir$ is modified as $k_\ir(\bfx,t) = k_\ir\left(1 - H \frac{\delta \phi(\bfx,t)}{\dot{\bar \phi}(t)} \right)$.
Accordingly, Eq.~(\ref{eq:one_pt_g_app}) is modified as 
\begin{align}
  \expval{\delta \phi(\bfx,t)}_{\text{inh}}
  &=  -\frac{1}{2} \int^t_{t_i} \dd t' g_0(t;t')\left[ V_\tho(t') \int^{k_\uv \frac{a(t')}{a_i}}_{k_\ir\left(1 - H \frac{\delta \phi(\bfx,t')}{\dot{\bar \phi}(t')} \right)} \frac{k^2 \dd k }{2\pi^2} |u_{k}(t')|^2 + 2 V_{c,\fo}(t') \right] \nonumber \\
  &= \expval{\delta \phi(\bfx,t)} -\frac{1}{2} \int^t_{t_i} \dd t' g_0(t;t') V_\tho(t') H \frac{\delta \phi(\bfx,t')}{\dot{\bar \phi}(t')} \mathcal P_{\delta \phi,\tre}(k_\ir,t').
  \label{eq:one_pt_g_app2}
\end{align}
If we define
\begin{align}
  \delta \phi_{\bfq,\text{inh}}(\eta) = \int \dd^3 x\, \ee^{-i\bfq \cdot \bfx }\expval{\delta \phi(\bfx,t)}_{\text{inh}},
\end{align}
we can find that the two-point correlation function in Eq.~(\ref{eq:del_phi_cons}) is added by 
\begin{align}
\expval{\delta \phi_{\bfq}(\eta) \delta \phi_{\bfq,\text{inh}}(\eta) + \delta \phi_{\bfq,\text{inh}}(\eta) \delta \phi_{\bfq}(\eta)} = (2\pi)^3 \delta(\bfq + \bfq') \frac{2\pi^2}{q^3}\mathcal P_{\delta \phi,\tre}(q,\eta) \, 2\frac{\langle \delta \dot \phi(\eta) \rangle_\ir}{\dot {\bar \phi}(\eta)}.
\end{align}
This cancels the $\langle \delta \dot \phi(\eta) \rangle_\ir$ term in Eq.~(\ref{eq:del_phi_cons}), which is consistent with the separate universe picture.

\section{Comparison with the previous paper}
\label{app:diff_previous}

In the main text, we have seen that the two-vertex contribution has the UV divergence (Eq.~(\ref{eq:p_b_ren})), while Ref.~\cite{Ballesteros:2024zdp} does not find it in the two-vertex contribution.
In addition, we have seen that the renormalization can be done without the $i\varepsilon$ prescription (see also footnote~\ref{ft:i_eps}), while Ref.~\cite{Ballesteros:2024zdp} claims that the $i\varepsilon$ prescription is needed for the UV convergence of the loop integrals in the setup of SR $\to$ USR $\to$ SR.
In this appendix, we clarify the origin of these discrepancies. 

To compare the results, we take a similar setup as in Ref.~\cite{Ballesteros:2024zdp}, where $V_{\no}|_{n\geq 3}$ becomes nonzero only in $\eta_1 < \eta < \eta_1 + \Delta \eta_1$ and $(\eta_1 + \Delta \eta_1<)\eta_2 < \eta < \eta_2 + \Delta \eta_2$.
If we send $\Delta \eta_1 \to 0$ and $\Delta \eta_2 \to 0$, we can reproduce the case in the reference. 
Then, let us focus on the two-vertex contribution (Eq.~(\ref{eq:2vx})), which Ref.~\cite{Ballesteros:2024zdp} claims the $i\varepsilon$ is needed for:
\begin{align}
  \mathcal P_{\delta \phi, 2\vx}(q,\eta) 
  &= -\frac{q^3}{\pi^2} \int^\eta_{\eta_1} \dd \eta' \int^{\eta'}_{\eta_1} \dd \eta'' a^4(\eta')  a^4(\eta'') V_\tho(\eta') V_\tho(\eta'') \Re\left[ [u_q(\eta) u_q^*(\eta')- u^*_q(\eta) u_q(\eta')] 
   u_q(\eta) u_q^*(\eta'') \phantom{\int \frac{\dd^3 k}{(2\pi)^3}} \right. \nonumber \\
  &\qquad \qquad \qquad \left. \times \int \frac{\dd^3 k}{(2\pi)^3} u_k(\eta')u^*_k(\eta'')u_{|\bfq - \bfk|}(\eta')u^*_{|\bfq - \bfk|}(\eta'') \right] \nonumber \\
  &= \frac{2q^3}{\pi^2} \int^\eta_{\eta_1} \dd \eta' \int^{\eta'}_{\eta_1} \dd \eta'' a^4(\eta')  a^4(\eta'') V_\tho(\eta') V_\tho(\eta'') \Im[u_q(\eta) u_q^*(\eta')] 
   \phantom{\int \frac{\dd^3 k}{(2\pi)^3}}  \nonumber \\
  &\qquad \qquad \qquad \times \int \frac{\dd^3 k}{(2\pi)^3} \Im [ u_q(\eta) u_q^*(\eta'') u_k(\eta')u^*_k(\eta'')u_{|\bfq - \bfk|}(\eta')u^*_{|\bfq - \bfk|}(\eta'')].
\end{align}
Taking the limit of $q \ll k$, we obtain 
\begin{align}
  \mathcal P_{\delta \phi, 2\vx}(q,\eta) 
  &= \frac{2q^3}{\pi^2} \int^\eta_{\eta_1} \dd \eta' \int^{\eta'}_{\eta_1} \dd \eta'' a^4(\eta')  a^4(\eta'') V_\tho(\eta') V_\tho(\eta'') \Im[u_q(\eta) u_q^*(\eta')] 
   \phantom{\int \frac{\dd^3 k}{(2\pi)^3}}  \nonumber \\
  &\qquad \qquad \qquad \times  \int^{k_\uv \frac{a(\eta'')}{a_1}}_{k_\ir} \frac{ k^2\dd k}{2\pi^2} \Im [ u_q(\eta) u_q^*(\eta'') (u_k(\eta')u^*_k(\eta''))^2],
\end{align}
where $a_1 = a(\eta_1)$.
Similar to the reference, let us focus on the contribution from $\eta_2 < \eta' < \eta_2 + \Delta \eta_2$ and $\eta_1 < \eta'' < \eta_1 + \Delta \eta_1$ and substitute the subhorizon-limit expression of $u_k$, given by Eq.~(\ref{eq:u_k_subh_limit}):
\begin{align}
  \mathcal P^\uv_{\delta \phi, 2\vx}(q,\eta) 
  &= \frac{2q^3}{\pi^2} \int^\eta_{\eta_2} \dd \eta' \int^{\eta_1+\Delta \eta_1}_{\eta_1} \dd \eta'' a^2(\eta')  a^2(\eta'') V_\tho(\eta') V_\tho(\eta'') \Im[u_q(\eta) u_q^*(\eta')] 
   \phantom{\int \frac{\dd^3 k}{(2\pi)^3}}  \nonumber \\
  &\qquad \qquad \qquad \times \Im \left[ u_q(\eta) u_q^*(\eta'')  \int^{k_\uv \frac{a(\eta'')}{a_1}} \frac{\dd k}{8\pi^2} \ee^{-2ik(\eta'-\eta'')} \right] \nonumber \\
  &= \frac{2q^3}{\pi^2} \int^\eta_{\eta_2} \dd \eta' \int^{\eta_1+\Delta \eta_1}_{\eta_1} \dd \eta'' a^2(\eta')  a^2(\eta'') V_\tho(\eta') V_\tho(\eta'') \Im[u_q(\eta) u_q^*(\eta')] 
   \phantom{\int \frac{\dd^3 k}{(2\pi)^3}}  \nonumber \\
  &\qquad \qquad \qquad \times \Im \left[ u_q(\eta) u_q^*(\eta'')  \frac{i}{16\pi^2 (\eta'-\eta'')} \ee^{-2i k_\uv \frac{a(\eta'')}{a_1} (\eta'-\eta'')} \right] \nonumber \\
  &= \frac{2q^3}{\pi^2} \int^\eta_{\eta_2} \dd \eta' \int^{\Delta \eta_2 + \eta_2 - \eta_1}_{\eta_2 - \eta_1 - \Delta \eta_1} \dd \eta_- a^2(\eta')  a^2(\eta'') V_\tho(\eta') V_\tho(\eta'') \Im[u_q(\eta) u_q^*(\eta')] 
   \phantom{\int \frac{\dd^3 k}{(2\pi)^3}}  \nonumber \\
  &\qquad \qquad \qquad \times \Im \left[ u_q(\eta) u_q^*(\eta'')  \frac{i}{16\pi^2 \eta_-} \ee^{-2i k_\uv \frac{a(\eta'')}{a_1} \eta_-} \right],
  \label{eq:p_uv_2vx_app}
\end{align}
where $\eta_- = \eta' -\eta''$ and we have neglected the terms independent of $k_\uv$.
In the limit of $k_\uv \to \infty$, this contribution goes to zero after the $\eta_-$ integral due to the Riemann-Lebesgue lemma.
The main difference from the main text is that the lower bound of the integral $\eta_-$ is nonzero and fixed.
If $k_\uv$ is sufficiently large, the $\ee^{-2i k_\uv \frac{a(\eta'')}{a_1} \eta_-}$ in the integrand rapidly oscillates within the interval of the $\eta_-$ integral.
This rapid oscillation suppresses the integral when 
\begin{align}
k_\uv \gg \max\left[\frac{1}{\Delta \eta_1}, \frac{a_1}{a(\eta_2)\Delta \eta_2} \right].
\label{eq:kuv_limit}
\end{align}
From this, we can see that $i\varepsilon$ prescription is not required for the convergence of Eq.~(\ref{eq:p_uv_2vx_app}) in contrast to the reference.

This inconsistency comes from the difference in the relation between $\Delta \eta$ and $k_\uv$.
In Ref.~\cite{Ballesteros:2024zdp}, the sudden transition limit is taken: 
\begin{align}
  \Delta \eta_1 \ll \frac{a_1}{a(\eta_1 +\Delta \eta_1) k_\uv}, \ \Delta \eta_2 \ll \frac{a_1}{a(\eta_2+\Delta \eta_2)k_\uv}.
\end{align}
This is opposite to the limit of Eq.~(\ref{eq:kuv_limit}).
Within this inequality, the $k_\uv$ dependence does not disappear in Eq.~(\ref{eq:p_uv_2vx_app}).
Specifically, the $k_\uv$ dependence of Eq.~(\ref{eq:p_uv_2vx_app}) becomes 
\begin{align}
  \mathcal P^\uv_{\delta \phi, 2\vx}(q,\eta)|_{\Delta \eta_1, \Delta \eta_2 \to 0} \propto \ee^{-2i k_\uv(\eta_2 - \eta_1)}.
   \label{eq:p_uv_2vx_app2}
\end{align}
In the reference, the $i\varepsilon$ is introduced to the time to remove this contribution. 
Also, the contributions from the same time domains $\eta_1 < \eta'' < \eta' < \eta_1+ \Delta \eta_1$ or $\eta_2 < \eta'' < \eta' < \eta_2+ \Delta \eta_2$ are ignored in the reference by introducing the UV cutoff to the time integral ($\int^{\eta'} \dd \eta'' \to \int^{\eta' - \frac{a_1}{a(\eta') k_\uv}} \dd \eta''$) within this limit.
This is why UV divergences in the two-vertex contribution do not appear in the reference as opposed to Sec.~\ref{sec:renorm} in this paper. 
Note that, in the opposite limit Eq.~(\ref{eq:kuv_limit}), the introduction of the UV cutoff for the time integral does not remove the contributions from the same time domains (see footnote~\ref{foot:time_uv_cut}), though we have seen in the main text that the UV cutoff for the time integral is unnecessary for the UV renormalization.

If we take $k_\uv \to \infty$ after the potential is fixed (that is, $\Delta \eta_1$ and $\Delta \eta_2$ are fixed in the above case), we naturally get the limit of Eq.~(\ref{eq:kuv_limit}).
Also, since the UV cutoff scale determines the shortest timescale we consider, we need to satisfy Eq.~(\ref{eq:kuv_limit}) to follow the transition between the SR and the non-SR periods properly. 
Given these, we believe that the correct limit is Eq.~(\ref{eq:kuv_limit}).

%%%%%%%%%%%%%%%%%%%%%%%%%%%%%%%%%%%
%%%%%%%%%%%%%%%%%%%%%%%%%%%%%%%%%%%
\small
\bibliographystyle{apsrev4-1}
\bibliography{draft_sh_one_loop_backre_and_reno}

%merlin.mbs apsrev4-1.bst 2010-07-25 4.21a (PWD, AO, DPC) hacked
%Control: key (0)
%Control: author (72) initials jnrlst
%Control: editor formatted (1) identically to author
%Control: production of article title (-1) disabled
%Control: page (0) single
%Control: year (1) truncated
%Control: production of eprint (0) enabled
\begin{thebibliography}{76}%
\makeatletter
\providecommand \@ifxundefined [1]{%
 \@ifx{#1\undefined}
}%
\providecommand \@ifnum [1]{%
 \ifnum #1\expandafter \@firstoftwo
 \else \expandafter \@secondoftwo
 \fi
}%
\providecommand \@ifx [1]{%
 \ifx #1\expandafter \@firstoftwo
 \else \expandafter \@secondoftwo
 \fi
}%
\providecommand \natexlab [1]{#1}%
\providecommand \enquote  [1]{``#1''}%
\providecommand \bibnamefont  [1]{#1}%
\providecommand \bibfnamefont [1]{#1}%
\providecommand \citenamefont [1]{#1}%
\providecommand \href@noop [0]{\@secondoftwo}%
\providecommand \href [0]{\begingroup \@sanitize@url \@href}%
\providecommand \@href[1]{\@@startlink{#1}\@@href}%
\providecommand \@@href[1]{\endgroup#1\@@endlink}%
\providecommand \@sanitize@url [0]{\catcode `\\12\catcode `\$12\catcode
  `\&12\catcode `\#12\catcode `\^12\catcode `\_12\catcode `\%12\relax}%
\providecommand \@@startlink[1]{}%
\providecommand \@@endlink[0]{}%
\providecommand \url  [0]{\begingroup\@sanitize@url \@url }%
\providecommand \@url [1]{\endgroup\@href {#1}{\urlprefix }}%
\providecommand \urlprefix  [0]{URL }%
\providecommand \Eprint [0]{\href }%
\providecommand \doibase [0]{http://dx.doi.org/}%
\providecommand \selectlanguage [0]{\@gobble}%
\providecommand \bibinfo  [0]{\@secondoftwo}%
\providecommand \bibfield  [0]{\@secondoftwo}%
\providecommand \translation [1]{[#1]}%
\providecommand \BibitemOpen [0]{}%
\providecommand \bibitemStop [0]{}%
\providecommand \bibitemNoStop [0]{.\EOS\space}%
\providecommand \EOS [0]{\spacefactor3000\relax}%
\providecommand \BibitemShut  [1]{\csname bibitem#1\endcsname}%
\let\auto@bib@innerbib\@empty
%</preamble>
\bibitem [{\citenamefont {Bardeen}(1980)}]{Bardeen:1980kt}%
  \BibitemOpen
  \bibfield  {author} {\bibinfo {author} {\bibfnamefont {J.~M.}\ \bibnamefont
  {Bardeen}},\ }\href {\doibase 10.1103/PhysRevD.22.1882} {\bibfield  {journal}
  {\bibinfo  {journal} {Phys. Rev. D}\ }\textbf {\bibinfo {volume} {22}},\
  \bibinfo {pages} {1882} (\bibinfo {year} {1980})}\BibitemShut {NoStop}%
\bibitem [{\citenamefont {Kodama}\ and\ \citenamefont
  {Sasaki}(1984)}]{Kodama:1984ziu}%
  \BibitemOpen
  \bibfield  {author} {\bibinfo {author} {\bibfnamefont {H.}~\bibnamefont
  {Kodama}}\ and\ \bibinfo {author} {\bibfnamefont {M.}~\bibnamefont
  {Sasaki}},\ }\href {\doibase 10.1143/PTPS.78.1} {\bibfield  {journal}
  {\bibinfo  {journal} {Prog. Theor. Phys. Suppl.}\ }\textbf {\bibinfo {volume}
  {78}},\ \bibinfo {pages} {1} (\bibinfo {year} {1984})}\BibitemShut {NoStop}%
\bibitem [{\citenamefont {Senatore}\ and\ \citenamefont
  {Zaldarriaga}(2010)}]{Senatore:2009cf}%
  \BibitemOpen
  \bibfield  {author} {\bibinfo {author} {\bibfnamefont {L.}~\bibnamefont
  {Senatore}}\ and\ \bibinfo {author} {\bibfnamefont {M.}~\bibnamefont
  {Zaldarriaga}},\ }\href {\doibase 10.1007/JHEP12(2010)008} {\bibfield
  {journal} {\bibinfo  {journal} {JHEP}\ }\textbf {\bibinfo {volume} {12}},\
  \bibinfo {pages} {008} (\bibinfo {year} {2010})},\ \Eprint
  {http://arxiv.org/abs/0912.2734} {arXiv:0912.2734 [hep-th]} \BibitemShut
  {NoStop}%
\bibitem [{\citenamefont {Senatore}\ and\ \citenamefont
  {Zaldarriaga}(2013)}]{Senatore:2012nq}%
  \BibitemOpen
  \bibfield  {author} {\bibinfo {author} {\bibfnamefont {L.}~\bibnamefont
  {Senatore}}\ and\ \bibinfo {author} {\bibfnamefont {M.}~\bibnamefont
  {Zaldarriaga}},\ }\href {\doibase 10.1007/JHEP01(2013)109} {\bibfield
  {journal} {\bibinfo  {journal} {JHEP}\ }\textbf {\bibinfo {volume} {01}},\
  \bibinfo {pages} {109} (\bibinfo {year} {2013})},\ \Eprint
  {http://arxiv.org/abs/1203.6354} {arXiv:1203.6354 [hep-th]} \BibitemShut
  {NoStop}%
\bibitem [{\citenamefont {Pimentel}\ \emph {et~al.}(2012)\citenamefont
  {Pimentel}, \citenamefont {Senatore},\ and\ \citenamefont
  {Zaldarriaga}}]{Pimentel:2012tw}%
  \BibitemOpen
  \bibfield  {author} {\bibinfo {author} {\bibfnamefont {G.~L.}\ \bibnamefont
  {Pimentel}}, \bibinfo {author} {\bibfnamefont {L.}~\bibnamefont {Senatore}},
  \ and\ \bibinfo {author} {\bibfnamefont {M.}~\bibnamefont {Zaldarriaga}},\
  }\href {\doibase 10.1007/JHEP07(2012)166} {\bibfield  {journal} {\bibinfo
  {journal} {JHEP}\ }\textbf {\bibinfo {volume} {07}},\ \bibinfo {pages} {166}
  (\bibinfo {year} {2012})},\ \Eprint {http://arxiv.org/abs/1203.6651}
  {arXiv:1203.6651 [hep-th]} \BibitemShut {NoStop}%
\bibitem [{\citenamefont {Lyth}\ \emph {et~al.}(2005)\citenamefont {Lyth},
  \citenamefont {Malik},\ and\ \citenamefont {Sasaki}}]{Lyth:2004gb}%
  \BibitemOpen
  \bibfield  {author} {\bibinfo {author} {\bibfnamefont {D.~H.}\ \bibnamefont
  {Lyth}}, \bibinfo {author} {\bibfnamefont {K.~A.}\ \bibnamefont {Malik}}, \
  and\ \bibinfo {author} {\bibfnamefont {M.}~\bibnamefont {Sasaki}},\ }\href
  {\doibase 10.1088/1475-7516/2005/05/004} {\bibfield  {journal} {\bibinfo
  {journal} {JCAP}\ }\textbf {\bibinfo {volume} {05}},\ \bibinfo {pages} {004}
  (\bibinfo {year} {2005})},\ \Eprint {http://arxiv.org/abs/astro-ph/0411220}
  {arXiv:astro-ph/0411220} \BibitemShut {NoStop}%
\bibitem [{\citenamefont {Chapline}(1975)}]{Chapline:1975ojl}%
  \BibitemOpen
  \bibfield  {author} {\bibinfo {author} {\bibfnamefont {G.~F.}\ \bibnamefont
  {Chapline}},\ }\href {\doibase 10.1038/253251a0} {\bibfield  {journal}
  {\bibinfo  {journal} {Nature}\ }\textbf {\bibinfo {volume} {253}},\ \bibinfo
  {pages} {251} (\bibinfo {year} {1975})}\BibitemShut {NoStop}%
\bibitem [{\citenamefont {Ivanov}\ \emph {et~al.}(1994)\citenamefont {Ivanov},
  \citenamefont {Naselsky},\ and\ \citenamefont {Novikov}}]{Ivanov:1994pa}%
  \BibitemOpen
  \bibfield  {author} {\bibinfo {author} {\bibfnamefont {P.}~\bibnamefont
  {Ivanov}}, \bibinfo {author} {\bibfnamefont {P.}~\bibnamefont {Naselsky}}, \
  and\ \bibinfo {author} {\bibfnamefont {I.}~\bibnamefont {Novikov}},\ }\href
  {\doibase 10.1103/PhysRevD.50.7173} {\bibfield  {journal} {\bibinfo
  {journal} {Phys. Rev. D}\ }\textbf {\bibinfo {volume} {50}},\ \bibinfo
  {pages} {7173} (\bibinfo {year} {1994})}\BibitemShut {NoStop}%
\bibitem [{\citenamefont {Yokoyama}(1997)}]{Yokoyama:1995ex}%
  \BibitemOpen
  \bibfield  {author} {\bibinfo {author} {\bibfnamefont {J.}~\bibnamefont
  {Yokoyama}},\ }\href@noop {} {\bibfield  {journal} {\bibinfo  {journal}
  {Astron. Astrophys.}\ }\textbf {\bibinfo {volume} {318}},\ \bibinfo {pages}
  {673} (\bibinfo {year} {1997})},\ \Eprint
  {http://arxiv.org/abs/astro-ph/9509027} {arXiv:astro-ph/9509027} \BibitemShut
  {NoStop}%
\bibitem [{\citenamefont {Garcia-Bellido}\ \emph {et~al.}(1996)\citenamefont
  {Garcia-Bellido}, \citenamefont {Linde},\ and\ \citenamefont
  {Wands}}]{GarciaBellido:1996qt}%
  \BibitemOpen
  \bibfield  {author} {\bibinfo {author} {\bibfnamefont {J.}~\bibnamefont
  {Garcia-Bellido}}, \bibinfo {author} {\bibfnamefont {A.~D.}\ \bibnamefont
  {Linde}}, \ and\ \bibinfo {author} {\bibfnamefont {D.}~\bibnamefont
  {Wands}},\ }\href {\doibase 10.1103/PhysRevD.54.6040} {\bibfield  {journal}
  {\bibinfo  {journal} {Phys. Rev.}\ }\textbf {\bibinfo {volume} {D54}},\
  \bibinfo {pages} {6040} (\bibinfo {year} {1996})},\ \Eprint
  {http://arxiv.org/abs/astro-ph/9605094} {arXiv:astro-ph/9605094 [astro-ph]}
  \BibitemShut {NoStop}%
%%CITATION = ASTRO-PH/9605094;%%
\bibitem [{\citenamefont {Afshordi}\ \emph {et~al.}(2003)\citenamefont
  {Afshordi}, \citenamefont {McDonald},\ and\ \citenamefont
  {Spergel}}]{Afshordi:2003zb}%
  \BibitemOpen
  \bibfield  {author} {\bibinfo {author} {\bibfnamefont {N.}~\bibnamefont
  {Afshordi}}, \bibinfo {author} {\bibfnamefont {P.}~\bibnamefont {McDonald}},
  \ and\ \bibinfo {author} {\bibfnamefont {D.~N.}\ \bibnamefont {Spergel}},\
  }\href {\doibase 10.1086/378763} {\bibfield  {journal} {\bibinfo  {journal}
  {Astrophys. J. Lett.}\ }\textbf {\bibinfo {volume} {594}},\ \bibinfo {pages}
  {L71} (\bibinfo {year} {2003})},\ \Eprint
  {http://arxiv.org/abs/astro-ph/0302035} {arXiv:astro-ph/0302035} \BibitemShut
  {NoStop}%
\bibitem [{\citenamefont {Frampton}\ \emph {et~al.}(2010)\citenamefont
  {Frampton}, \citenamefont {Kawasaki}, \citenamefont {Takahashi},\ and\
  \citenamefont {Yanagida}}]{Frampton:2010sw}%
  \BibitemOpen
  \bibfield  {author} {\bibinfo {author} {\bibfnamefont {P.~H.}\ \bibnamefont
  {Frampton}}, \bibinfo {author} {\bibfnamefont {M.}~\bibnamefont {Kawasaki}},
  \bibinfo {author} {\bibfnamefont {F.}~\bibnamefont {Takahashi}}, \ and\
  \bibinfo {author} {\bibfnamefont {T.~T.}\ \bibnamefont {Yanagida}},\ }\href
  {\doibase 10.1088/1475-7516/2010/04/023} {\bibfield  {journal} {\bibinfo
  {journal} {JCAP}\ }\textbf {\bibinfo {volume} {1004}},\ \bibinfo {pages}
  {023} (\bibinfo {year} {2010})},\ \Eprint {http://arxiv.org/abs/1001.2308}
  {arXiv:1001.2308 [hep-ph]} \BibitemShut {NoStop}%
%%CITATION = ARXIV:1001.2308;%%
\bibitem [{\citenamefont {Belotsky}\ \emph {et~al.}(2014)\citenamefont
  {Belotsky}, \citenamefont {Dmitriev}, \citenamefont {Esipova}, \citenamefont
  {Gani}, \citenamefont {Grobov}, \citenamefont {Khlopov}, \citenamefont
  {Kirillov}, \citenamefont {Rubin},\ and\ \citenamefont
  {Svadkovsky}}]{Belotsky:2014kca}%
  \BibitemOpen
  \bibfield  {author} {\bibinfo {author} {\bibfnamefont {K.~M.}\ \bibnamefont
  {Belotsky}}, \bibinfo {author} {\bibfnamefont {A.~D.}\ \bibnamefont
  {Dmitriev}}, \bibinfo {author} {\bibfnamefont {E.~A.}\ \bibnamefont
  {Esipova}}, \bibinfo {author} {\bibfnamefont {V.~A.}\ \bibnamefont {Gani}},
  \bibinfo {author} {\bibfnamefont {A.~V.}\ \bibnamefont {Grobov}}, \bibinfo
  {author} {\bibfnamefont {M.~Y.}\ \bibnamefont {Khlopov}}, \bibinfo {author}
  {\bibfnamefont {A.~A.}\ \bibnamefont {Kirillov}}, \bibinfo {author}
  {\bibfnamefont {S.~G.}\ \bibnamefont {Rubin}}, \ and\ \bibinfo {author}
  {\bibfnamefont {I.~V.}\ \bibnamefont {Svadkovsky}},\ }\href {\doibase
  10.1142/S0217732314400057} {\bibfield  {journal} {\bibinfo  {journal} {Mod.
  Phys. Lett. A}\ }\textbf {\bibinfo {volume} {29}},\ \bibinfo {pages}
  {1440005} (\bibinfo {year} {2014})},\ \Eprint
  {http://arxiv.org/abs/1410.0203} {arXiv:1410.0203 [astro-ph.CO]} \BibitemShut
  {NoStop}%
\bibitem [{\citenamefont {Carr}\ \emph {et~al.}(2016)\citenamefont {Carr},
  \citenamefont {Kuhnel},\ and\ \citenamefont {Sandstad}}]{Carr:2016drx}%
  \BibitemOpen
  \bibfield  {author} {\bibinfo {author} {\bibfnamefont {B.}~\bibnamefont
  {Carr}}, \bibinfo {author} {\bibfnamefont {F.}~\bibnamefont {Kuhnel}}, \ and\
  \bibinfo {author} {\bibfnamefont {M.}~\bibnamefont {Sandstad}},\ }\href
  {\doibase 10.1103/PhysRevD.94.083504} {\bibfield  {journal} {\bibinfo
  {journal} {Phys. Rev.}\ }\textbf {\bibinfo {volume} {D94}},\ \bibinfo {pages}
  {083504} (\bibinfo {year} {2016})},\ \Eprint
  {http://arxiv.org/abs/1607.06077} {arXiv:1607.06077 [astro-ph.CO]}
  \BibitemShut {NoStop}%
%%CITATION = ARXIV:1607.06077;%%
\bibitem [{\citenamefont {Inomata}\ \emph {et~al.}(2017)\citenamefont
  {Inomata}, \citenamefont {Kawasaki}, \citenamefont {Mukaida}, \citenamefont
  {Tada},\ and\ \citenamefont {Yanagida}}]{Inomata:2017okj}%
  \BibitemOpen
  \bibfield  {author} {\bibinfo {author} {\bibfnamefont {K.}~\bibnamefont
  {Inomata}}, \bibinfo {author} {\bibfnamefont {M.}~\bibnamefont {Kawasaki}},
  \bibinfo {author} {\bibfnamefont {K.}~\bibnamefont {Mukaida}}, \bibinfo
  {author} {\bibfnamefont {Y.}~\bibnamefont {Tada}}, \ and\ \bibinfo {author}
  {\bibfnamefont {T.~T.}\ \bibnamefont {Yanagida}},\ }\href {\doibase
  10.1103/PhysRevD.96.043504} {\bibfield  {journal} {\bibinfo  {journal} {Phys.
  Rev.}\ }\textbf {\bibinfo {volume} {D96}},\ \bibinfo {pages} {043504}
  (\bibinfo {year} {2017})},\ \Eprint {http://arxiv.org/abs/1701.02544}
  {arXiv:1701.02544 [astro-ph.CO]} \BibitemShut {NoStop}%
%%CITATION = ARXIV:1701.02544;%%
\bibitem [{\citenamefont {Espinosa}\ \emph {et~al.}(2018)\citenamefont
  {Espinosa}, \citenamefont {Racco},\ and\ \citenamefont
  {Riotto}}]{Espinosa:2017sgp}%
  \BibitemOpen
  \bibfield  {author} {\bibinfo {author} {\bibfnamefont {J.~R.}\ \bibnamefont
  {Espinosa}}, \bibinfo {author} {\bibfnamefont {D.}~\bibnamefont {Racco}}, \
  and\ \bibinfo {author} {\bibfnamefont {A.}~\bibnamefont {Riotto}},\ }\href
  {\doibase 10.1103/PhysRevLett.120.121301} {\bibfield  {journal} {\bibinfo
  {journal} {Phys. Rev. Lett.}\ }\textbf {\bibinfo {volume} {120}},\ \bibinfo
  {pages} {121301} (\bibinfo {year} {2018})},\ \Eprint
  {http://arxiv.org/abs/1710.11196} {arXiv:1710.11196 [hep-ph]} \BibitemShut
  {NoStop}%
\bibitem [{\citenamefont {Bird}\ \emph {et~al.}(2016)\citenamefont {Bird},
  \citenamefont {Cholis}, \citenamefont {Mu{\~n}oz}, \citenamefont
  {Ali-Ha{\"i}moud}, \citenamefont {Kamionkowski}, \citenamefont {Kovetz},
  \citenamefont {Raccanelli},\ and\ \citenamefont {Riess}}]{Bird:2016dcv}%
  \BibitemOpen
  \bibfield  {author} {\bibinfo {author} {\bibfnamefont {S.}~\bibnamefont
  {Bird}}, \bibinfo {author} {\bibfnamefont {I.}~\bibnamefont {Cholis}},
  \bibinfo {author} {\bibfnamefont {J.~B.}\ \bibnamefont {Mu{\~n}oz}}, \bibinfo
  {author} {\bibfnamefont {Y.}~\bibnamefont {Ali-Ha{\"i}moud}}, \bibinfo
  {author} {\bibfnamefont {M.}~\bibnamefont {Kamionkowski}}, \bibinfo {author}
  {\bibfnamefont {E.~D.}\ \bibnamefont {Kovetz}}, \bibinfo {author}
  {\bibfnamefont {A.}~\bibnamefont {Raccanelli}}, \ and\ \bibinfo {author}
  {\bibfnamefont {A.~G.}\ \bibnamefont {Riess}},\ }\href {\doibase
  10.1103/PhysRevLett.116.201301} {\bibfield  {journal} {\bibinfo  {journal}
  {Phys. Rev. Lett.}\ }\textbf {\bibinfo {volume} {116}},\ \bibinfo {pages}
  {201301} (\bibinfo {year} {2016})},\ \Eprint
  {http://arxiv.org/abs/1603.00464} {arXiv:1603.00464 [astro-ph.CO]}
  \BibitemShut {NoStop}%
%%CITATION = ARXIV:1603.00464;%%
\bibitem [{\citenamefont {Clesse}\ and\ \citenamefont
  {Garc{\'i}a-Bellido}(2016)}]{Clesse:2016vqa}%
  \BibitemOpen
  \bibfield  {author} {\bibinfo {author} {\bibfnamefont {S.}~\bibnamefont
  {Clesse}}\ and\ \bibinfo {author} {\bibfnamefont {J.}~\bibnamefont
  {Garc{\'i}a-Bellido}},\ }\href {\doibase 10.1016/j.dark.2016.10.002}
  {\bibfield  {journal} {\bibinfo  {journal} {Phys. Dark Univ.}\ }\textbf
  {\bibinfo {volume} {10}},\ \bibinfo {pages} {002} (\bibinfo {year} {2016})},\
  \Eprint {http://arxiv.org/abs/1603.05234} {arXiv:1603.05234 [astro-ph.CO]}
  \BibitemShut {NoStop}%
%%CITATION = ARXIV:1603.05234;%%
\bibitem [{\citenamefont {Sasaki}\ \emph {et~al.}(2016)\citenamefont {Sasaki},
  \citenamefont {Suyama}, \citenamefont {Tanaka},\ and\ \citenamefont
  {Yokoyama}}]{Sasaki:2016jop}%
  \BibitemOpen
  \bibfield  {author} {\bibinfo {author} {\bibfnamefont {M.}~\bibnamefont
  {Sasaki}}, \bibinfo {author} {\bibfnamefont {T.}~\bibnamefont {Suyama}},
  \bibinfo {author} {\bibfnamefont {T.}~\bibnamefont {Tanaka}}, \ and\ \bibinfo
  {author} {\bibfnamefont {S.}~\bibnamefont {Yokoyama}},\ }\href {\doibase
  10.1103/PhysRevLett.117.061101} {\bibfield  {journal} {\bibinfo  {journal}
  {Phys. Rev. Lett.}\ }\textbf {\bibinfo {volume} {117}},\ \bibinfo {pages}
  {061101} (\bibinfo {year} {2016})},\ \Eprint
  {http://arxiv.org/abs/1603.08338} {arXiv:1603.08338 [astro-ph.CO]}
  \BibitemShut {NoStop}%
%%CITATION = ARXIV:1603.08338;%%
\bibitem [{\citenamefont {Eroshenko}(2018)}]{Eroshenko:2016hmn}%
  \BibitemOpen
  \bibfield  {author} {\bibinfo {author} {\bibfnamefont {Y.~N.}\ \bibnamefont
  {Eroshenko}},\ }\href {\doibase 10.1088/1742-6596/1051/1/012010} {\bibfield
  {journal} {\bibinfo  {journal} {J. Phys. Conf. Ser.}\ }\textbf {\bibinfo
  {volume} {1051}},\ \bibinfo {pages} {012010} (\bibinfo {year} {2018})},\
  \Eprint {http://arxiv.org/abs/1604.04932} {arXiv:1604.04932 [astro-ph.CO]}
  \BibitemShut {NoStop}%
\bibitem [{\citenamefont {Sasaki}\ \emph {et~al.}(2018)\citenamefont {Sasaki},
  \citenamefont {Suyama}, \citenamefont {Tanaka},\ and\ \citenamefont
  {Yokoyama}}]{Sasaki:2018dmp}%
  \BibitemOpen
  \bibfield  {author} {\bibinfo {author} {\bibfnamefont {M.}~\bibnamefont
  {Sasaki}}, \bibinfo {author} {\bibfnamefont {T.}~\bibnamefont {Suyama}},
  \bibinfo {author} {\bibfnamefont {T.}~\bibnamefont {Tanaka}}, \ and\ \bibinfo
  {author} {\bibfnamefont {S.}~\bibnamefont {Yokoyama}},\ }\href {\doibase
  10.1088/1361-6382/aaa7b4} {\bibfield  {journal} {\bibinfo  {journal} {Class.
  Quant. Grav.}\ }\textbf {\bibinfo {volume} {35}},\ \bibinfo {pages} {063001}
  (\bibinfo {year} {2018})},\ \Eprint {http://arxiv.org/abs/1801.05235}
  {arXiv:1801.05235 [astro-ph.CO]} \BibitemShut {NoStop}%
%%CITATION = ARXIV:1801.05235;%%
\bibitem [{\citenamefont {Carr}\ \emph {et~al.}(2021)\citenamefont {Carr},
  \citenamefont {Kohri}, \citenamefont {Sendouda},\ and\ \citenamefont
  {Yokoyama}}]{Carr:2020gox}%
  \BibitemOpen
  \bibfield  {author} {\bibinfo {author} {\bibfnamefont {B.}~\bibnamefont
  {Carr}}, \bibinfo {author} {\bibfnamefont {K.}~\bibnamefont {Kohri}},
  \bibinfo {author} {\bibfnamefont {Y.}~\bibnamefont {Sendouda}}, \ and\
  \bibinfo {author} {\bibfnamefont {J.}~\bibnamefont {Yokoyama}},\ }\href
  {\doibase 10.1088/1361-6633/ac1e31} {\bibfield  {journal} {\bibinfo
  {journal} {Rept. Prog. Phys.}\ }\textbf {\bibinfo {volume} {84}},\ \bibinfo
  {pages} {116902} (\bibinfo {year} {2021})},\ \Eprint
  {http://arxiv.org/abs/2002.12778} {arXiv:2002.12778 [astro-ph.CO]}
  \BibitemShut {NoStop}%
\bibitem [{\citenamefont {Green}\ and\ \citenamefont
  {Kavanagh}(2021)}]{Green:2020jor}%
  \BibitemOpen
  \bibfield  {author} {\bibinfo {author} {\bibfnamefont {A.~M.}\ \bibnamefont
  {Green}}\ and\ \bibinfo {author} {\bibfnamefont {B.~J.}\ \bibnamefont
  {Kavanagh}},\ }\href {\doibase 10.1088/1361-6471/abc534} {\bibfield
  {journal} {\bibinfo  {journal} {J. Phys. G}\ }\textbf {\bibinfo {volume}
  {48}},\ \bibinfo {pages} {043001} (\bibinfo {year} {2021})},\ \Eprint
  {http://arxiv.org/abs/2007.10722} {arXiv:2007.10722 [astro-ph.CO]}
  \BibitemShut {NoStop}%
\bibitem [{\citenamefont {Escriv\`a}\ \emph {et~al.}(2022)\citenamefont
  {Escriv\`a}, \citenamefont {Kuhnel},\ and\ \citenamefont
  {Tada}}]{Escriva:2022duf}%
  \BibitemOpen
  \bibfield  {author} {\bibinfo {author} {\bibfnamefont {A.}~\bibnamefont
  {Escriv\`a}}, \bibinfo {author} {\bibfnamefont {F.}~\bibnamefont {Kuhnel}}, \
  and\ \bibinfo {author} {\bibfnamefont {Y.}~\bibnamefont {Tada}},\ }\href
  {\doibase 10.1016/B978-0-32-395636-9.00012-8} {\  (\bibinfo {year} {2022}),\
  10.1016/B978-0-32-395636-9.00012-8},\ \Eprint
  {http://arxiv.org/abs/2211.05767} {arXiv:2211.05767 [astro-ph.CO]}
  \BibitemShut {NoStop}%
\bibitem [{\citenamefont {Kristiano}\ and\ \citenamefont
  {Yokoyama}(2024{\natexlab{a}})}]{Kristiano:2022maq}%
  \BibitemOpen
  \bibfield  {author} {\bibinfo {author} {\bibfnamefont {J.}~\bibnamefont
  {Kristiano}}\ and\ \bibinfo {author} {\bibfnamefont {J.}~\bibnamefont
  {Yokoyama}},\ }\href {\doibase 10.1103/PhysRevLett.132.221003} {\bibfield
  {journal} {\bibinfo  {journal} {Phys. Rev. Lett.}\ }\textbf {\bibinfo
  {volume} {132}},\ \bibinfo {pages} {221003} (\bibinfo {year}
  {2024}{\natexlab{a}})},\ \Eprint {http://arxiv.org/abs/2211.03395}
  {arXiv:2211.03395 [hep-th]} \BibitemShut {NoStop}%
\bibitem [{\citenamefont {Cheng}\ \emph {et~al.}(2022)\citenamefont {Cheng},
  \citenamefont {Lee},\ and\ \citenamefont {Ng}}]{Cheng:2021lif}%
  \BibitemOpen
  \bibfield  {author} {\bibinfo {author} {\bibfnamefont {S.-L.}\ \bibnamefont
  {Cheng}}, \bibinfo {author} {\bibfnamefont {D.-S.}\ \bibnamefont {Lee}}, \
  and\ \bibinfo {author} {\bibfnamefont {K.-W.}\ \bibnamefont {Ng}},\ }\href
  {\doibase 10.1016/j.physletb.2022.136956} {\bibfield  {journal} {\bibinfo
  {journal} {Phys. Lett. B}\ }\textbf {\bibinfo {volume} {827}},\ \bibinfo
  {pages} {136956} (\bibinfo {year} {2022})},\ \Eprint
  {http://arxiv.org/abs/2106.09275} {arXiv:2106.09275 [astro-ph.CO]}
  \BibitemShut {NoStop}%
\bibitem [{\citenamefont {Kinney}(1997)}]{Kinney:1997ne}%
  \BibitemOpen
  \bibfield  {author} {\bibinfo {author} {\bibfnamefont {W.~H.}\ \bibnamefont
  {Kinney}},\ }\href {\doibase 10.1103/PhysRevD.56.2002} {\bibfield  {journal}
  {\bibinfo  {journal} {Phys. Rev. D}\ }\textbf {\bibinfo {volume} {56}},\
  \bibinfo {pages} {2002} (\bibinfo {year} {1997})},\ \Eprint
  {http://arxiv.org/abs/hep-ph/9702427} {arXiv:hep-ph/9702427} \BibitemShut
  {NoStop}%
\bibitem [{\citenamefont {Inoue}\ and\ \citenamefont
  {Yokoyama}(2002)}]{Inoue:2001zt}%
  \BibitemOpen
  \bibfield  {author} {\bibinfo {author} {\bibfnamefont {S.}~\bibnamefont
  {Inoue}}\ and\ \bibinfo {author} {\bibfnamefont {J.}~\bibnamefont
  {Yokoyama}},\ }\href {\doibase 10.1016/S0370-2693(01)01369-7} {\bibfield
  {journal} {\bibinfo  {journal} {Phys. Lett. B}\ }\textbf {\bibinfo {volume}
  {524}},\ \bibinfo {pages} {15} (\bibinfo {year} {2002})},\ \Eprint
  {http://arxiv.org/abs/hep-ph/0104083} {arXiv:hep-ph/0104083} \BibitemShut
  {NoStop}%
\bibitem [{\citenamefont {Kinney}(2005)}]{Kinney:2005vj}%
  \BibitemOpen
  \bibfield  {author} {\bibinfo {author} {\bibfnamefont {W.~H.}\ \bibnamefont
  {Kinney}},\ }\href {\doibase 10.1103/PhysRevD.72.023515} {\bibfield
  {journal} {\bibinfo  {journal} {Phys. Rev. D}\ }\textbf {\bibinfo {volume}
  {72}},\ \bibinfo {pages} {023515} (\bibinfo {year} {2005})},\ \Eprint
  {http://arxiv.org/abs/gr-qc/0503017} {arXiv:gr-qc/0503017} \BibitemShut
  {NoStop}%
\bibitem [{\citenamefont {Martin}\ \emph {et~al.}(2013)\citenamefont {Martin},
  \citenamefont {Motohashi},\ and\ \citenamefont {Suyama}}]{Martin:2012pe}%
  \BibitemOpen
  \bibfield  {author} {\bibinfo {author} {\bibfnamefont {J.}~\bibnamefont
  {Martin}}, \bibinfo {author} {\bibfnamefont {H.}~\bibnamefont {Motohashi}}, \
  and\ \bibinfo {author} {\bibfnamefont {T.}~\bibnamefont {Suyama}},\ }\href
  {\doibase 10.1103/PhysRevD.87.023514} {\bibfield  {journal} {\bibinfo
  {journal} {Phys. Rev. D}\ }\textbf {\bibinfo {volume} {87}},\ \bibinfo
  {pages} {023514} (\bibinfo {year} {2013})},\ \Eprint
  {http://arxiv.org/abs/1211.0083} {arXiv:1211.0083 [astro-ph.CO]} \BibitemShut
  {NoStop}%
\bibitem [{\citenamefont {Riotto}(2023{\natexlab{a}})}]{Riotto:2023hoz}%
  \BibitemOpen
  \bibfield  {author} {\bibinfo {author} {\bibfnamefont {A.}~\bibnamefont
  {Riotto}},\ }\href@noop {} {\  (\bibinfo {year} {2023}{\natexlab{a}})},\
  \Eprint {http://arxiv.org/abs/2301.00599} {arXiv:2301.00599 [astro-ph.CO]}
  \BibitemShut {NoStop}%
\bibitem [{\citenamefont {Choudhury}\ \emph {et~al.}(2024)\citenamefont
  {Choudhury}, \citenamefont {Gangopadhyay},\ and\ \citenamefont
  {Sami}}]{Choudhury:2023vuj}%
  \BibitemOpen
  \bibfield  {author} {\bibinfo {author} {\bibfnamefont {S.}~\bibnamefont
  {Choudhury}}, \bibinfo {author} {\bibfnamefont {M.~R.}\ \bibnamefont
  {Gangopadhyay}}, \ and\ \bibinfo {author} {\bibfnamefont {M.}~\bibnamefont
  {Sami}},\ }\href {\doibase 10.1140/epjc/s10052-024-13218-2} {\bibfield
  {journal} {\bibinfo  {journal} {Eur. Phys. J. C}\ }\textbf {\bibinfo {volume}
  {84}},\ \bibinfo {pages} {884} (\bibinfo {year} {2024})},\ \Eprint
  {http://arxiv.org/abs/2301.10000} {arXiv:2301.10000 [astro-ph.CO]}
  \BibitemShut {NoStop}%
\bibitem [{\citenamefont {Kristiano}\ and\ \citenamefont
  {Yokoyama}(2024{\natexlab{b}})}]{Kristiano:2023scm}%
  \BibitemOpen
  \bibfield  {author} {\bibinfo {author} {\bibfnamefont {J.}~\bibnamefont
  {Kristiano}}\ and\ \bibinfo {author} {\bibfnamefont {J.}~\bibnamefont
  {Yokoyama}},\ }\href {\doibase 10.1103/PhysRevD.109.103541} {\bibfield
  {journal} {\bibinfo  {journal} {Phys. Rev. D}\ }\textbf {\bibinfo {volume}
  {109}},\ \bibinfo {pages} {103541} (\bibinfo {year} {2024}{\natexlab{b}})},\
  \Eprint {http://arxiv.org/abs/2303.00341} {arXiv:2303.00341 [hep-th]}
  \BibitemShut {NoStop}%
\bibitem [{\citenamefont {Riotto}(2023{\natexlab{b}})}]{Riotto:2023gpm}%
  \BibitemOpen
  \bibfield  {author} {\bibinfo {author} {\bibfnamefont {A.}~\bibnamefont
  {Riotto}},\ }\href@noop {} {\  (\bibinfo {year} {2023}{\natexlab{b}})},\
  \Eprint {http://arxiv.org/abs/2303.01727} {arXiv:2303.01727 [astro-ph.CO]}
  \BibitemShut {NoStop}%
\bibitem [{\citenamefont {Firouzjahi}(2023)}]{Firouzjahi:2023aum}%
  \BibitemOpen
  \bibfield  {author} {\bibinfo {author} {\bibfnamefont {H.}~\bibnamefont
  {Firouzjahi}},\ }\href {\doibase 10.1088/1475-7516/2023/10/006} {\bibfield
  {journal} {\bibinfo  {journal} {JCAP}\ }\textbf {\bibinfo {volume} {10}},\
  \bibinfo {pages} {006} (\bibinfo {year} {2023})},\ \Eprint
  {http://arxiv.org/abs/2303.12025} {arXiv:2303.12025 [astro-ph.CO]}
  \BibitemShut {NoStop}%
\bibitem [{\citenamefont {Motohashi}\ and\ \citenamefont
  {Tada}(2023)}]{Motohashi:2023syh}%
  \BibitemOpen
  \bibfield  {author} {\bibinfo {author} {\bibfnamefont {H.}~\bibnamefont
  {Motohashi}}\ and\ \bibinfo {author} {\bibfnamefont {Y.}~\bibnamefont
  {Tada}},\ }\href {\doibase 10.1088/1475-7516/2023/08/069} {\bibfield
  {journal} {\bibinfo  {journal} {JCAP}\ }\textbf {\bibinfo {volume} {08}},\
  \bibinfo {pages} {069} (\bibinfo {year} {2023})},\ \Eprint
  {http://arxiv.org/abs/2303.16035} {arXiv:2303.16035 [astro-ph.CO]}
  \BibitemShut {NoStop}%
\bibitem [{\citenamefont {Firouzjahi}\ and\ \citenamefont
  {Riotto}(2024)}]{Firouzjahi:2023ahg}%
  \BibitemOpen
  \bibfield  {author} {\bibinfo {author} {\bibfnamefont {H.}~\bibnamefont
  {Firouzjahi}}\ and\ \bibinfo {author} {\bibfnamefont {A.}~\bibnamefont
  {Riotto}},\ }\href {\doibase 10.1088/1475-7516/2024/02/021} {\bibfield
  {journal} {\bibinfo  {journal} {JCAP}\ }\textbf {\bibinfo {volume} {02}},\
  \bibinfo {pages} {021} (\bibinfo {year} {2024})},\ \Eprint
  {http://arxiv.org/abs/2304.07801} {arXiv:2304.07801 [astro-ph.CO]}
  \BibitemShut {NoStop}%
\bibitem [{\citenamefont {Franciolini}\ \emph {et~al.}(2024)\citenamefont
  {Franciolini}, \citenamefont {Iovino}, \citenamefont {Taoso},\ and\
  \citenamefont {Urbano}}]{Franciolini:2023agm}%
  \BibitemOpen
  \bibfield  {author} {\bibinfo {author} {\bibfnamefont {G.}~\bibnamefont
  {Franciolini}}, \bibinfo {author} {\bibfnamefont {A.}~\bibnamefont {Iovino},
  \bibfnamefont {Junior.}}, \bibinfo {author} {\bibfnamefont {M.}~\bibnamefont
  {Taoso}}, \ and\ \bibinfo {author} {\bibfnamefont {A.}~\bibnamefont
  {Urbano}},\ }\href {\doibase 10.1103/PhysRevD.109.123550} {\bibfield
  {journal} {\bibinfo  {journal} {Phys. Rev. D}\ }\textbf {\bibinfo {volume}
  {109}},\ \bibinfo {pages} {123550} (\bibinfo {year} {2024})},\ \Eprint
  {http://arxiv.org/abs/2305.03491} {arXiv:2305.03491 [astro-ph.CO]}
  \BibitemShut {NoStop}%
\bibitem [{\citenamefont {Tasinato}(2023)}]{Tasinato:2023ukp}%
  \BibitemOpen
  \bibfield  {author} {\bibinfo {author} {\bibfnamefont {G.}~\bibnamefont
  {Tasinato}},\ }\href {\doibase 10.1103/PhysRevD.108.043526} {\bibfield
  {journal} {\bibinfo  {journal} {Phys. Rev. D}\ }\textbf {\bibinfo {volume}
  {108}},\ \bibinfo {pages} {043526} (\bibinfo {year} {2023})},\ \Eprint
  {http://arxiv.org/abs/2305.11568} {arXiv:2305.11568 [hep-th]} \BibitemShut
  {NoStop}%
\bibitem [{\citenamefont {Cheng}\ \emph {et~al.}(2024)\citenamefont {Cheng},
  \citenamefont {Lee},\ and\ \citenamefont {Ng}}]{Cheng:2023ikq}%
  \BibitemOpen
  \bibfield  {author} {\bibinfo {author} {\bibfnamefont {S.-L.}\ \bibnamefont
  {Cheng}}, \bibinfo {author} {\bibfnamefont {D.-S.}\ \bibnamefont {Lee}}, \
  and\ \bibinfo {author} {\bibfnamefont {K.-W.}\ \bibnamefont {Ng}},\ }\href
  {\doibase 10.1088/1475-7516/2024/03/008} {\bibfield  {journal} {\bibinfo
  {journal} {JCAP}\ }\textbf {\bibinfo {volume} {03}},\ \bibinfo {pages} {008}
  (\bibinfo {year} {2024})},\ \Eprint {http://arxiv.org/abs/2305.16810}
  {arXiv:2305.16810 [astro-ph.CO]} \BibitemShut {NoStop}%
\bibitem [{\citenamefont {Fumagalli}(2023)}]{Fumagalli:2023hpa}%
  \BibitemOpen
  \bibfield  {author} {\bibinfo {author} {\bibfnamefont {J.}~\bibnamefont
  {Fumagalli}},\ }\href@noop {} {\  (\bibinfo {year} {2023})},\ \Eprint
  {http://arxiv.org/abs/2305.19263} {arXiv:2305.19263 [astro-ph.CO]}
  \BibitemShut {NoStop}%
\bibitem [{\citenamefont {Maity}\ \emph {et~al.}(2024)\citenamefont {Maity},
  \citenamefont {Ragavendra}, \citenamefont {Sethi},\ and\ \citenamefont
  {Sriramkumar}}]{Maity:2023qzw}%
  \BibitemOpen
  \bibfield  {author} {\bibinfo {author} {\bibfnamefont {S.}~\bibnamefont
  {Maity}}, \bibinfo {author} {\bibfnamefont {H.~V.}\ \bibnamefont
  {Ragavendra}}, \bibinfo {author} {\bibfnamefont {S.~K.}\ \bibnamefont
  {Sethi}}, \ and\ \bibinfo {author} {\bibfnamefont {L.}~\bibnamefont
  {Sriramkumar}},\ }\href {\doibase 10.1088/1475-7516/2024/05/046} {\bibfield
  {journal} {\bibinfo  {journal} {JCAP}\ }\textbf {\bibinfo {volume} {05}},\
  \bibinfo {pages} {046} (\bibinfo {year} {2024})},\ \Eprint
  {http://arxiv.org/abs/2307.13636} {arXiv:2307.13636 [astro-ph.CO]}
  \BibitemShut {NoStop}%
\bibitem [{\citenamefont {Tada}\ \emph {et~al.}(2024)\citenamefont {Tada},
  \citenamefont {Terada},\ and\ \citenamefont {Tokuda}}]{Tada:2023rgp}%
  \BibitemOpen
  \bibfield  {author} {\bibinfo {author} {\bibfnamefont {Y.}~\bibnamefont
  {Tada}}, \bibinfo {author} {\bibfnamefont {T.}~\bibnamefont {Terada}}, \ and\
  \bibinfo {author} {\bibfnamefont {J.}~\bibnamefont {Tokuda}},\ }\href
  {\doibase 10.1007/JHEP01(2024)105} {\bibfield  {journal} {\bibinfo  {journal}
  {JHEP}\ }\textbf {\bibinfo {volume} {01}},\ \bibinfo {pages} {105} (\bibinfo
  {year} {2024})},\ \Eprint {http://arxiv.org/abs/2308.04732} {arXiv:2308.04732
  [hep-th]} \BibitemShut {NoStop}%
\bibitem [{\citenamefont {Firouzjahi}(2024)}]{Firouzjahi:2023bkt}%
  \BibitemOpen
  \bibfield  {author} {\bibinfo {author} {\bibfnamefont {H.}~\bibnamefont
  {Firouzjahi}},\ }\href {\doibase 10.1103/PhysRevD.109.043514} {\bibfield
  {journal} {\bibinfo  {journal} {Phys. Rev. D}\ }\textbf {\bibinfo {volume}
  {109}},\ \bibinfo {pages} {043514} (\bibinfo {year} {2024})},\ \Eprint
  {http://arxiv.org/abs/2311.04080} {arXiv:2311.04080 [astro-ph.CO]}
  \BibitemShut {NoStop}%
\bibitem [{\citenamefont {Davies}\ \emph {et~al.}(2024)\citenamefont {Davies},
  \citenamefont {Iacconi},\ and\ \citenamefont {Mulryne}}]{Davies:2023hhn}%
  \BibitemOpen
  \bibfield  {author} {\bibinfo {author} {\bibfnamefont {M.~W.}\ \bibnamefont
  {Davies}}, \bibinfo {author} {\bibfnamefont {L.}~\bibnamefont {Iacconi}}, \
  and\ \bibinfo {author} {\bibfnamefont {D.~J.}\ \bibnamefont {Mulryne}},\
  }\href {\doibase 10.1088/1475-7516/2024/04/050} {\bibfield  {journal}
  {\bibinfo  {journal} {JCAP}\ }\textbf {\bibinfo {volume} {04}},\ \bibinfo
  {pages} {050} (\bibinfo {year} {2024})},\ \Eprint
  {http://arxiv.org/abs/2312.05694} {arXiv:2312.05694 [astro-ph.CO]}
  \BibitemShut {NoStop}%
\bibitem [{\citenamefont {Iacconi}\ \emph {et~al.}(2024)\citenamefont
  {Iacconi}, \citenamefont {Mulryne},\ and\ \citenamefont
  {Seery}}]{Iacconi:2023ggt}%
  \BibitemOpen
  \bibfield  {author} {\bibinfo {author} {\bibfnamefont {L.}~\bibnamefont
  {Iacconi}}, \bibinfo {author} {\bibfnamefont {D.}~\bibnamefont {Mulryne}}, \
  and\ \bibinfo {author} {\bibfnamefont {D.}~\bibnamefont {Seery}},\ }\href
  {\doibase 10.1088/1475-7516/2024/06/062} {\bibfield  {journal} {\bibinfo
  {journal} {JCAP}\ }\textbf {\bibinfo {volume} {06}},\ \bibinfo {pages} {062}
  (\bibinfo {year} {2024})},\ \Eprint {http://arxiv.org/abs/2312.12424}
  {arXiv:2312.12424 [astro-ph.CO]} \BibitemShut {NoStop}%
\bibitem [{\citenamefont {Saburov}\ and\ \citenamefont
  {Ketov}(2024)}]{Saburov:2024und}%
  \BibitemOpen
  \bibfield  {author} {\bibinfo {author} {\bibfnamefont {S.}~\bibnamefont
  {Saburov}}\ and\ \bibinfo {author} {\bibfnamefont {S.~V.}\ \bibnamefont
  {Ketov}},\ }\href {\doibase 10.3390/universe10090354} {\bibfield  {journal}
  {\bibinfo  {journal} {Universe}\ }\textbf {\bibinfo {volume} {10}},\ \bibinfo
  {pages} {354} (\bibinfo {year} {2024})},\ \Eprint
  {http://arxiv.org/abs/2402.02934} {arXiv:2402.02934 [gr-qc]} \BibitemShut
  {NoStop}%
\bibitem [{\citenamefont {Ballesteros}\ and\ \citenamefont
  {Egea}(2024)}]{Ballesteros:2024zdp}%
  \BibitemOpen
  \bibfield  {author} {\bibinfo {author} {\bibfnamefont {G.}~\bibnamefont
  {Ballesteros}}\ and\ \bibinfo {author} {\bibfnamefont {J.~G.}\ \bibnamefont
  {Egea}},\ }\href {\doibase 10.1088/1475-7516/2024/07/052} {\bibfield
  {journal} {\bibinfo  {journal} {JCAP}\ }\textbf {\bibinfo {volume} {07}},\
  \bibinfo {pages} {052} (\bibinfo {year} {2024})},\ \Eprint
  {http://arxiv.org/abs/2404.07196} {arXiv:2404.07196 [astro-ph.CO]}
  \BibitemShut {NoStop}%
\bibitem [{\citenamefont {Kristiano}\ and\ \citenamefont
  {Yokoyama}(2024{\natexlab{c}})}]{Kristiano:2024vst}%
  \BibitemOpen
  \bibfield  {author} {\bibinfo {author} {\bibfnamefont {J.}~\bibnamefont
  {Kristiano}}\ and\ \bibinfo {author} {\bibfnamefont {J.}~\bibnamefont
  {Yokoyama}},\ }\href {\doibase 10.1088/1475-7516/2024/10/036} {\bibfield
  {journal} {\bibinfo  {journal} {JCAP}\ }\textbf {\bibinfo {volume} {10}},\
  \bibinfo {pages} {036} (\bibinfo {year} {2024}{\natexlab{c}})},\ \Eprint
  {http://arxiv.org/abs/2405.12145} {arXiv:2405.12145 [astro-ph.CO]}
  \BibitemShut {NoStop}%
\bibitem [{\citenamefont {Kristiano}\ and\ \citenamefont
  {Yokoyama}(2024{\natexlab{d}})}]{Kristiano:2024ngc}%
  \BibitemOpen
  \bibfield  {author} {\bibinfo {author} {\bibfnamefont {J.}~\bibnamefont
  {Kristiano}}\ and\ \bibinfo {author} {\bibfnamefont {J.}~\bibnamefont
  {Yokoyama}},\ }\href@noop {} {\  (\bibinfo {year} {2024}{\natexlab{d}})},\
  \Eprint {http://arxiv.org/abs/2405.12149} {arXiv:2405.12149 [astro-ph.CO]}
  \BibitemShut {NoStop}%
\bibitem [{\citenamefont {Kawaguchi}\ \emph
  {et~al.}(2024{\natexlab{a}})\citenamefont {Kawaguchi}, \citenamefont
  {Tsujikawa},\ and\ \citenamefont {Yamada}}]{Kawaguchi:2024rsv}%
  \BibitemOpen
  \bibfield  {author} {\bibinfo {author} {\bibfnamefont {R.}~\bibnamefont
  {Kawaguchi}}, \bibinfo {author} {\bibfnamefont {S.}~\bibnamefont
  {Tsujikawa}}, \ and\ \bibinfo {author} {\bibfnamefont {Y.}~\bibnamefont
  {Yamada}},\ }\href {\doibase 10.1007/JHEP12(2024)095} {\bibfield  {journal}
  {\bibinfo  {journal} {JHEP}\ }\textbf {\bibinfo {volume} {12}},\ \bibinfo
  {pages} {095} (\bibinfo {year} {2024}{\natexlab{a}})},\ \Eprint
  {http://arxiv.org/abs/2407.19742} {arXiv:2407.19742 [hep-th]} \BibitemShut
  {NoStop}%
\bibitem [{\citenamefont {Fumagalli}(2025)}]{Fumagalli:2024jzz}%
  \BibitemOpen
  \bibfield  {author} {\bibinfo {author} {\bibfnamefont {J.}~\bibnamefont
  {Fumagalli}},\ }\href {\doibase 10.1007/JHEP01(2025)108} {\bibfield
  {journal} {\bibinfo  {journal} {JHEP}\ }\textbf {\bibinfo {volume} {01}},\
  \bibinfo {pages} {108} (\bibinfo {year} {2025})},\ \Eprint
  {http://arxiv.org/abs/2408.08296} {arXiv:2408.08296 [astro-ph.CO]}
  \BibitemShut {NoStop}%
\bibitem [{\citenamefont {Sheikhahmadi}\ and\ \citenamefont
  {Nassiri-Rad}(2024)}]{Sheikhahmadi:2024peu}%
  \BibitemOpen
  \bibfield  {author} {\bibinfo {author} {\bibfnamefont {H.}~\bibnamefont
  {Sheikhahmadi}}\ and\ \bibinfo {author} {\bibfnamefont {A.}~\bibnamefont
  {Nassiri-Rad}},\ }\href@noop {} {\  (\bibinfo {year} {2024})},\ \Eprint
  {http://arxiv.org/abs/2411.18525} {arXiv:2411.18525 [astro-ph.CO]}
  \BibitemShut {NoStop}%
\bibitem [{\citenamefont {Frolovsky}\ and\ \citenamefont
  {Ketov}(2025)}]{Frolovsky:2025qre}%
  \BibitemOpen
  \bibfield  {author} {\bibinfo {author} {\bibfnamefont {D.}~\bibnamefont
  {Frolovsky}}\ and\ \bibinfo {author} {\bibfnamefont {S.~V.}\ \bibnamefont
  {Ketov}},\ }\href@noop {} {\  (\bibinfo {year} {2025})},\ \Eprint
  {http://arxiv.org/abs/2502.00628} {arXiv:2502.00628 [gr-qc]} \BibitemShut
  {NoStop}%
\bibitem [{\citenamefont {Sasaki}\ and\ \citenamefont
  {Tanaka}(1998)}]{Sasaki:1998ug}%
  \BibitemOpen
  \bibfield  {author} {\bibinfo {author} {\bibfnamefont {M.}~\bibnamefont
  {Sasaki}}\ and\ \bibinfo {author} {\bibfnamefont {T.}~\bibnamefont
  {Tanaka}},\ }\href {\doibase 10.1143/PTP.99.763} {\bibfield  {journal}
  {\bibinfo  {journal} {Prog. Theor. Phys.}\ }\textbf {\bibinfo {volume}
  {99}},\ \bibinfo {pages} {763} (\bibinfo {year} {1998})},\ \Eprint
  {http://arxiv.org/abs/gr-qc/9801017} {arXiv:gr-qc/9801017} \BibitemShut
  {NoStop}%
\bibitem [{\citenamefont {Wands}\ \emph {et~al.}(2000)\citenamefont {Wands},
  \citenamefont {Malik}, \citenamefont {Lyth},\ and\ \citenamefont
  {Liddle}}]{Wands:2000dp}%
  \BibitemOpen
  \bibfield  {author} {\bibinfo {author} {\bibfnamefont {D.}~\bibnamefont
  {Wands}}, \bibinfo {author} {\bibfnamefont {K.~A.}\ \bibnamefont {Malik}},
  \bibinfo {author} {\bibfnamefont {D.~H.}\ \bibnamefont {Lyth}}, \ and\
  \bibinfo {author} {\bibfnamefont {A.~R.}\ \bibnamefont {Liddle}},\ }\href
  {\doibase 10.1103/PhysRevD.62.043527} {\bibfield  {journal} {\bibinfo
  {journal} {Phys. Rev. D}\ }\textbf {\bibinfo {volume} {62}},\ \bibinfo
  {pages} {043527} (\bibinfo {year} {2000})},\ \Eprint
  {http://arxiv.org/abs/astro-ph/0003278} {arXiv:astro-ph/0003278} \BibitemShut
  {NoStop}%
\bibitem [{\citenamefont {Lyth}\ and\ \citenamefont
  {Wands}(2003)}]{Lyth:2003im}%
  \BibitemOpen
  \bibfield  {author} {\bibinfo {author} {\bibfnamefont {D.~H.}\ \bibnamefont
  {Lyth}}\ and\ \bibinfo {author} {\bibfnamefont {D.}~\bibnamefont {Wands}},\
  }\href {\doibase 10.1103/PhysRevD.68.103515} {\bibfield  {journal} {\bibinfo
  {journal} {Phys. Rev. D}\ }\textbf {\bibinfo {volume} {68}},\ \bibinfo
  {pages} {103515} (\bibinfo {year} {2003})},\ \Eprint
  {http://arxiv.org/abs/astro-ph/0306498} {arXiv:astro-ph/0306498} \BibitemShut
  {NoStop}%
\bibitem [{\citenamefont {Inomata}(2024)}]{Inomata:2024lud}%
  \BibitemOpen
  \bibfield  {author} {\bibinfo {author} {\bibfnamefont {K.}~\bibnamefont
  {Inomata}},\ }\href {\doibase 10.1103/PhysRevLett.133.141001} {\bibfield
  {journal} {\bibinfo  {journal} {Phys. Rev. Lett.}\ }\textbf {\bibinfo
  {volume} {133}},\ \bibinfo {pages} {141001} (\bibinfo {year} {2024})},\
  \Eprint {http://arxiv.org/abs/2403.04682} {arXiv:2403.04682 [astro-ph.CO]}
  \BibitemShut {NoStop}%
\bibitem [{\citenamefont {Baumann}\ and\ \citenamefont
  {Green}(2011)}]{Baumann:2011su}%
  \BibitemOpen
  \bibfield  {author} {\bibinfo {author} {\bibfnamefont {D.}~\bibnamefont
  {Baumann}}\ and\ \bibinfo {author} {\bibfnamefont {D.}~\bibnamefont
  {Green}},\ }\href {\doibase 10.1088/1475-7516/2011/09/014} {\bibfield
  {journal} {\bibinfo  {journal} {JCAP}\ }\textbf {\bibinfo {volume} {09}},\
  \bibinfo {pages} {014} (\bibinfo {year} {2011})},\ \Eprint
  {http://arxiv.org/abs/1102.5343} {arXiv:1102.5343 [hep-th]} \BibitemShut
  {NoStop}%
\bibitem [{\citenamefont {Pajer}\ \emph {et~al.}(2017)\citenamefont {Pajer},
  \citenamefont {Pimentel},\ and\ \citenamefont {Van~Wijck}}]{Pajer:2016ieg}%
  \BibitemOpen
  \bibfield  {author} {\bibinfo {author} {\bibfnamefont {E.}~\bibnamefont
  {Pajer}}, \bibinfo {author} {\bibfnamefont {G.~L.}\ \bibnamefont {Pimentel}},
  \ and\ \bibinfo {author} {\bibfnamefont {J.~V.~S.}\ \bibnamefont
  {Van~Wijck}},\ }\href {\doibase 10.1088/1475-7516/2017/06/009} {\bibfield
  {journal} {\bibinfo  {journal} {JCAP}\ }\textbf {\bibinfo {volume} {06}},\
  \bibinfo {pages} {009} (\bibinfo {year} {2017})},\ \Eprint
  {http://arxiv.org/abs/1609.06993} {arXiv:1609.06993 [hep-th]} \BibitemShut
  {NoStop}%
\bibitem [{\citenamefont {Inomata}\ \emph {et~al.}(2023)\citenamefont
  {Inomata}, \citenamefont {Braglia}, \citenamefont {Chen},\ and\ \citenamefont
  {Renaux-Petel}}]{Inomata:2022yte}%
  \BibitemOpen
  \bibfield  {author} {\bibinfo {author} {\bibfnamefont {K.}~\bibnamefont
  {Inomata}}, \bibinfo {author} {\bibfnamefont {M.}~\bibnamefont {Braglia}},
  \bibinfo {author} {\bibfnamefont {X.}~\bibnamefont {Chen}}, \ and\ \bibinfo
  {author} {\bibfnamefont {S.}~\bibnamefont {Renaux-Petel}},\ }\href {\doibase
  10.1088/1475-7516/2023/04/011} {\bibfield  {journal} {\bibinfo  {journal}
  {JCAP}\ }\textbf {\bibinfo {volume} {04}},\ \bibinfo {pages} {011} (\bibinfo
  {year} {2023})},\ \bibinfo {note} {[Erratum: JCAP 09, E01 (2023)]},\ \Eprint
  {http://arxiv.org/abs/2211.02586} {arXiv:2211.02586 [astro-ph.CO]}
  \BibitemShut {NoStop}%
\bibitem [{\citenamefont {Weinberg}(2005)}]{Weinberg:2005vy}%
  \BibitemOpen
  \bibfield  {author} {\bibinfo {author} {\bibfnamefont {S.}~\bibnamefont
  {Weinberg}},\ }\href {\doibase 10.1103/PhysRevD.72.043514} {\bibfield
  {journal} {\bibinfo  {journal} {Phys. Rev. D}\ }\textbf {\bibinfo {volume}
  {72}},\ \bibinfo {pages} {043514} (\bibinfo {year} {2005})},\ \Eprint
  {http://arxiv.org/abs/hep-th/0506236} {arXiv:hep-th/0506236} \BibitemShut
  {NoStop}%
\bibitem [{\citenamefont {Arroja}\ and\ \citenamefont
  {Tanaka}(2011)}]{Arroja:2011yj}%
  \BibitemOpen
  \bibfield  {author} {\bibinfo {author} {\bibfnamefont {F.}~\bibnamefont
  {Arroja}}\ and\ \bibinfo {author} {\bibfnamefont {T.}~\bibnamefont
  {Tanaka}},\ }\href {\doibase 10.1088/1475-7516/2011/05/005} {\bibfield
  {journal} {\bibinfo  {journal} {JCAP}\ }\textbf {\bibinfo {volume} {05}},\
  \bibinfo {pages} {005} (\bibinfo {year} {2011})},\ \Eprint
  {http://arxiv.org/abs/1103.1102} {arXiv:1103.1102 [astro-ph.CO]} \BibitemShut
  {NoStop}%
\bibitem [{\citenamefont {Kawaguchi}\ \emph
  {et~al.}(2024{\natexlab{b}})\citenamefont {Kawaguchi}, \citenamefont
  {Tsujikawa},\ and\ \citenamefont {Yamada}}]{Kawaguchi:2024lsw}%
  \BibitemOpen
  \bibfield  {author} {\bibinfo {author} {\bibfnamefont {R.}~\bibnamefont
  {Kawaguchi}}, \bibinfo {author} {\bibfnamefont {S.}~\bibnamefont
  {Tsujikawa}}, \ and\ \bibinfo {author} {\bibfnamefont {Y.}~\bibnamefont
  {Yamada}},\ }\href {\doibase 10.1016/j.physletb.2024.138962} {\bibfield
  {journal} {\bibinfo  {journal} {Phys. Lett. B}\ }\textbf {\bibinfo {volume}
  {856}},\ \bibinfo {pages} {138962} (\bibinfo {year} {2024}{\natexlab{b}})},\
  \Eprint {http://arxiv.org/abs/2403.16022} {arXiv:2403.16022 [hep-th]}
  \BibitemShut {NoStop}%
\bibitem [{\citenamefont {Giddings}\ and\ \citenamefont
  {Sloth}(2011)}]{Giddings:2010nc}%
  \BibitemOpen
  \bibfield  {author} {\bibinfo {author} {\bibfnamefont {S.~B.}\ \bibnamefont
  {Giddings}}\ and\ \bibinfo {author} {\bibfnamefont {M.~S.}\ \bibnamefont
  {Sloth}},\ }\href {\doibase 10.1088/1475-7516/2011/01/023} {\bibfield
  {journal} {\bibinfo  {journal} {JCAP}\ }\textbf {\bibinfo {volume} {01}},\
  \bibinfo {pages} {023} (\bibinfo {year} {2011})},\ \Eprint
  {http://arxiv.org/abs/1005.1056} {arXiv:1005.1056 [hep-th]} \BibitemShut
  {NoStop}%
\bibitem [{\citenamefont {Byrnes}\ \emph {et~al.}(2010)\citenamefont {Byrnes},
  \citenamefont {Gerstenlauer}, \citenamefont {Hebecker}, \citenamefont
  {Nurmi},\ and\ \citenamefont {Tasinato}}]{Byrnes:2010yc}%
  \BibitemOpen
  \bibfield  {author} {\bibinfo {author} {\bibfnamefont {C.~T.}\ \bibnamefont
  {Byrnes}}, \bibinfo {author} {\bibfnamefont {M.}~\bibnamefont
  {Gerstenlauer}}, \bibinfo {author} {\bibfnamefont {A.}~\bibnamefont
  {Hebecker}}, \bibinfo {author} {\bibfnamefont {S.}~\bibnamefont {Nurmi}}, \
  and\ \bibinfo {author} {\bibfnamefont {G.}~\bibnamefont {Tasinato}},\ }\href
  {\doibase 10.1088/1475-7516/2010/08/006} {\bibfield  {journal} {\bibinfo
  {journal} {JCAP}\ }\textbf {\bibinfo {volume} {08}},\ \bibinfo {pages} {006}
  (\bibinfo {year} {2010})},\ \Eprint {http://arxiv.org/abs/1005.3307}
  {arXiv:1005.3307 [hep-th]} \BibitemShut {NoStop}%
\bibitem [{\citenamefont {Inomata}\ \emph {et~al.}(2021)\citenamefont
  {Inomata}, \citenamefont {McDonough},\ and\ \citenamefont
  {Hu}}]{Inomata:2021uqj}%
  \BibitemOpen
  \bibfield  {author} {\bibinfo {author} {\bibfnamefont {K.}~\bibnamefont
  {Inomata}}, \bibinfo {author} {\bibfnamefont {E.}~\bibnamefont {McDonough}},
  \ and\ \bibinfo {author} {\bibfnamefont {W.}~\bibnamefont {Hu}},\ }\href
  {\doibase 10.1103/PhysRevD.104.123553} {\bibfield  {journal} {\bibinfo
  {journal} {Phys. Rev. D}\ }\textbf {\bibinfo {volume} {104}},\ \bibinfo
  {pages} {123553} (\bibinfo {year} {2021})},\ \Eprint
  {http://arxiv.org/abs/2104.03972} {arXiv:2104.03972 [astro-ph.CO]}
  \BibitemShut {NoStop}%
\bibitem [{\citenamefont {Inomata}\ \emph {et~al.}(2022)\citenamefont
  {Inomata}, \citenamefont {McDonough},\ and\ \citenamefont
  {Hu}}]{Inomata:2021tpx}%
  \BibitemOpen
  \bibfield  {author} {\bibinfo {author} {\bibfnamefont {K.}~\bibnamefont
  {Inomata}}, \bibinfo {author} {\bibfnamefont {E.}~\bibnamefont {McDonough}},
  \ and\ \bibinfo {author} {\bibfnamefont {W.}~\bibnamefont {Hu}},\ }\href
  {\doibase 10.1088/1475-7516/2022/02/031} {\bibfield  {journal} {\bibinfo
  {journal} {JCAP}\ }\textbf {\bibinfo {volume} {02}},\ \bibinfo {pages} {031}
  (\bibinfo {year} {2022})},\ \Eprint {http://arxiv.org/abs/2110.14641}
  {arXiv:2110.14641 [astro-ph.CO]} \BibitemShut {NoStop}%
\bibitem [{\citenamefont {Iacconi}\ and\ \citenamefont
  {Mulryne}(2023)}]{Iacconi:2023slv}%
  \BibitemOpen
  \bibfield  {author} {\bibinfo {author} {\bibfnamefont {L.}~\bibnamefont
  {Iacconi}}\ and\ \bibinfo {author} {\bibfnamefont {D.~J.}\ \bibnamefont
  {Mulryne}},\ }\href {\doibase 10.1088/1475-7516/2023/09/033} {\bibfield
  {journal} {\bibinfo  {journal} {JCAP}\ }\textbf {\bibinfo {volume} {09}},\
  \bibinfo {pages} {033} (\bibinfo {year} {2023})},\ \Eprint
  {http://arxiv.org/abs/2304.14260} {arXiv:2304.14260 [astro-ph.CO]}
  \BibitemShut {NoStop}%
\bibitem [{\citenamefont {Fumagalli}\ \emph {et~al.}(2024)\citenamefont
  {Fumagalli}, \citenamefont {Bhattacharya}, \citenamefont {Peloso},
  \citenamefont {Renaux-Petel},\ and\ \citenamefont
  {Witkowski}}]{Fumagalli:2023loc}%
  \BibitemOpen
  \bibfield  {author} {\bibinfo {author} {\bibfnamefont {J.}~\bibnamefont
  {Fumagalli}}, \bibinfo {author} {\bibfnamefont {S.}~\bibnamefont
  {Bhattacharya}}, \bibinfo {author} {\bibfnamefont {M.}~\bibnamefont
  {Peloso}}, \bibinfo {author} {\bibfnamefont {S.}~\bibnamefont
  {Renaux-Petel}}, \ and\ \bibinfo {author} {\bibfnamefont {L.~T.}\
  \bibnamefont {Witkowski}},\ }\href {\doibase 10.1088/1475-7516/2024/04/029}
  {\bibfield  {journal} {\bibinfo  {journal} {JCAP}\ }\textbf {\bibinfo
  {volume} {04}},\ \bibinfo {pages} {029} (\bibinfo {year} {2024})},\ \Eprint
  {http://arxiv.org/abs/2307.08358} {arXiv:2307.08358 [astro-ph.CO]}
  \BibitemShut {NoStop}%
\bibitem [{\citenamefont {Caravano}\ \emph
  {et~al.}(2024{\natexlab{a}})\citenamefont {Caravano}, \citenamefont
  {Inomata},\ and\ \citenamefont {Renaux-Petel}}]{Caravano:2024tlp}%
  \BibitemOpen
  \bibfield  {author} {\bibinfo {author} {\bibfnamefont {A.}~\bibnamefont
  {Caravano}}, \bibinfo {author} {\bibfnamefont {K.}~\bibnamefont {Inomata}}, \
  and\ \bibinfo {author} {\bibfnamefont {S.}~\bibnamefont {Renaux-Petel}},\
  }\href {\doibase 10.1103/PhysRevLett.133.151001} {\bibfield  {journal}
  {\bibinfo  {journal} {Phys. Rev. Lett.}\ }\textbf {\bibinfo {volume} {133}},\
  \bibinfo {pages} {151001} (\bibinfo {year} {2024}{\natexlab{a}})},\ \Eprint
  {http://arxiv.org/abs/2403.12811} {arXiv:2403.12811 [astro-ph.CO]}
  \BibitemShut {NoStop}%
\bibitem [{\citenamefont {Caravano}\ \emph
  {et~al.}(2024{\natexlab{b}})\citenamefont {Caravano}, \citenamefont
  {Franciolini},\ and\ \citenamefont {Renaux-Petel}}]{Caravano:2024moy}%
  \BibitemOpen
  \bibfield  {author} {\bibinfo {author} {\bibfnamefont {A.}~\bibnamefont
  {Caravano}}, \bibinfo {author} {\bibfnamefont {G.}~\bibnamefont
  {Franciolini}}, \ and\ \bibinfo {author} {\bibfnamefont {S.}~\bibnamefont
  {Renaux-Petel}},\ }\href@noop {} {\  (\bibinfo {year}
  {2024}{\natexlab{b}})},\ \Eprint {http://arxiv.org/abs/2410.23942}
  {arXiv:2410.23942 [astro-ph.CO]} \BibitemShut {NoStop}%
\bibitem [{\citenamefont {Tanaka}\ and\ \citenamefont
  {Urakawa}(2013)}]{Tanaka:2013caa}%
  \BibitemOpen
  \bibfield  {author} {\bibinfo {author} {\bibfnamefont {T.}~\bibnamefont
  {Tanaka}}\ and\ \bibinfo {author} {\bibfnamefont {Y.}~\bibnamefont
  {Urakawa}},\ }\href {\doibase 10.1088/0264-9381/30/23/233001} {\bibfield
  {journal} {\bibinfo  {journal} {Class. Quant. Grav.}\ }\textbf {\bibinfo
  {volume} {30}},\ \bibinfo {pages} {233001} (\bibinfo {year} {2013})},\
  \Eprint {http://arxiv.org/abs/1306.4461} {arXiv:1306.4461 [hep-th]}
  \BibitemShut {NoStop}%
\bibitem [{\citenamefont {Maldacena}(2003)}]{Maldacena:2002vr}%
  \BibitemOpen
  \bibfield  {author} {\bibinfo {author} {\bibfnamefont {J.~M.}\ \bibnamefont
  {Maldacena}},\ }\href {\doibase 10.1088/1126-6708/2003/05/013} {\bibfield
  {journal} {\bibinfo  {journal} {JHEP}\ }\textbf {\bibinfo {volume} {05}},\
  \bibinfo {pages} {013} (\bibinfo {year} {2003})},\ \Eprint
  {http://arxiv.org/abs/astro-ph/0210603} {arXiv:astro-ph/0210603} \BibitemShut
  {NoStop}%
\bibitem [{\citenamefont {Hinterbichler}\ \emph {et~al.}(2012)\citenamefont
  {Hinterbichler}, \citenamefont {Hui},\ and\ \citenamefont
  {Khoury}}]{Hinterbichler:2012nm}%
  \BibitemOpen
  \bibfield  {author} {\bibinfo {author} {\bibfnamefont {K.}~\bibnamefont
  {Hinterbichler}}, \bibinfo {author} {\bibfnamefont {L.}~\bibnamefont {Hui}},
  \ and\ \bibinfo {author} {\bibfnamefont {J.}~\bibnamefont {Khoury}},\ }\href
  {\doibase 10.1088/1475-7516/2012/08/017} {\bibfield  {journal} {\bibinfo
  {journal} {JCAP}\ }\textbf {\bibinfo {volume} {08}},\ \bibinfo {pages} {017}
  (\bibinfo {year} {2012})},\ \Eprint {http://arxiv.org/abs/1203.6351}
  {arXiv:1203.6351 [hep-th]} \BibitemShut {NoStop}%
\bibitem [{\citenamefont {Peskin}\ and\ \citenamefont
  {Schroeder}(1995)}]{Peskin:1995ev}%
  \BibitemOpen
  \bibfield  {author} {\bibinfo {author} {\bibfnamefont {M.~E.}\ \bibnamefont
  {Peskin}}\ and\ \bibinfo {author} {\bibfnamefont {D.~V.}\ \bibnamefont
  {Schroeder}},\ }\href {\doibase 10.1201/9780429503559} {\emph {\bibinfo
  {title} {{An Introduction to quantum field theory}}}}\ (\bibinfo  {publisher}
  {Addison-Wesley},\ \bibinfo {address} {Reading, USA},\ \bibinfo {year}
  {1995})\BibitemShut {NoStop}%
\end{thebibliography}%

\end{document}